\documentclass[11pt]{article}

\usepackage{fullpage}
\usepackage{latexsym}
\usepackage{amssymb,amsfonts,amsmath,amsthm}
\usepackage{graphicx}
\usepackage{subfigure}

\def\01{\{0,1\}}
\newcommand{\ceil}[1]{\lceil{#1}\rceil}

\newcommand{\eps}{\epsilon}
\newcommand{\IP}{\mbox{\rm IP}}
\newcommand{\EQ}{\mbox{\rm EQ}}
\newcommand{\INT}{\mbox{\rm INT}}
\newcommand{\Exp}{{\mathbb{E}}}
\newcommand{\ket}[1]{|#1\rangle}
\newcommand{\bra}[1]{\langle#1|}
\newcommand{\ketbra}[2]{|#1\rangle\langle#2|}
\newcommand{\norm}[1]{{\left\|{#1}\right\|}}
\newcommand{\inp}[2]{\langle{#1}|{#2}\rangle} 
\newcommand{\inpc}[2]{\langle{#1},{#2}\rangle} 
\newcommand{\Tr}{\mbox{\rm Tr}}
\newcommand{\rank}{\mbox{\rm rank}}

\newtheorem{theorem}{Theorem}
\newtheorem{lemma}[theorem]{Lemma}

\renewcommand{\qed}{\hfill{\rule{2mm}{2mm}}}
\renewenvironment{proof}[1][]{\begin{trivlist}
\item[\hspace{\labelsep}{\bf\noindent Proof#1:\/}] }{\qed\end{trivlist}}

\begin{document}
\title{Non-locality and Communication Complexity}
\author{Harry Buhrman\thanks{CWI and University of Amsterdam. Partially supported by a Vici grant from the Netherlands Organization for Scientific Research (NWO), and by the European Commission under the Integrated Project Qubit Applications (QAP) funded by the IST directorate as Contract Number 015848.}
\and
Richard Cleve\thanks{Institute for Quantum Computing and School of Computer Science, University of Waterloo, and Perimeter Institute for Theoretical Physics.
Partially supported by Canada's NSERC, CIFAR, QuantumWorks, MITACS, and the U.S. ARO.}
\and
Serge Massar\thanks{Laboratoire d'Information Quantique, CP 225,
 Universit\'{e} Libre de Bruxelles (U.L.B.),
 Boulevard du Triomphe, B-1050 Bruxelles, Belgium.
 Partially supported by the Interuniversity Attraction Poles Programme -
Belgian State - Belgian Science Policy under grant IAP6-10 and by the EU project QAP contract 015848.}
\and
Ronald de Wolf\thanks{CWI Amsterdam. Partially supported by a Vidi grant from the Netherlands Organization for Scientific Research (NWO), and by the European Commission under the Integrated Project Qubit Applications (QAP) funded by the IST directorate as Contract Number 015848.}
}
\maketitle

\begin{abstract}
  Quantum information processing is the emerging field that defines
  and realizes
  computing devices that make use of quantum mechanical principles,
  like the superposition principle, entanglement, and interference.
Until recently the common notion of computing was based
on classical mechanics, and did not take into account all the
 possibilities that physically-realizable
  computing devices offer in principle. The field gained momentum
  after Peter Shor developed an efficient algorithm for factoring
  numbers, demonstrating the potential computing powers that quantum
  computing devices can unleash.

  In this review we study the information counterpart of computing. It
  was realized early on by Holevo, that quantum bits, the quantum
  mechanical counterpart of classical bits, cannot be used for
  efficient transformation of information, in the sense that arbitrary
  $k$-bit messages can not be compressed into messages of $k-1$ qubits.

The abstract form of the
distributed computing setting is called \emph{communication complexity}.
It studies the amount of information, in terms of bits or in our case qubits, that two
spatially separated
  computing devices need to exchange in order to perform some computational task.
  Surprisingly, quantum mechanics can be used to obtain dramatic advantages for such tasks.

  We review the area of quantum communication complexity, and show how
  it connects the foundational physics questions regarding non-locality
  with those of communication complexity studied in
  theoretical computer science. The first examples exhibiting the
  advantage of the use of qubits in distributed information-processing
  tasks were based on non-locality tests. However, by now the field has produced
  strong and interesting quantum protocols and algorithms of its own
  that demonstrate that entanglement, although it cannot be used to
  replace communication, can be used to reduce the
  communication \emph{exponentially}. In turn, these new advances yield a new outlook on the foundations of physics, and could even yield new proposals for experiments that test the foundations of physics.

\end{abstract}

\tableofcontents

\section{Introduction}

\subsection{Background}

During the last decades of the twentieth
century it was realized that information processing at the quantum level could offer tremendous advantages over conventional ``classical'' information processing. Quantum information admits extremely efficient
algorithms, such as Shor's factoring algorithm~\cite{shor:factoring}, and qualitatively superior cryptographic
protocols, such as the BB84 key distribution protocol~\cite{bb84}. Many other works contributed to put this field on solid foundations. Quantum error-correcting codes and fault-tolerant quantum computation showed that these beautiful ideas could in principle be realized experimentally. These codes, combined with Holevo's Theorem, Schumacher compression, and entanglement distillation (which are analogs of Shannon's noiseless coding theorem) gave us the foundations of an information theory
pertaining to quantum systems in terms of quantum bits, or \textit{qubits},
and entanglement that is measured (in the bipartite case) in entanglement bits, or \textit{ebits}.
These discoveries generated huge excitement. By now quantum
information has become a well-established field, and there are many
reviews and textbooks to which we refer the reader for background
information. See for example~\cite{nielsen&chuang:qc}.

In view of the advantages that quantum information offers for computation and cryptography, it is natural to
enquire whether quantum information is also a superior
medium for efficient communication. In this article we will review progress on this specific question,  and its relation to the problem of quantum non-locality which has fascinated physicists for decades.

On the face of it, there are important reasons for doubting that quantum information provides such a communication
efficiency advantage.
Many years before the ``quantum information'' discipline took hold on
a large scale, Holevo~\cite{holevo} proved an important theorem about the classical
information capacity of quantum channels.
Holevo's Theorem---as it is now called---states
that, for any classical message, the cost of transmitting it from one party (Alice)
to another party (Bob) in terms of
quantum bits (\textit{qu}bits) is the same as the cost of transmitting it in terms of
classical bits. If the task requires $k$ bits on average, then it also
requires $k$ qubits on average. The latter consequence of Holevo's Theorem can be proven quite simply using a different approach~\cite{nayak:qfa}, and this proof is reproduced in
Appendix~\ref{nayakbound}.
Thus one would naively expect that quantum information cannot provide a communication efficiency advantage.
This intuition turns out to be wrong. Tremendous communication savings are
possible with the use of quantum information, as explained in the next section.

\subsection{Communication complexity}

To understand why quantum information can provide a communication advantage
without contradicting Holevo's Theorem, it is
necessary to consider more precisely the various scenarios that can be
associated with ``communication''.

\begin{figure}[h]
\setlength{\unitlength}{100000sp}
\vspace{2mm}
\begin{center}
\begin{picture}(180,70)(0,0)
\put(60,35){\circle{30}}
\put(51.5,33){Alice}
\put(58,64){$x \in \{0,1\}^n$}
\put(60,60){\vector(0,-1){10}}
\put(140,35){\circle{30}}
\put(133.5,33){Bob}
\put(138,2){$x$}
\put(140,20){\vector(0,-1){10}}
\put(75,35){\vector(1,0){50}}
\put(0,64){Input:}
\put(0,2){Output:}
\put(76,27){communication}
\end{picture}
\end{center}
\caption{The basic \textit{communication} scenario: Alice receives an
$n$-bit string $x$ as input and sends one message to Bob, who must output $x$.
For this task, a quantum message is no more efficient than a classical message.}\label{fig:comm}
\end{figure}
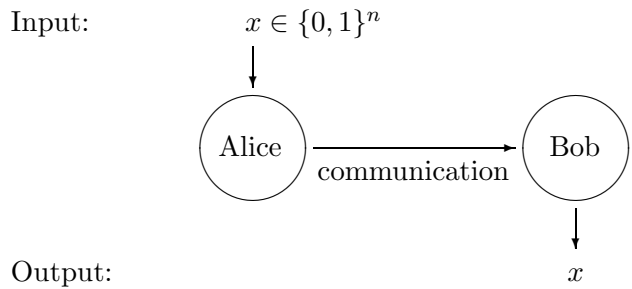

The simplest scenario, corresponding to the case covered by Holevo's Theorem,
is illustrated in Fig.~\ref{fig:comm}.
There are two parties that we refer to as Alice and Bob.
Alice has an $n$-bit string $x$ that she would like to convey to Bob
by sending one message.
Here it is indeed true, by Holevo's Theorem~\cite{holevo}, that quantum
messages are no more efficient than classical messages.
Alice must send $n$ qubits to accomplish this specific task.

A variant of the communication scenario is where Bob's goal is not
to determine Alice's data~$x$, but to determine
some information that is a \emph{function} of $x$ in a way that may depend on
other data~$y$ that resides with Bob (while $y$ is unknown to Alice).
Such a scenario could occur when Alice and Bob each begin with
$n$-bit strings, $x$ and $y$, respectively (Alice knows $x$ but not $y$
and Bob knows $y$ but not $x$), and the goal is for Bob to determine
the value of some function $f(x,y)$ (where $f$ is known to both parties).
An example where such a scenario could arise is where Alice and Bob are
interested in scheduling an appointment.
Alice's schedule could be represented by~$x$ and Bob's by~$y$: if there are $n$ time-slots,
then we can set the $i$th bit of $x$ to 1 if Alice is available in time-slot~$i$, and similarly for~$y$.
How much communication is required for Bob to find a time when they
are both available (i.e., an $i$ such that $x_i=y_i=1$)? We shall see that, for \textit{this} communication scenario, quantum
information enables Alice and Bob to accomplish the task with less
(asymptotically less in the number of time-slots) qubit communication than would be required by
any protocol that is restricted to classical bit communication.

\begin{figure}[h]
\setlength{\unitlength}{100000sp}
\vspace{2mm}
\begin{center}
\begin{picture}(180,70)(0,0)
\put(60,35){\circle{30}}
\put(51.5,33){Alice}
\put(58,64){$x \in \{0,1\}^n$}
\put(60,60){\vector(0,-1){10}}
\put(140,35){\circle{30}}
\put(133.5,33){Bob}
\put(138,64){$y \in \{0,1\}^n$}
\put(130,2){$f(x,y)$}
\put(140,60){\vector(0,-1){10}}
\put(140,20){\vector(0,-1){10}}
\put(75,39){\vector(1,0){50}}
\put(125,35){\vector(-1,0){50}}
\put(75,31){\vector(1,0){50}}
\put(76,24){communication}
\put(0,64){Inputs:}
\put(0,2){Output:}
\end{picture}
\end{center}
\caption{The basic \textit{communication complexity} scenario: Alice and Bob
receive $n$-bit strings, $x$ and $y$ respectively, as input and their goal to
compute some function of these values $f(x,y)$, as Bob's output.
There are tasks of this form where communication in terms of quantum messages
is much more efficient than communication in terms of classical messages.
The number of qubits can be exponentially smaller than the number of
bits. Note that in this framework we do not take into account the time
and other resources that Alice and Bob spend locally
(although in practice it turns out
that their local computations are almost always efficient).}\label{fig:cc}
\end{figure}
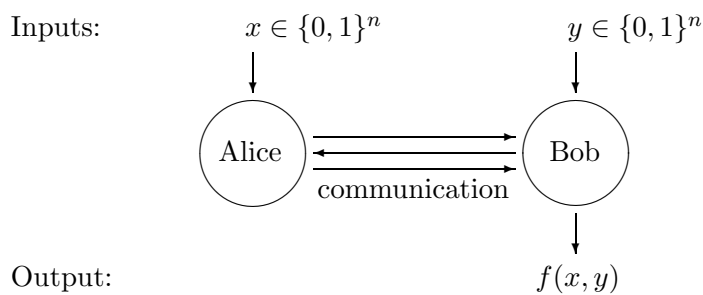

This kind of scenario, illustrated in Fig.~\ref{fig:cc} (for general
functions or relations $f$ on $\{0,1\}^n \times \{0,1\}^n$) is known as
\textit{communication complexity}.
It has been extensively studied in the classical case. Indeed, whereas the trivial solution to this problem is for Alice to send Bob her input $x$, and for Bob to compute $f(x,y)$, it is often possible for Bob to compute $f$ with much less than $n$ bits of classical communication.
These savings in classical communication are very interesting both from a practical and a conceptual point of view.
Section~\ref{sec:cc} outlines
several of the key results in the area, and we refer the reader to the
textbooks~\cite{kushilevitz&nisan:cc,Hromkovic97} for further information.

When Alice and Bob can communicate qubits, further reductions in the amount of communication are possible, sometimes even exponential reductions. This remarkable situation is clearly worthy of further study. It is one of the main subjects covered by the present review, and we will see many examples later.

\subsection{Quantum non-locality}

Long before the work on quantum communication complexity mentioned in the previous
section, physicists investigating the foundations of quantum mechanics studied the scenario where local measurements are carried out on two entangled particles. Such entangled states can (at least in principle) be easily produced by having the particles interact together for some time, and then sending the particles away to far-off locations. Local measurements are then carried out on the particles.
This scenario was first studied by Einstein, Podolsky, and Rosen~\cite{epr} and immediately afterwards by Schr{\"o}dinger~\cite{schrodinger:35,schrodinger:36} (who coined the word {\em entanglement}). In these works it was realized that the results of the local measurements would exhibit very interesting correlations. For instance, for some pairs of the measurements, the results may be always the same; for other pairs of measurements, the results may be always opposite, etc.

Nevertheless, one can easily show---this follows immediately from the structure of quantum mechanics---that the parties carrying out the measurements cannot use the entangled particles to communicate to each other.
More precisely, if two physically separated parties, Alice and Bob,
initially possess entangled particles and then Alice is given an
arbitrary bit $x$, there is no way for Alice to manipulate her
particles in order to convey any information about $x$ to Bob when he performs measurements on his particles.

Given that these correlations cannot be used for communication, one would naively expect that if a (quantum or classical) model can reproduce these correlations,
then it is not necessary for that model to use communication.
This is indeed the case in the quantum scenario where, having established the
entanglement through some interaction in the past, no communication is needed
at the time of the measurement.
But if one wants to reproduce these correlations in a purely classical model,
then classical communication between the parties is required at the moment of
the measurements!
This situation is even more surprising if the particles are widely separated
from each other and the measurements take place during a very short time interval,
so short that the two measurement events are space-like separated.
In this case the communication would have to occur faster than the speed of
light!

This remarkable feature of quantum mechanics was discovered by Bell~\cite{bell:epr}, and is now known as ``quantum non-locality''. It has been the subject of much further theoretical and experimental study since. Indeed it is one of the most surprising and counter-intuitive features of quantum mechanics. Bell's Theorem shows that Einstein's program of trying to rationalize quantum mechanics by reducing it to classical mechanics is futile and doomed to failure, as it cannot be done without giving up another cornerstone of twentieth century physics (discovered by Einstein himself), namely the fact that information cannot travel faster than the speed of light. More recently, another reason why such a reduction is doomed emerged through the study of quantum information. Namely we expect any such classical description of quantum mechanics to be exponentially inefficient, i.e., to use \emph{exponentially} more resources than the quantum theory. We will discuss quantum non-locality extensively in the present review, focusing on its connection to communication complexity.

\subsection{Unity of quantum communication complexity and quantum non-locality}

The reason why in this review we deal with quantum communication complexity and
quantum non-locality together is that these two topics are intimately related.
Indeed they can be formulated in a unified way, and furthermore many questions can
be mapped from one topic to the other. In fact, during the past dozen years an
intense cross-fertilization has occurred between these two fields, which has
considerably enriched both of them.

To see the unity between the two subjects, recall that in both cases the parties, Alice and Bob, are given some inputs, $x$ and $y$. In one case these inputs correspond to the arguments of the function that must be computed. In the other case these inputs correspond to a description of the measurements that must be carried out on the particles (the ``measurement settings''). And in both cases Alice and Bob must provide an output, $a$ and $b$. In communication complexity we require that $b=f(x,y)$ and $a$ is irrelevant; in non-locality we are interested in the correlations between $a$, $b$ and $x$, $y$ (for instance we request that $a=b$ when $x$ and $y$ have certain values and that $a\neq b$ when $x$ and $y$ have some other values). We can unify these descriptions by saying that the aim in both cases is to produce a joint probability distribution
$$P(a,b|x,y)$$
 of the outputs given the inputs, such that $P(a,b|x,y)$ has certain desirable properties.

\subsection{Resources}

In both communication and non-locality, the basic question one wants to answer is: what is the minimum amount of resources necessary to reproduce the distribution $P(a,b|x,y)$, and how does this amount change when one changes the model, i.e., when one changes the type of resource that can be used. There are in fact many different types of resources that can be compared, and we now briefly review them. We will come back to them in more detail in the body of the review.

\begin{itemize}
\item \textit{Quantum communication.} The parties are allowed to send each other quantum states. One quantifies the amount of communication by the number of \textit{qubits} sent.
\item \textit{Classical communication.} The parties are allowed to send each other classical communication. One quantifies the amount of communication by the number of \textit{bits} sent.
\item \textit{Entanglement.} The parties share entangled states. One
  quantifies the amount of entanglement by the number of qubits that
  the state locally consists of.
  For example  we frequently use
  maximally entangled states of 2 qubits, called \textit{ebits} (also known as EPR pairs after~\cite{epr}),
 $\frac{1}{\sqrt{2}}(\ket{0}\ket{0} + \ket{1}\ket{1})$ or
  something that can be obtained from this with local operations.

\item \textit{Shared randomness.} The parties have randomness, i.e., they are allowed to toss coins. In the case of \emph{shared} randomness, the parties both share the same string of coins. This could for instance be implemented by having the parties toss the coins beforehand, at some earlier time when they are together, and then use the coins later when they need to solve the communication complexity problem.
\item \textit{Local randomness.} The parties have randomness,
  i.e., they are allowed to toss coins. In the case of \emph{local} randomness
  the coins are tossed locally, and the string of outcomes of the
  coins for Alice is independent of the string of outcomes of the coins for Bob.
\end{itemize}

The rational for measuring classical information in terms of bits is Shannon's noiseless coding theorem~\cite{shannon:communication}, which states that, asymptotically, the information produced by a stochastic source can be encoded in a number of bits equal to the entropy of the source. This is paralleled in the quantum case by Schumacher compression~\cite{schumacher:95}, which states that, asymptotically, the information produced by a stochastic quantum source can be encoded into a number of qubits equal to the von Neumann entropy of the source. And it is paralleled in the case of entanglement, by entanglement distillations, namely the fact that pure two-party entangled states can, asymptotically in the number of copies of the state, be converted into the number of ebits equal to the von Neumann entropy of the reduced density matrix of each party~\cite{bbps:entanglementconcentration}. In the context of communication complexity, however, we are not dealing with the asymptotic limit of large amounts of communication or large amounts of entanglement. Thus whereas in most cases we will keep the basic concepts of bits, qubits and ebits, it could be relevant in specific cases to consider variants on these resources, such as trits, non-maximally entangled states, etc.

The above resources have been ordered (more or less) from the strongest to the weakest.
Indeed most of these resources imply the ones below them. For instance one can send classical information using qubits; one can use quantum communication to distribute entanglement; one can measure the entangled particles to produce shared randomness, etc.
The only case where the ordering is not so clear is between classical
communication and entanglement. Indeed if two parties share an
entangled state, they cannot use it to communicate (as discussed
above). But on the other hand (as discussed below)
sharing $n$ ebits may allow one to save an exponentially large (in
$n$) amount of bits in some communication scenarios (whereas in all
other cases, $n$ uses of one resource allows one to implement $n$ uses
of the resources below it).

There are also a number of nontrivial ways in which these resources can be substituted one for the other. \textit{Quantum teleportation} allows one to substitute one ebit and two bits of classical communication for one qubit of quantum communication~\cite{teleporting}. \textit{Dense coding} shows that sharing one ebit and then communicating one qubit allows one to communicate two bits~\cite{superdense}. \textit{Newman's Theorem} states that in the context of communication complexity, having shared randomness can save only a small amount of communication compared to having local randomness~\cite{newman:random}.

In addition we will at some points in this review  consider other additional (more specialized or more exotic) resources. For instance one can consider
\begin{itemize}
\item \textit{One-way classical or quantum communication.} Alice is allowed to communicate to Bob, but Bob is not allowed to communicate back to Alice.
\item \textit{Simultaneous Message Passing model.} In this model there is a third party, called the Referee, and messages are only allowed from Alice to the Referee and from Bob to the Referee.  It is the Referee who has to compute the value of the function $f(x,y)$.
\item \textit{Multipartite entanglement.} Sometimes one is interested in non-locality or communication complexity between more than two parties. Contrary to bipartite entanglement where it is sufficient to consider ebits, there are many kinds of multiparticle entanglement (such as GHZ states, W states, etc.) which could be useful for solving different communication problems.
\item \textit{Non-local (or PR) boxes.} This exotic resource is intermediate between an ebit and a bit. Indeed, it is a resource which does not enable the parties to communicate (in the same way that entanglement does not allow communication). But to be produced physically it requires a bit of communication between the parties at the moment it is used (contrary to entanglement which once established requires no more communication). Its study provides a deeper understanding of the power and limitations of quantum entanglement in communication complexity.
\end{itemize}

\subsection{Basic scenarios}

The basic question  asked in communication complexity and quantum non-locality
is to understand how much of these resources are required in different situations.

Thus classical communication complexity~\cite{kushilevitz&nisan:cc} is basically concerned with understanding how much classical communication is required to compute the value of a function $f(x,y)$, possibly using (shared or local) randomness.

In quantum communication complexity the parties are trying to
compute the value of $f$, but may now use quantum resources. In the
\emph{quantum communication model}, introduced by Yao~\cite{yao:qcircuit}, they can communicate qubits,
and in the \emph{entanglement} model, introduced by Cleve and Buhrman~\cite{cleve&buhrman:subs},
the parties share entangled particles and are allowed to communicate classical bits. 
When one extends the quantum communication model of Yao such that the parties also share entangled particles,
quantum teleportation shows that these two models are essentially
equivalent: one qubit in the first model can be replaced by two bits and one ebit
in the entanglement, and conversely one bit can be simulated by one qubit. It is, however, a challenging open problem
whether the quantum communication model, \emph{without} shared entanglement, is
essentially equivalent to the entanglement model.

\textit{Non-locality}, although at first sight a very different topic, is also concerned with comparing resources. Indeed the basic question in this area is to compare:
\begin{itemize}
\item The correlations that can be obtained if the parties share entanglement and carry out local measurements on their particles, but are not allowed any communication.
\item The correlations that can be obtained if the parties have shared randomness, but are not allowed any communication. This is known in the physics literature as a \textit{local hidden variable model}.
\end{itemize}

Bell's Theorem states that these two scenarios are not equivalent: shared randomness alone is not sufficient to reproduce the quantum correlations.

\subsection{Mappings between communication complexity and non-locality}

Thus quantum communication complexity, classical communication complexity, and non-locality can be put in a unified framework in which similar kinds of resources are compared.
In addition, in some cases there exist mappings between
quantum communication complexity scenarios and non-locality scenarios.

The most simple such mapping occurs
in the entanglement model if the parties can solve the communication complexity problem more efficiently using entanglement than without entanglement, and if this can be done by measuring their entangled particles \textit{before} they communicate to each other. Then it immediately follows that the correlations obtained by measuring their entangled particles (but without communicating), cannot be realized in a local hidden variable model.

Conversely it is possible to map any non-locality experiment to a communication complexity problem in the entanglement model. This was the approach used in the original paper~\cite{cleve&buhrman:subs}. It mapped the non-local correlations that arise in the GHZ paradox to a communication complexity problem. This approach has since been generalized~\cite{PhysRevLett.92.127901}, although in the resulting communication complexity problem the function $f(x,y)$ is only computed successfully by the parties with non-zero probability.

Another mapping can occur in the quantum communication model when
\textit{one-way} quantum communication from Alice to Bob is more efficient than classical communication. Then it is often
possible to construct from the communication complexity problem a nontrivial non-locality scenario. This approach has yielded some very interesting non-locality scenarios which we will describe in detail below.

\subsection{Summary of the review}

In this review we will present some of the main results obtained so far in the field of quantum communication complexity. We start by introducing quantum non-locality in Section~\ref{sec:nl}, focusing on its relation with communication complexity. We present simple examples such as the GHZ paradox, the CHSH example, the magic square game, but rephrasing them in the language of data processing. Next we present quantum communication complexity in Section~\ref{sec:cc}, illustrating it with examples such as the distributed Deutsch-Jozsa  problem, the intersection problem, Raz's problem, and the hidden matching problem. In Section~\ref{sec:NLandCC} we unite these two approaches, showing how some of the examples  from quantum communication complexity can be used to derive new non-locality games. In section~\ref{sec:SMP} we discuss another model of communication complexity, the simultaneous message passing model, and show how classical communication, entanglement, quantum communication can be traded one for the other in this model. In Section~\ref{sec:otherQNL} we discuss several additional aspects of quantum non-locality, such as non-local boxes, Tsirelson bounds, and simulation of quantum correlations using classical resources. Finally we consider in Section~\ref{sec:impl} experimental issues, in particular the detection loophole, and present  the outlook for future experiments. We conclude by discussing some open questions in the field.
The interested reader can also consult the earlier review~\cite{brassard:qcc} which covers some of the material presented here.

\section{Simple Non-locality Examples}\label{sec:nl}

The idea of non-locality was originally concerned with the possibility that
quantum mechanics is actually a \emph{classical theory} that depends on
``hidden variables'' whose values might be discovered in the future as
part of some successor theory to quantum mechanics.
Bell~\cite{bell:epr} proposed a hypothetical experiment for ruling out
such classical theories under the assumption that measurements of
quantum systems can occur at different points in space-time, and
information cannot be transmitted faster than the speed of light.

Another way of interpreting Bell's experiment is as a method for
two (or more) cooperating distributed parties to compute some sort of
input-output relation, where each party receives input data and must
produce output data consistent with the relation.
In Bell's experiment, there is such a task that cannot be accomplished
in a setting where the information processing resources are all classical.
In contrast, the task \textit{can} be accomplished if the parties share
prior entanglement.

Since Bell's seminal work, the concept of quantum non-locality has been extensively studied, by physicists, philosophers, and more recently by computer scientists. Some of the important early advances have been the Clauser-Horn-Shimony-Holt (CHSH) inequality \cite{chsh} which allows Bell's surprising predictions to be tested even in the presence of noise; and the GHZ-Mermin scenario~\cite{greenberger:ghz,mermin:reality} which was the first "pseudo-telepathy" game. More recently there has been a more or less systematic enumeration of Bell inequalities for small number of settings and/or outcomes (see, e.g., \cite{cglmp,collins2004rtq,werner2001amb,Zukowski:all-Bell}); the study of the statistical power of non-locality tests~\cite{vandam2005ssn}; an understanding of the limits to quantum non-locality (Tsirelson-type bounds)~\cite{Tsirelson80} as compared to the larger world of correlations obeying only the no-signalling conditions (e.g., non-local boxes); investigations of the power of non-locality in cryptographic settings~\cite{bhk:crypto}, etc.

In the next paragraphs we review various non-locality scenarios, casting them in the
language of data processing. The reader wishing to complement this overview could consult two recent reviews, 
written more from physics~\cite{werner2001bia} and computer science~\cite{brassard2005qpt} perspectives.

\subsection{GHZ: Greenberger-Horne-Zeilinger and Mermin}

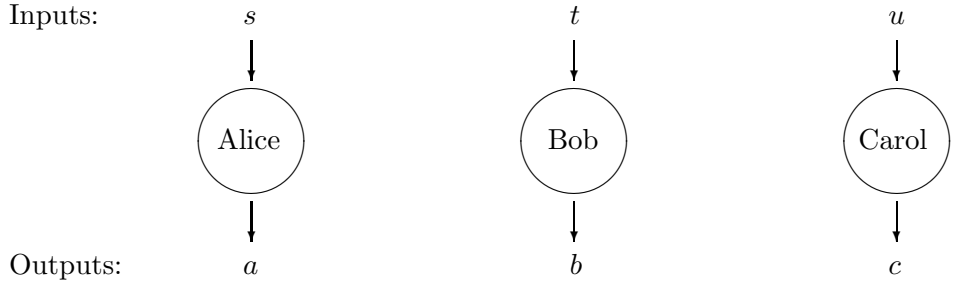
\begin{figure}[h]
\setlength{\unitlength}{100000sp}
\vspace{2mm}
\begin{center}
\begin{picture}(280,70)(0,0)
\put(60,35){\circle{30}}
\put(51.5,33){Alice}
\put(58,64){$s$}
\put(58,2){$a$}
\put(60,60){\vector(0,-1){10}}
\put(60,20){\vector(0,-1){10}}
\put(140,35){\circle{30}}
\put(133.5,33){Bob}
\put(139,64){$t$}
\put(139,2){$b$}
\put(140,60){\vector(0,-1){10}}
\put(140,20){\vector(0,-1){10}}
\put(220,35){\circle{30}}
\put(210.5,33){Carol}
\put(218,64){$u$}
\put(218,2){$c$}
\put(220,60){\vector(0,-1){10}}
\put(220,20){\vector(0,-1){10}}
\put(0,64){Inputs:}
\put(0,2){Outputs:}
\end{picture}
\end{center}
\caption{The general form of a \textit{non-locality} scenario involving
three parties: Alice, Bob, and Carol receive inputs $s$, $t$, $u$ respectively,
and are required to produce outputs $a$, $b$, $c$, respectively, satisfying
certain conditions.
Once the inputs are received, no communication is permitted between the
parties.
For the specific GHZ scenario, it is possible to accomplish the task if the
parties are in possession of a tripartite entangled state.
Without the prior entanglement, it is impossible to accomplish the
task.}\label{fig:ghz}
\end{figure}

The following scenario essentially underlies those
of~\cite{greenberger:ghz,mermin:reality}, but is cast in the language
of data processing.
The basic structure is illustrated in Fig.~\ref{fig:ghz}.
Three physically separated parties---call them Alice, Bob, and
Carol---receive \textit{input} bits $s$, $t$, and $u$, respectively,
which are arbitrary subject to the condition that
$s \oplus t \oplus u = 0$ ($\oplus$ denotes \textit{exclusive or}, which is the sum of its arguments in modulo~2 arithmetic).
Once they receive their input data, they are forbidden from having
any communication between them.
Their goal is to produce \textit{output} bits $a$, $b$, and $c$,
respectively, such that
\begin{equation}\label{eq:ghz}
a \oplus b \oplus c =
\begin{cases}
0 & \mbox{if $stu = 000$} \\
1 & \mbox{if $stu \in \{011, 101, 110\}$.}
\end{cases}
\end{equation}
Note that the task that the three parties are trying to accomplish is
the computation of a relation, where there are three input bits ($stu$)
and three output bits ($abc$).
The task is nontrivial in light of the fact that the input bits
are distributed among the parties so that each party is given the value
of only one of them; the output bits are also distributed.

The first observation is that \textit{with classical resources} there
must be communication among the three parties to succeed.
To see why this is so, first consider deterministic strategies (later
we will analyze the case of probabilistic strategies, where the parties
behave stochastically, i.e., they can flip coins).
Since Alice cannot receive any information from Bob or Carol,
her output bit $a$ can depend only on the value of her input bit $s$.
Let $a_0$ (respectively $a_1$) be Alice's output when her input bit is $0$ (respectively $1$).
Similarly, let $b_0, b_1$ and $c_0, c_1$
be Bob and Carol's outputs for their respective input values.
Note that the six bits $a_0, a_1, b_0, b_1, c_0, c_1$ completely
characterize any deterministic strategy of Alice, Bob, and Carol.
The conditions of the problem translate into the equations
\begin{eqnarray}\label{eq:ghzequations}
a_0 \oplus b_0 \oplus c_0 & = & 0, \nonumber \\
a_0 \oplus b_1 \oplus c_1 & = & 1, \nonumber \\
a_1 \oplus b_0 \oplus c_1 & = & 1, \nonumber \\
a_1 \oplus b_1 \oplus c_0 & = & 1.
\end{eqnarray}
It is impossible to satisfy all four equations simultaneously.
This is because summing the four equations modulo two, yields $0 = 1$ (recall that $1+1=0$ modulo 2).
Therefore, for any strategy, there exists an input configuration
$stu \in \{000, 011, 101, 110\}$ for which it fails. Note however that
for any \emph{three} out of the four equations from~(\ref{eq:ghzequations})
there \emph{is} a strategy that satisfies these three equations perfectly.

To see why probabilistic strategies cannot succeed either, note that any
such strategy can be modeled as a deterministic strategy where Alice, Bob,
and Carol have access to a random variable $r$ (for example, $r$ could be
the outcomes of a sequence of uniformly distributed random bits).
This $r$ is sometimes referred to as a ``local hidden variable''.
It is assumed that the testing procedure does not have access to $r$, so
that the input bits ($stu$) are uncorrelated
with $r$.
The intuitive way of thinking about this scenario is that the three parties
get together before the game starts, randomly select $r$, and then each
party secretly keeps a copy of this information.
An example of a probabilistic strategy is for $r \in \{0,1\}^2$ to be two
uniformly random bits that specify which three of the four equations
in~(\ref{eq:ghzequations}) are satisfied.
This probabilistic strategy succeeds with probability $3/4$.
We next show that this success probability is optimal.

Suppose that the input data $s,t,u$ is uniformly distributed over $\{000,011,101,110\}$.
Then the success probability that any randomized protocol achieves
is
\begin{equation}
\sum_r q_r \frac{1}{4} \sum_{s,t,u} P(s,t,u,r),
\end{equation}
where $q_r$ is the probability (of the shared randomness) that the
parties flip $r$, and $P(s,t,u,r)=1$ if the deterministic protocol
corresponding to $r$ is correct on input $stu$ and $P(s,t,u,r)=0$ otherwise.
Clearly this is bounded above by
\begin{equation}
\max_{r} \frac{1}{4} \sum_{s,t,u} P(s,t,u,r),
\end{equation}
which by the above discussion is at most $3/4$.

Now consider the same problem, but where Alice, Bob, and Carol have an
additional resource: each is supplied with a qubit, where the state of the
combined 3-qubit system is%
\footnote{This is an \textit{entangled} state that is equivalent to the
so-called GHZ state $\frac{1}{\sqrt 2}\ket{000} + \frac{1}{\sqrt 2}\ket{111}$
(under local unitary operations).}
\begin{equation}
\textstyle{\frac{1}{2}}\ket{000} - \textstyle{\frac{1}{2}}\ket{011} -
\textstyle{\frac{1}{2}}\ket{101} - \textstyle{\frac{1}{2}}\ket{110}.
\end{equation}
The parties are allowed to apply unitary transformations and
perform measurements on their individual qubits, but communication
between the parties is still forbidden.
It turns out that now the parties \textit{can} produce $a$, $b$, $c$
satisfying Eq.~(\ref{eq:ghz}). This is achieved by the procedure
that follows.

The procedure for Alice is to measure her qubit in the computational
basis (consisting of $\ket{0}$ and $\ket{1}$) if her input bit~$s$ is~0, and to measure her qubit in the
Hadamard basis  (consisting of $H\ket{0}=\frac{1}{\sqrt{2}}(\ket{0}+\ket{1})$
and $H\ket{1}=\frac{1}{\sqrt{2}}(\ket{0}-\ket{1})$) if her input bit is~1.
In either case, she sets her output bit $a$ to the outcome of her measurement.
The procedures for Bob and Carol are similar to that of Alice, but
with Bob's bits being $s$ and $b$, and Carol's bits being $u$ and $c$.

To see why the described procedure \textit{always} produces output
bits $abc$ satisfying Eq.~(\ref{eq:ghz}), consider the various cases
of the input possibilities $stu$.
In the case where $stu = 000$, the state is measured in the computational
basis, so clearly the outcomes are from $\{000,011,101,110\}$, and hence
satisfy $a \oplus b \oplus c = 0$.
The case where $stu = 011$ can be analyzed by assuming that a Hadamard
transform is applied to the last two qubits of the state prior to a
measurement in the computational basis.
Since
\begin{eqnarray}
\lefteqn{\left(I \otimes H \otimes  H\right)
(\textstyle{\frac{1}{2}}\ket{000} - \textstyle{\frac{1}{2}}
\ket{011} - \textstyle{\frac{1}{2}}\ket{101} - \textstyle{\frac{1}{2}}\ket{110})}
\nonumber \\
& = &
\left(I \otimes H \otimes  H\right)
\left(\textstyle{\frac{1}{2}}\ket{0}(\ket{00} - \ket{11})
-\textstyle{\frac{1}{2}}\ket{1}(\ket{01} + \ket{10})\right) \nonumber \\
& = &
\textstyle{\frac{1}{2}}\ket{0}(\ket{01} + \ket{10})
-\textstyle{\frac{1}{2}}\ket{1}(\ket{00} - \ket{11}) \nonumber \\
& = & \textstyle{\frac{1}{2}}\ket{001} + \textstyle{\frac{1}{2}}
\ket{010} - \textstyle{\frac{1}{2}}\ket{100} + \textstyle{\frac{1}{2}}\ket{111},
\end{eqnarray}
$a \oplus b \oplus c = 1$, as required, in this case.
The remaining cases where $stu = 101$ and $110$ are similar by the
symmetry of the entangled state and protocol.

We have shown that the entangled state enables the three parties to
correlate their output bits with their inputs bits in a manner that is
impossible to achieve with classical resources, unless there is
communication among the parties.
It should be noted that, in accomplishing this task using the entangled
state, no actual communication occurs among the parties.
In particular, the output bits $a$, $b$, and $c$ individually contain
no information about $stu$; they are uniformly distributed in all cases.
It is only the trivariate correlations among $a$, $b$, and $c$ that are
related to the input data $stu$.

\subsection{CHSH: Clauser-Horne-Shimony-Holt}\label{sec:CHSH}

\begin{figure}[h]
\setlength{\unitlength}{100000sp}
\vspace{2mm}
\begin{center}
\begin{picture}(180,70)(0,0)
\put(60,35){\circle{30}}
\put(51.5,33){Alice}
\put(58,64){$s$}
\put(58,2){$a$}
\put(60,60){\vector(0,-1){10}}
\put(60,20){\vector(0,-1){10}}
\put(140,35){\circle{30}}
\put(133.5,33){Bob}
\put(139,64){$t$}
\put(139,2){$b$}
\put(140,60){\vector(0,-1){10}}
\put(140,20){\vector(0,-1){10}}
\put(0,64){Inputs:}
\put(0,2){Outputs:}
\end{picture}
\end{center}
\caption{The non-locality scenario involving two parties:
Alice and Bob receive inputs $s$ and $t$ respectively,
and are required to produce outputs $a$ and $b$ respectively,
satisfying certain conditions.
Once the inputs are received, no communication is permitted between
the parties.
For the specific CHSH scenario, it is possible to accomplish the task
with probability $\cos^2(\pi/8) = 0.853\ldots$ if the parties are in
possession of an ebit.
Without the prior entanglement, the highest possible success probability
is $3/4$.}\label{fig:chsh}
\end{figure}
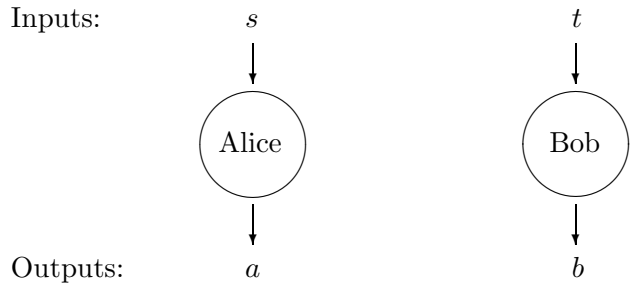

The following scenario essentially underlies that of~\cite{chsh} but is
cast in the language of data processing.
The basic structure is illustrated in Fig.~\ref{fig:chsh}.
Alice and Bob receive input bits $s$ and $t$, respectively, and,
after this, they are forbidden from communicating with each other.
Their goal is to produce output bits $a$ and $b$, respectively, such that
\begin{equation}\label{eq:chsh}
a \oplus b = s \wedge t,
\end{equation}
(`$\wedge$' is the \textit{logical and}, which is 1 if all its arguments are 1, and which is 0 otherwise)
or, failing that, to satisfy this condition with as high a probability as possible.
To analyze the situation in terms of classical information, first again consider
the case of deterministic strategies.
For these, Alice's output bit depends solely on her input bit
$s$ and similarly for Bob.
Let $a_0, a_1$ be the two possibilities for Alice and $b_0, b_1$ be the
two possibilities for Bob.
These four bits completely characterize any deterministic strategy.
Condition~(\ref{eq:chsh}) translates into the equations
\begin{eqnarray}\label{eq:eqchsh}
a_0 \oplus b_0 & = & 0, \nonumber \\
a_0 \oplus b_1 & = & 0, \nonumber \\
a_1 \oplus b_0 & = & 0, \nonumber \\
a_1 \oplus b_1 & = & 1.
\end{eqnarray}
It is impossible to satisfy all four equations simultaneously (since summing
them modulo 2 yields $0 = 1$).
Therefore it is impossible to satisfy Condition~(\ref{eq:chsh}) absolutely.
By using a probabilistic strategy, Alice and Bob can satisfy
Condition~(\ref{eq:chsh}) with probability $3/4$.
For such a strategy, we allow Alice and Bob to have \textit{a priori} classical
random variables, whose distribution is independent of that of the
inputs $s$ and $t$.
Note that \emph{any} three of the four equations of~(\ref{eq:eqchsh}) can be
simultaneously satisfied.
The probabilistic classical strategy works as follows.
Alice and Bob have uniformly-distributed random bits that are used
to specify which of the four equations of~(\ref{eq:eqchsh}) is violated,
and then play the strategy that satisfies the other three perfectly.
It is easy to see that (a) for any input $st$, the resulting outputs satisfy
Condition~(\ref{eq:chsh}) with probability $3/4$, and (b) this is optimal in
that no probabilistic strategy can attain a success probability greater
than $3/4$.

Now consider the same problem but where Alice and Bob are each supplied with
a qubit where the state of the two-qubit system is initialized to
\begin{equation}
\textstyle{\frac{1}{\sqrt 2}}(\ket{00} - \ket{11}).
\end{equation}
It turns out that now the parties can produce data that satisfies
Condition~(\ref{eq:chsh}) with probability $\cos^2(\pi/8) = 0.853\ldots$,
which is higher than what is possible in the classical case.
This is achieved by the following procedures.
Denote the unitary operation that rotates the qubit by angle $\theta$ by $R(\theta)=\left(
\begin{array}{lr}
\cos\theta & -\sin\theta \\
\sin\theta & \cos\theta
\end{array}
\right)$ (where we have written it out in the computational basis).
Alice applies one of two rotations
on her qubit, depending on her input
bit $s$: if $s=0$ the rotation is $R(-\pi/16)$; if $s=1$ the rotation is
$R(3\pi/16)$.
Then Alice measures her qubit in the computational basis and sets her
output bit $a$ to the result.
Bob's procedure is the same, depending on his input bit $t$.
It is straightforward to calculate that, if Alice rotates by $\theta_1$ and
Bob rotates by $\theta_2$, then the entangled state becomes
\begin{equation}
\textstyle{\frac{1}{\sqrt{2}}}
(\cos(\theta_1 + \theta_2)(\ket{00} - \ket{11})
+ \sin(\theta_1 + \theta_2)(\ket{01} + \ket{10})).
\end{equation}
After the measurements, the probability that $a \oplus b = 0$ is
$\cos^2(\theta_1 + \theta_2)$.
It is now a straightforward exercise to verify that Condition~\ref{eq:chsh}
is satisfied with probability $\cos^2(\pi/8)$ for all four input
possibilities.

\subsection{Tsirelson's upper bound for CHSH}
\label{Tsirelson}

Although the protocol in the previous subsection using entanglement has a
higher success probability ($\cos^2(\pi/8) = 0.853\ldots$) than any classical
protocol ($3/4$), it still does not succeed with probability~1.
This raises the question of whether there is a different strategy using
entanglement that always succeeds---or, failing that, whose success
probability exceeds $\cos^2(\pi/8)$.
Tsirelson~\cite{Tsirelson80} first showed that the above quantum
protocol is optimal in that it is impossible to exceed success
probability $\cos^2(\pi/8)$, regardless of the strategy---including
any amount of prior entanglement---the parties start with.
What follows is a simple proof of this result.

Consider an arbitrary bipartite entangled state $\ket{\psi}_{AB}$.
An arbitrary strategy for Alice that uses this entangled state can be
represented by two observables%
\footnote{An observable is a Hermitian operator. One associates to an observable a projective measurement, with one projector for each of the eigenspaces of the observable.}
$A_0$ and $A_1$, each with eigenvalues in $\{+1,-1\}$.
When Alice's input bit is $0$, she obtains her output bit by applying
the projective measurement corresponding to the eigenspaces of $A_0$ to
the component of $\ket{\psi}_{AB}$ in her possession.
The $+1$-eigenspace of $A_0$ corresponds to output bit $0$,
while the $-1$-eigenspace corresponds to the output bit $1$.
When her input bit is $1$, she applies the  measurement
corresponding to $A_1$.
Similarly, an arbitrary strategy for Bob can be represented by two
observables $B_0$ and $B_1$.

At this point, the reader might object that $\ket{\psi}_{AB}$, $A_0$, $A_1$,
$B_0$, and $B_1$ do not capture every possible strategy of Alice and Bob,
since they need not be limited to applying projective measurements.
Although non-projective measurements may be used, such measurements can
always be simulated by projective measurements in a larger Hilbert space.
Thus, no generality has been lost because any strategy can be converted
to the above form.

Since the observables have eigenvalues in $\{+1,-1\}$ rather than
$\{0,1\}$, it is more convenient here to think of Alice and Bob's
output bits in these terms as $a' = (-1)^a$ and $b' = (-1)^b$,
respectively.
Then the protocol succeeds on input $st$ if and only if
$(-1)^{s \wedge t} \cdot a' \cdot b' = 1$.

If $s$ and $t$ are randomly chosen according the uniform distribution,
then the expected value of $(-1)^{s \wedge t} \cdot a' \cdot b'$ is
\begin{equation}
\bra{\psi}_{AB}
\left({\textstyle \frac{1}{4}}A_0 \otimes B_0
+ {\textstyle \frac{1}{4}}A_0 \otimes B_1
+ {\textstyle \frac{1}{4}}A_1 \otimes B_0
- {\textstyle \frac{1}{4}}A_1 \otimes B_1\right)
\ket{\psi}_{AB},
\end{equation}
and is therefore upper bounded by the largest eigenvalue of
\begin{equation}
M = {\textstyle \frac{1}{4}}A_0 \otimes B_0
+ {\textstyle \frac{1}{4}}A_0 \otimes B_1
+ {\textstyle \frac{1}{4}}A_1 \otimes B_0
- {\textstyle \frac{1}{4}}A_1 \otimes B_1.
\end{equation}
It is straightforward to calculate that
\begin{equation}
M^2 = {\textstyle \frac{1}{4}}I
- {\textstyle \frac{1}{16}}(A_0 A_1) \otimes (B_0 B_1)
+ {\textstyle \frac{1}{16}}(A_0 A_1) \otimes (B_1 B_0)
+ {\textstyle \frac{1}{16}}(A_1 A_0) \otimes (B_0 B_1)
- {\textstyle \frac{1}{16}}(A_1 A_0) \otimes (B_1 B_0),
\end{equation}
from which we can upper bound the maximum eigenvalue of $M^2$ by the sum
of the maximum eigenvalue in each term, obtaining
$\frac{1}{4}+\frac{1}{16}+\frac{1}{16}+\frac{1}{16}+\frac{1}{16} = \frac{1}{2}$.
It follows that the largest eigenvalue of $M$ itself is at most
$1/\sqrt{2}$, which therefore upper bounds the expected value of
$(-1)^{s \wedge t} \cdot a' \cdot b'$.
This translates into an upper bound of
$(1+1/\sqrt{2})/2 = \cos^2(\pi/8)$ for the success probability of
the actual protocol (where Alice and Bob output bits $a$ and $b$).
This completes the proof of Tsirelson's upper bound for CHSH.

\subsection{Magic square game}

In one respect the GHZ example is more striking than the CHSH example:
in the former case, the protocol with entanglement \textit{always} succeeds,
while in the latter case the protocol with entanglement merely succeeds
with higher probability.
However, the GHZ example involves three parties, whereas the CHSH example
only involves two.
Is there a \textit{two-party} scenario where the quantum protocol always
succeeds, whereas the best classical success probability is bounded below~$1$?
The answer is affirmative, see for instance \cite{cabello01A,cabello01B,cabello2005}. A particularly elegant example is
the following game, which has been referred to as the \emph{magic square game}~\cite{Aravind02}.

To define this game, consider the problem of labeling the entries of
a $3 \times 3$ matrix with bits so that the parity of each row is
even, whereas the parity of each column is odd.
It is not hard to see that this is impossible%
\footnote{As before, we can express a valid solution in terms of equations,
in this case six of them (where arithmetic is modulo 2):
$m_{11} + m_{12} + m_{13} = 0$,
$m_{21} + m_{22} + m_{23} = 0$,
$m_{31} + m_{32} + m_{33} = 0$,
$m_{11} + m_{21} + m_{31} = 1$,
$m_{12} + m_{22} + m_{32} = 1$,
$m_{13} + m_{23} + m_{33} = 1$.
Adding these equations modulo 2 yields $0 = 1$.}.
The two matrices
\begin{center}
\begin{tabular}{|c|c|c|}
\hline
0 & 0 & 0 \\
\hline
0 & 0 & 0 \\
\hline
1 & 1 & 0 \\
\hline
\end{tabular}
\hspace*{40mm}
\begin{tabular}{|c|c|c|}
\hline
0 & 0 & 0 \\
\hline
0 & 0 & 0 \\
\hline
1 & 1 & 1 \\
\hline
\end{tabular}
\end{center}
each satisfy five out of the six constraints.
For the first matrix, all rows have even parity,
but only the first two columns have odd parity.
For the second matrix, the first two rows have even
parity, and all columns have odd parity.

Bearing the above in mind, consider the game where Alice receives
$s \in \{1,2,3\}$ as input (specifying the number of a row), and Bob receives
$t \in \{1,2,3\}$ as input (specifying the number of a column).
Their goal is to each produce $3$-bit outputs, $a_1a_2a_3$ for Alice
and $b_1b_2b_3$ for Bob, with these properties:
\begin{enumerate}
\item They satisfy the row/column parity constraints.
Namely,
$a_1 \oplus a_2 \oplus a_3 = 0$ and $b_1 \oplus b_2 \oplus b_3 = 1$.
\item They are consistent where the row intersects the column.
Namely, $a_t = b_s$.
\end{enumerate}
As usual, Alice and Bob are forbidden from communicating once the
game starts, so Alice does not know what $t$ is and Bob does not
know what $s$ is.
We shall observe that, classically, the best success probability
possible is $8/9$, whereas there is a quantum strategy that always
succeeds.

An example of a strategy that attains success probability $8/9$
(when the input $st$ is uniformly distributed) is
where Alice plays according to the rows of the first matrix above
and Bob plays according the columns of the second matrix above.
This succeeds in all cases, except where $s = t = 3$.
To see why this is optimal, note that for any other
classical strategy, it is possible to represent it as two matrices
as above but with different entries.
Alice plays according to the rows of the first matrix and Bob
plays according to the columns of the second matrix.
We can assume that the rows of Alice's matrix all have even
parity; if she outputs a row with odd parity then they immediately
lose, regardless of Bob's output.
Similarly, we can assume that all columns of Bob's matrix have
odd parity.%
\footnote{In fact, the game can be simplified so that Alice and Bob
each output just two bits, since the parity constraint determines
the third bit.}
Considering such a pair matrices, the players lose at each entry
where they differ.
There must be such an entry, since otherwise it would be possible
to have all rows even and all columns odd with one matrix.
Thus, when the input $st$ is chosen uniformly from
$\{1,2,3\}\times\{1,2,3\}$, the success probability is at most~$8/9$.

The quantum strategy for this game is based on the following
observation due to Mermin~\cite{Mermin90,Mermin93}.
Let $I$, $X$, $Y$, $Z$ denote the $2 \times 2$ Pauli matrices:
\begin{equation}\label{eqpaulis}
I = \left(\begin{array}{rr}1 & 0\\ 0&1\end{array}\right), \
X = \left(\begin{array}{rr}0&1\\ 1&0\end{array}\right), \
Y = \left(\begin{array}{rr}0&-i\\i&0\end{array}\right), \mbox{ and }
Z = \left(\begin{array}{rr}1&0\\ 0&-1\end{array}\right).
\end{equation}
Each is an observable with eigenvalues in $\{+1,-1\}$.
Consider the following table of two-qubit observables that are each
a tensor product of two Pauli matrices:
\begin{center}
\begin{tabular}{|c|c|c|}
\hline
$X \otimes X$ & $Y \otimes Z$ & $Z \otimes Y$ \\
\hline
$Y \otimes Y$ & $Z \otimes X$ & $X \otimes Z$ \\
\hline
$Z \otimes Z$ & $X \otimes Y$ & $Y \otimes X$ \\
\hline
\end{tabular}
\end{center}
For our present purposes, the noteworthy property is that the
observables along each row commute and their product is
$I \otimes I$, and the observables along each column
commute and their product is $- I \otimes I$.
This implies that, for any two-qubit state, performing the
three measurements along any row results in three
$\{+1,-1\}$-valued bits whose product is $+1$.
Also, performing the three measurements along any column
results in three $\{+1,-1\}$-valued bits whose product is $-1$.
 This
can be seen more easily when one simultaneously diagonalizes the three
commuting observables. They will have $1$ and $-1$ eigenvalues on the
diagonal. Each consecutive observable will project the state onto a possible
refinement of the current eigenspace the state lies in. This will
yield that the product of the outcomes of the three observables will
be $1$ in case  the observables belong to a  row of the matrix, because
the product of the row observables is $I \otimes I$,
and $-1$ when they belong to a column, since the product of the
observables for each column is $- I \otimes I$.

We can now describe the quantum protocol.
It uses two pairs of entangled qubits, each of which is in initial state
$\frac{1}{\sqrt 2}(\ket{01}-\ket{10})$.
Alice, on input $s$, applies three two-qubit measurements
corresponding to the observables in row $s$ of the above table.
For each measurement, if the result is $+1$, she outputs $0$
and if the result is $-1$, she outputs $1$.
Similarly, Bob, on input $t$, applies the measurements
corresponding to the observables in column $t$, and
converts the outcomes into bits in the same manner.

We have already established that Alice and Bob's output bits
satisfy the required parity constraints.
It remains to show that Alice and Bob's output bits that
correspond to where the row meets the column are the same.
For that measurement, Alice and Bob are measuring with respect
to the same observable in the above table. Because all the observables
in each row and in each column commute, we may assume that the place
where they intersect is the first observable applied.
Those bits are obtained by Alice and Bob each measuring
$\frac{1}{2}(\ket{01}-\ket{10})(\ket{01}-\ket{10})$
with respect to the observable in entry $(s,t)$ of the table.
To show that their measurements will agree for all cases of $st$,
we consider the individual Pauli measurements on the individual
entangled pairs of the form $\frac{1}{\sqrt 2}(\ket{01}-\ket{10})$.
Let $a'$ and $b'$ denote the outcomes of the first measurement
(in terms of bits), and $a''$ and $b''$ denote the outcomes
of the second.
Since the measurement associated with the tensor product of two
observables is operationally equivalent to measuring each individual
observable and taking the product of the results, we have that
$a_t = a' \oplus a''$ and $b_s = b' \oplus b''$.
It is straightforward to verify that if the same measurement from
$\{X,Y,Z\}$ is applied to each qubit of
$\frac{1}{\sqrt 2}(\ket{01}-\ket{10})$
then the outcomes will be distinct.
Therefore, $a' \oplus b' = 1$ and $a'' \oplus b'' = 1$, from
which it follows that
\begin{eqnarray}
a_t \oplus b_s & = & (a' \oplus a'') \oplus (b' \oplus b'') \nonumber \\
& = & (a' \oplus b') \oplus (a'' \oplus b'') \nonumber \\
& = & 1 \oplus 1 \nonumber \\
& = & 0,
\end{eqnarray}
so $a_t = b_s$.
This completes the analysis of the magic square game.

\section{Communication Complexity}\label{sec:cc}
In the last section we considered scenarios without communication.
Here we will extend the non-locality setting to one where
the parties (Alice and Bob) are allowed to send information to each other in the form
of bits or qubits. They can still have shared randomness and may share
an entangled quantum state.  We are now
interested in the \emph{minimum number} of bits or qubits that are needed in
order to compute a function that depends on the inputs of all the
parties.

 The ability to send information to
each other departs from the setting of non-locality.  We will see that
entanglement can be used to reduce (for certain functions) the
communication drastically compared to when the parties share just
classical resources. Accordingly, while entanglement cannot be
used for signalling, it can be used to significantly reduce the communication needed for certain tasks.
In later sections we will see how some of the ideas
and protocols developed in the setting of communication complexity
can be used to formulate new non-locality games.

Communication complexity has been studied extensively in the area of
theoretical computer science and has deep connections
with seemingly unrelated areas, such as VLSI design, circuit lower
bounds, lower bounds on branching programs, size of data structures, and bounds on the length
of logical proof systems, to name just a few.
We refer to the textbooks~\cite{kushilevitz&nisan:cc,Hromkovic97} for more details.

\subsection{The setting}\label{sec:setting}

First we sketch the setting for classical communication complexity.
Alice and Bob want to compute some function $f: \mathcal{D}\rightarrow\01$,
where $\mathcal{D}\subseteq X\times Y$.
If the domain $\mathcal{D}$ equals $X\times Y$ then $f$ is called
a \emph{total} function, otherwise it is a \emph{promise} function.
Alice receives input $x\in X$, Bob receives input $y\in Y$, with
$(x,y)\in\mathcal{D}$.
A typical situation, illustrated in Fig.~\ref{fig:cc}, is where $X=Y=\01^n$,
so both Alice and Bob receive an $n$-bit input string.
As the value $f(x,y)$ will generally depend on both $x$ and $y$,
some communication between Alice and Bob is required in order
for them to be able to compute $f(x,y)$.
We are interested in the \emph{minimal} amount of communication they need.

A communication \emph{protocol} is a distributed algorithm where
first Alice does some individual computation, and then sends a message
(of one or more bits) to Bob, then Bob does some computation and sends
a message to Alice, etc. Each message is called a \emph{round}.
After one or more rounds the protocol terminates and outputs some value,
which must be known to both players.
The \emph{cost} of a protocol is the total number of bits communicated
on the worst-case input.
A \emph{deterministic} protocol for $f$ always has to output
the right value $f(x,y)$ for all $(x,y)\in\mathcal{D}$.
In a \emph{bounded-error} protocol, Alice and Bob may
flip coins and  the protocol has to output the right
value $f(x,y)$ with probability $\geq 2/3$ for all $(x,y)\in\mathcal{D}$.
We could either allow Alice and Bob to toss coins
individually (local randomness, or ``private coin'') or jointly (shared randomness, or ``public coin''). 
The later is analogous to the local hidden variables in non-locality games.
A public coin can simulate a private coin and is potentially more powerful.
However, Newman's theorem~\cite{newman:random} says that having a public coin can save
at most $O(\log n)$ bits of communication, compared to a protocol with a private coin.

Some often studied functions are:
\begin{itemize}
\item \emph{Equality:} $\EQ(x,y)=1$ if $x=y$, and $\EQ(x,y)=0$ otherwise
\item \emph{Inner product:} $\IP(x,y)=\sum_{i=1}^n x_i y_i\pmod 2$
(for $x,y\in\01^n$, $x_i$ is the $i$th bit of $x$)
\item \emph{Intersection:} $\INT(x,y)=1$ if there is an $i$ where $x_i=y_i=1$,
and $\INT(x,y)=0$ otherwise
(viewing $x$ as corresponding to the set $\{i : x_i=1\}$ and similarly for $y$,
$\INT(x,y)$ says whether the sets $x$ and $y$ intersect).
A variant of this problem asks to actually find an $i$ where
$x_i=y_i=1$, or to output that none such $i$ exists.
\end{itemize}
Let us first consider the equality problem, which will recur throughout the text.
The goal for Alice is to determine whether her $n$-bit input is the same as Bob's or not.
It is not hard to show that in the deterministic case, $n$ bits of communication are needed (see Section~\ref{ssecrectangles} of the appendix for a proof),
so Bob might as well send his string to Alice after which Alice announces the answer to Bob with one more bit.

To illustrate the power of randomness, let us give a simple yet efficient
bounded-error protocol for the equality problem.
Alice and Bob jointly toss a random string $r\in\01^n$.
Alice sends the bit $a=x\cdot r$ to Bob
(where `$\cdot$' is inner product mod 2).
Bob computes $b=y\cdot r$ and compares this with $a$.
If $x=y$ then $a=b$, but if $x\neq y$ then $a\neq b$ with probability 1/2.
Repeating this a few times, Alice and Bob can decide equality with small error
using $O(n)$  public coin flips and a constant amount of communication.

This protocol uses public coins, but note that Newman's theorem implies that there
exists an $O(\log n)$-bit protocol that uses a private coin.
Let us explicitly describe such a protocol. Alice views her $n$ bits as the coefficients
of a polynomial $p_x$ over some finite field $\mathbb{F}$ of about $3n$ elements:\footnote{For those not familiar with finite fields: it suffices to choose a prime number $p\approx 3n$ and do all additions and multiplications modulo this $p$.}
$p_x(t)=\sum_{i=1}^n x_i t^{i-1}$. She picks a random element $a\in\mathbb{F}$,
and sends Bob the pair $a,p_x(a)$, which she can do using $2\log(3n)$ bits.
Bob computes $p_y(a)$ and outputs 1 if $p_x(a)=p_y(a)$, and outputs 0 otherwise.
Clearly, if $x=y$ then Bob always outputs the correct answer 1.
However, if $x\neq y$ then the polynomial $p_x(t)-p_y(t)$ is a
polynomial in $t$ of degree at most $n-1$ that is not identically equal to 0.
Such a polynomial can be 0 on at most $n-1$ elements of $\mathbb{F}$. Hence with probability at least $2/3$,
the field element $a$ that Alice chose satisfies $p_x(a)\neq p_y(a)$,
and Bob will give the correct output 0 also in this case.

\subsection{The quantum question}

Now what happens if we give Alice and Bob a quantum computer
and allow them to send each other qubits and/or to make use of ebits
that they share at the start of the protocol?

Formally speaking, we can model a quantum protocol as follows.
The total state consists of 3 parts: Alice's private space,
the channel, and Bob's private space.
The starting state is $\ket{x}\ket{0}\ket{y}$:
Alice gets $x$, the channel is initialized to 0, and Bob gets $y$.
Now Alice applies a unitary transformation to her space and
the channel. This corresponds to her private computation as well
as to putting a message on the channel (the length of this message
is the number of channel-qubits affected by Alice's operation).
Then Bob applies a unitary transformation to his space and the channel, etc.
At the end of the protocol Alice or Bob makes a measurement
to determine the output of the protocol.
This model was introduced by Yao~\cite{yao:qcircuit}.

In the second model, introduced by Cleve and
Buhrman~\cite{cleve&buhrman:subs}, Alice and Bob share an unlimited
number of ebits at the start of the protocol,
but now they communicate via a \emph{classical} channel:
the channel has to be in a classical state throughout the protocol.
We only count the communication, not the number of ebits used.
Protocols of this kind can simulate protocols of the first kind with only a factor 2 overhead:
using teleportation, the parties can send each other a qubit using an ebit
and two classical bits of communication. Hence the qubit-protocols that we describe below also
immediately yield protocols that work with entanglement and a classical channel.
Note that an ebit can simulate a public coin toss:
if Alice and Bob each measure their half of the pair of qubits, they get the same random bit.

The third variant combines the strengths of the other two:
here Alice and Bob start out with an unlimited number of ebits
\emph{and} they are allowed to communicate qubits.
This third kind of communication complexity is in fact equivalent to the
second, up to a factor of 2, again by teleportation.

Before continuing to study this model, we first have to face
an important question, already mentioned in the introduction: \emph{is there anything to be gained here?}
At first sight, the following argument seems to rule out any significant gain.
Suppose that in the classical world $k$ bits
have to be communicated in order to compute $f$.
Since Holevo's theorem says that $k$ qubits cannot contain more
information than $k$ classical bits, it seems that the quantum
communication complexity should be roughly $k$ qubits as well
(maybe $k/2$ to account for superdense coding, but not less).
Surprisingly (and fortunately for us), this argument is false, and quantum
communication can sometimes be much less than classical communication
complexity. The information-theoretic argument via Holevo's theorem
fails, because Alice and Bob do not need to communicate the information
in the $k$ bits of the classical protocol;
they are only interested in the value $f(x,y)$, which is just 1 bit.
Below we will survey some of the main examples that have so far been found of
differences between quantum and classical communication complexity.

\subsection{The first examples}

Quantum communication complexity was introduced by Yao~\cite{yao:qcircuit}
and studied by Kremer~\cite{kremer:thesis}, but neither showed any
advantages of quantum over classical communication.
Cleve and Buhrman~\cite{cleve&buhrman:subs} introduced the variant
with classical communication and prior entanglement, and exhibited
the first quantum protocol provably better than any classical protocol.
It uses quantum entanglement to save 1 bit of classical communication.
This gap was extended by Buhrman, Cleve, and van Dam~\cite{bcd:qecc}
and, for arbitrary $k$ parties, by Buhrman, van Dam, H{\o}yer,
and Tapp~\cite{bdht:multiparty}.

\subsection{Distributed Deutsch-Jozsa}\label{secdistributeddj}

The first impressively large gaps between quantum and classical
communication complexity were exhibited by Buhrman, Cleve, and Wigderson~\cite{BuhrmanCleveWigderson98}.
Their protocols are distributed versions of known quantum
query algorithms, like the Deutsch-Jozsa~\cite{deutsch&jozsa} and Grover~\cite{grover:search} algorithms.

Let us start with the first one. It is actually explained most easily in a direct way,
without reference to the Deutsch-Jozsa algorithm (though that is where the idea came from).
The problem deals with a promise version of the equality problem.  Suppose the $n$-bit
inputs $x$ and $y$ are restricted to the following case:
\begin{quote}
{\bf DJ promise:}
either $x=y$, or $x$ and $y$ differ in exactly $n/2$ positions
\end{quote}
Note that this promise only makes sense if $n$ is an even number, otherwise $n/2$ would not be integer.
In fact it will be convenient to assume $n$ a power of 2.
Here is a simple quantum protocol to solve this promise version of equality using only $\log n$ qubits:
\begin{enumerate}
\item Alice sends Bob the $\log n$-qubit state $\frac{1}{\sqrt{n}}\sum_{i=1}^n(-1)^{x_i}\ket{i}$,
which she can prepare unitarily from $x$ and $\log n$ $\ket{0}$-qubits.
\item Bob applies the unitary map $\ket{i}\mapsto(-1)^{y_i}\ket{i}$ to the state, applies a Hadamard transform
to each qubit (for this it is convenient to view $i$ as a $\log n$-bit string), and measures the resulting $\log n$-qubit state.
\item Bob outputs 1 if the measurement gave $\ket{0^{\log n}}$ and outputs 0 otherwise.
\end{enumerate}
It is clear that this protocol only communicates $\log n$ qubits, but why does it work?
Note that the state that Bob measures is
$$
H^{\otimes\log n}\left(\frac{1}{\sqrt{n}}\sum_{i=1}^n(-1)^{x_i+y_i}\ket{i}\right)=
\frac{1}{n}\sum_{i=1}^n (-1)^{x_i+y_i}\sum_{j\in\01^{\log n}}(-1)^{i\cdot j}\ket{j}
$$
This superposition looks rather unwieldy, but consider the amplitude of the $\ket{0^{\log n}}$ basis state.
It is $\frac{1}{n}\sum_{i=1}^n (-1)^{x_i+y_i}$, which
is 1 if $x=y$ and $0$ otherwise because the promise now guarantees
that $x$ and $y$ differ in exactly $n/2$ of the bits!  Hence Bob will always give the correct answer.

What about efficient \emph{classical} protocols (without entanglement) for this problem?
Proving lower bounds on communication complexity often requires a very technical combinatorial
analysis.  Buhrman, Cleve, and Wigderson used a deep combinatorial result of Frankl and R\"{o}dl~\cite{frankl&rodl:forbidden}
to prove that every classical errorless protocol for this problem needs to send at least $0.007 n$ bits.
We give the details in Appendix~\ref{app:DJ}.

This $\log n$-qubits-vs-$0.007 n$-bits example was the first exponentially large
separation of quantum and classical communication complexity.
Notice, however, that the difference disappears if we move to the \emph{bounded-error} setting,
allowing the protocol to have some small error probability. We can
 use the randomized protocol for equality discussed above or even simpler:
Alice can just send a few $(i,x_i)$ pairs to Bob, who then compares the $x_i$'s with his $y_i$'s.
If $x=y$ he will not see a difference, but if $x$ and $y$ differ in $n/2$ positions,
then Bob will probably detect this. Hence $O(\log n)$ classical bits of communication
suffice in the bounded-error setting, in sharp contrast to the errorless setting.

\subsection{The Intersection problem}

Now consider the Intersection function, which is 1 if $x_i=y_i=1$ for at least one $i$.
Note that this is a decision problem of the appointment-scheduling problem mentioned in the introduction.
Buhrman, Cleve, and Wigderson~\cite{BuhrmanCleveWigderson98} also presented an efficient quantum protocol for this.
Their protocol is based on Lov Grover's famous quantum search algorithm~\cite{grover:search},
which we will briefly sketch here.

Suppose there is some $n$-bit string $z$ and we would like to find an index $i$ such that $z_i=1$.
We cannot ``look'' at $z$ directly, but we can apply the following unitary map:
$$
O_z:\ket{i}\mapsto(-1)^{z_i}\ket{i}.
$$
Grover's algorithm starts in a uniform superposition $\frac{1}{\sqrt{n}}\sum_{i=1}^n\ket{i}$
and then repeatedly applies the following unitary \emph{Grover iterate} to the state:
$$
G=H^{\otimes\log n}O_0H^{\otimes\log n}O_z,
$$
where $H^{\otimes\log n}$ is the $\log n$-qubit Hadamard transform, and $O_0$ is the unitary that puts a `$-$' in front
of the all-0 state. Suppose there are exactly $t$ solutions: $t$ indices $i$ where $z_i=1$.
We will not give the analysis here (see for instance~\cite{bhmt:countingj}),
but one can show that after about $\frac{\pi}{4}\sqrt{n/t}$ Grover-iterations, most of the amplitude of the state sits
on such solutions.  Measuring the state will now with high probability give us a solution.
Of course we may not know $t$ in advance, but there is a way to find a solution with high probability
using $O(\sqrt{n})$ Grover-iterates even in that case.

Now what about the Intersection problem?  Note that we just want to find a solution for
the string $z=x\wedge y$, which is the bit-wise AND of $x$ and $y$,
since  $z_i =1$ whenever both $x_i=1$ and $y_i=1$. The idea is now to let Alice run
Grover's algorithm to search for such a solution. Clearly, she can prepare the uniform starting state herself.
She can also apply $H$ and $O_0$ herself.  The only thing where she needs Bob's help, is in implementing
$O_z$. This they do as follows. Whenever Alice want to apply $O_z$ to a state
$$
\ket{\phi}=\sum_{i=1}^n\alpha_i\ket{i},
$$
she tags on her $x_i$ in an extra qubit and sends Bob the state
$$
\sum_{i=1}^n\alpha_i\ket{i}\ket{x_i}.
$$
Bob applies the unitary map
$$
\ket{i}\ket{x_i}\mapsto(-1)^{x_i\wedge y_i}\ket{i}\ket{x_i}
$$
and sends back the result. Alice sets the last qubit back to $\ket{0}$ (which she
can do unitarily because she has $x$), and now she has the state $O_z\ket{\phi}$!
Thus we can simulate $O_z$ using 2 messages of $\log(n)+1$ qubits each.
Thus Alice and Bob can run Grover's algorithm to find an intersection, using
$O(\sqrt{n})$ messages of $O(\log n)$ qubits each, for total communication of $O(\sqrt{n}\log n)$ qubits.
Later Aaronson and Ambainis~\cite{aaronson&ambainis:search} gave a more complicated protocol
that uses $O(\sqrt{n})$ qubits of communication.

What about lower bounds? It is a well-known result of classical communication complexity that
classical bounded-error protocols for the Intersection problem need about $n$ bits
of communication~\cite{ks:disj,razborov:disj}.
Thus we have a quadratic quantum-classical separation for this problem.
Could the separation be even bigger than quadratic? This question was open for quite a few
years after~\cite{BuhrmanCleveWigderson98} appeared, until finally Razborov~\cite{razborov:qdisj}
showed that any bounded-error quantum protocol for Intersection needs to communicate
about $\sqrt{n}$ qubits.
His proof is beautiful but deep and complicated.  We sketch it in Appendix~\ref{appquantumlowerbounds}.

\subsection{Raz's problem}\label{sec:ranraz}

Notice the contrast between the examples of the last two sections.
For the Distributed Deutsch-Jozsa problem we get an \emph{exponential}
quantum-classical separation, but the separation only holds
if we require the classical protocol to be errorless.
On the other hand, the gap for the disjointness function is only
\emph{quadratic}, but it holds even if we allow classical protocols
to have some error probability.

Raz~\cite{raz:qcc} exhibited a function where the quantum-classical
separation has both features: the quantum protocol is exponentially better than
the classical protocol, even if the latter is allowed some error probability.
Consider the following promise problem $\bf P$:
\begin{quote}
Alice receives a unit vector $v\in\mathbb{R}^m$ and a decomposition of the
corresponding space in two orthogonal subspaces $H^{(0)}$ and $H^{(1)}$.\\
Bob receives an $m\times m$ unitary transformation $U$.\\
Promise: $Uv$ is either ``close'' to $H^{(0)}$ or to $H^{(1)}$
(more precisely, letting $P$ be the projector on subspace $H$, a vector $v$ is close to $H$ if $\norm{Pv}^2 \geq 2/3$).\\
Question: which of the two?
\end{quote}
As stated, this is a problem with continuous input,
but it can be discretized in a natural way by approximating
each real number by $O(\log m)$ bits. Alice and Bob's input
is now $n=O(m^2\log m)$ bits long.
There is a simple yet efficient 2-round quantum protocol for this problem:
Alice views $v$ as a $\log m$-qubit vector and sends this to Bob.
Bob applies $U$ and sends back the result.
Alice then measures in which subspace $H^{(i)}$ the vector $Uv$
lies and outputs the resulting $i$.
This takes only $2\log m=O(\log n)$ qubits of communication.

The efficiency of this protocol comes from the fact that an
$m$-dimensional unit vector can be ``compressed'' or ``represented''
as a $\log m$-qubit state.
Similar compression is not possible with classical bits, which
suggests that any classical protocol for $\bf P$ will have to send
the vector $v$ more or less literally and
hence will require a lot of communication.
This turns out to be true but the proof (given in~\cite{raz:qcc})
is surprisingly hard. It shows that any bounded-error protocol for $\bf P$ needs
to send at least about $n^{1/4}/\log n$ bits.

\subsection{The Hidden Matching problem}\label{HMp}

Consider the following promise problem $\bf HM$ from~\cite{bjk:q1way}, for even integer $n$:
\begin{quote}
Alice receives a string $x\in \01^n$.\\
Bob receives a perfect matching $M$ on $\{1,\ldots,n\}$
(i.e., a partition into $n/2$ disjoint pairs $M=\{(i_1,j_1),\ldots,(i_{n/2},j_{n/2})\}$).\\
Question: output a triple $(i,j,x_i\oplus x_j)$ for some $(i,j)\in M$.
\end{quote}
This communication problem is not a function, but a \emph{relation}:
for each input-pair $x,M$ there are $n/2$ different correct answers
instead of only one: $(i,j, x_i\oplus y_i)$ is correct for each $(i,j) \in M$.
We consider \emph{one-way} protocols here, where Alice sends one message to Bob
and then Bob should produce a triple $(i,j,x_i\oplus x_j)$.

We now describe a quantum protocol where Alice sends only $O(\log n)$
qubits and Bob gives one of the correct answers with probability~1~\cite{bjk:q1way}.
Alice sends Bob the following $\log n$-qubit message:
$$
\frac{1}{\sqrt{n}}\sum_{i=1}^n(-1)^{x_i}\ket{i}.
$$
Bob views $M$ as an orthogonal decomposition of the space $\mathbb{C}^n$ into $n/2$
2-dimensional subspaces. For instance, the projector for the subspace corresponding
to $(i,j)\in M$ would be $P_{ij}=\ketbra{i}{i}+\ketbra{j}{j}$.  Bob applies this measurement on
the state he received, and obtains the label of some random $(i,j)\in M$ as well as the projected state
$$
\frac{1}{\sqrt{2}}\left((-1)^{x_i}\ket{i}+(-1)^{x_j}\ket{j}\right).
$$
An appropriate measurement on this state will give Bob the bit $x_i\oplus x_j$ with certainty,
and he can output the correct answer $(i,j,x_i\oplus x_j)$.

What about classical protocols? First note that the {\bf HM} problem can be solved by a short classical message from Bob to Alice: Bob sends Alice a pair $(i,j)\in M$ using $2 \log n$ bits, which allows Alice to compute $x_i\oplus x_j$.
But the situation is radically different if we consider
\emph{classical one-way communication from Alice to Bob only}.
Indeed, one can show that if Alice sends Bob pairs $(i,x_i)$ for
$O(\sqrt{n})$ randomly chosen $i$'s, then Bob probably received both
points from at least one pair in $M$\footnote{This is due to
  an effect called the ``birthday paradox'' or ``birthday problem''. It
  states that if we throw roughly $\sqrt{n}$ balls into $n$ bins at random,
then probably there will be a bin containing at least two balls.}\label{foot:birthday}.
This allows him to output a correct answer.  On the other hand, Bar-Yossef, Jayram, and Kerenidis~\cite{bjk:q1way}
proved that any classical  protocol solving the Hidden Matching
problem, even with small error probability and involving only  one-way communication from Alice to Bob needs messages
of length at least about $\sqrt{n}$.  Thus we have an exponential separation between classical one-way protocols
and quantum one-way protocols.

Variants of the Hidden Matching problem have been used recently to obtain other quantum-classical separations.
For example, Gavinsky et al.~\cite{gkkrw:1way} showed a $\log n$-qubits-versus-$\sqrt{n}$-classical-bits
separation for one-way protocols for a Boolean \emph{function} derived from the Hidden Matching problem
(while {\bf HM} itself is a \emph{relational} problem).
Gavinsky~\cite{gavinsky:interactionvsquantum} used another variant of {\bf HM} to exhibit a relational
problem where quantum one-way protocols are exponentially more efficient than classical \emph{two-way} protocols.%

\subsection{Inner product}
\label{sect:InnerProduct}

In the previous sections we gave examples of quantum-classical separations.
The parameters were different, but in each case we showed that there was a
quantum protocol for the problem at hand that required far less communication than the best classical protocols.
Could this always be the case?  Could quantum communication complexity be much more efficient
for \emph{every} communication complexity problem?
The answer to this is negative---in fact for most communication complexity problems,
quantum communication does not help much.

An important example is the inner product function ($\IP(x,y)=x\cdot y=\sum_{i=1}^n x_i y_i\pmod 2$).
All protocols, both classical and quantum, need to send about $n$ bits/qubits to solve this.
We will sketch the proof of~\cite{cdnt:ip} here for the
  case of errorless quantum protocols with qubit communication and
  without entanglement, the proof for the more general case of entanglement is slightly
  more complicated.
The proof uses the $\IP$-protocol to communicate Alice's $n$-bit
input to Bob, and then invokes Holevo's theorem to conclude that
many qubits must have been communicated in order to achieve this.
Suppose Alice and Bob have some protocol $P$ for $\IP$.
They can use this to compute the following mapping\footnote{
This is an
  oversimplification of matters: in order to get
  the map of Eq.~(\ref{state:clean}) one first  needs
  to construct a new protocol $P^{-1}$ which is the reverse of the original
  communication protocol $P$. This can be done without
  error because the original protocol is without error. Combining protocols
$P$ and  $P^{-1}$ one can obtain map~(\ref{state:clean}). If protocol
  $P$ uses $c$ qubits of communication, protocol $P^{-1}$ also uses $c$
  qubits, and the protocol for obtaining state~(\ref{state:clean}) uses $2c$ qubits. But the crucial point is that still at most $c$ qubits are
  sent from Alice to Bob, since $P^{-1}$ is the reverse of $P$.
  Holevo's theorem lower bounds the
  communication from Alice to Bob, and  hence we get a lower bound of
  $n$ qubits on $c$.}:
\begin{equation}\label{state:clean}
\ket{x}\ket{y}\mapsto\ket{x}(-1)^{x\cdot y}\ket{y}.
\end{equation}
Now suppose Alice starts with an arbitrary $n$-bit state $\ket{x}$
and Bob starts with the uniform superposition
$\frac{1}{\sqrt{2^n}}\sum_{y\in\01^n}\ket{y}$.
If they apply the above mapping, the final state becomes
$$
\ket{x}\frac{1}{\sqrt{2^n}}\sum_{y\in\01^n}(-1)^{x\cdot y}\ket{y}.
$$
If Bob applies a Hadamard transform to each of his $n$ qubits,
then he obtains the basis state $\ket{x}$, so Alice's $n$ classical
bits have been communicated to Bob.
Holevo's theorem now implies that the \IP-protocol must communicate $n$ qubits (which can trivially be achieved).
The same argument can, with a minor modification, be made to work even
if Alice and Bob share unlimited prior entanglement, yielding a lower
bound of $n/2$ qubits (which can trivially be achieved using dense coding).
With some more technical complication, the same idea gives an
$\frac{1}{2}(1-2\eps)^2n$ lower bound for
$\eps$-error protocols~\cite{cdnt:ip}.
The constant factor in this bound was subsequently improved to the optimal
$\frac{1}{2}$ by Nayak and Salzman~\cite{nayak&salzman:entanglement}.

\section{Non-locality and Communication Complexity}\label{sec:NLandCC}

\subsection{Converting communication complexity to non-locality}

In Section~\ref{sec:nl} we introduced several simple non-locality
scenarios. Then in Section~\ref{sec:cc} we introduced communication
complexity, and gave several problems for which there are large,
sometimes exponential, separations between the classical and quantum
communication complexity. In this section we shall put together these
two approaches, and derive from the communication complexity problems
new non-locality problems which are very hard, sometimes exponentially
hard, to solve in a classical model. In particular we shall present
non-locality problems based on the Distributed Deutsch-Jozsa problem
and on the Hidden Matching problem. In Section~\ref{sec:impl} we
shall come back to these non-locality problems, and will discuss
these newly developed tests in the context of experimental errors.

In this section we shall use the following mapping which, when applicable, is very powerful.

\begin{quote}
{\bf Mapping one-way quantum communication complexity to non-locality.}\\
Consider a communication complexity problem where the
number $q$ of qubits exchanged in the quantum communication model with \textit{one-way}
communication from Alice to Bob is less than the number $c$ of
bits required to solve the problem classically when the parties have
shared randomness;  and further suppose that---due to some symmetry
of the problem---it can be solved if Alice starts with an arbitrary
basis state $\ket{k}$ (the value of $k$ being known beforehand to
both Alice and Bob) as follows: she carries out a transformation $U_A(x)$ on this state (that depends on her input $x$ but does not depend on $k$), sends it to Bob who carries out a transformation $U_B(y)$ (that depends on his input $y$ but does not depend on $k$) and then measures in the computational basis. The probability of finding result $\ell$ is thus
$|\bra{\ell} U_B(y) U_A(x)\ket{k}|^2$. From the knowledge of $\ell$, $k$, and $y$, Bob can find the value of the function $f(x,y)$.

Now consider the following process: Alice and Bob share a maximally entangled state $|\psi\rangle = 2^{-q/2}\sum_{i=0}^{2^q-1} \ket{i}\ket{i}$; Alice carries out a local transformation $U_A(x)^T$ (where `$T$' means transposition in the $\ket{i}$-basis); she measures in the computational basis. Bob carries out the transformation $U_B(y)$; he measures in the computational basis. Suppose that Alice obtains outcome $k$ and Bob obtains outcome $\ell$.
The probability of finding these joint outcomes is
$P(k,\ell|x,y)=|\bra{\ell}\bra{k}U_B(y)  U_A(x)^T \ket{\psi}|^2=2^{-q}|\bra{\ell} U_B(y) U_A(x)\ket{k}|^2$
(the last equality is easy to check).
If Alice now sends to Bob the outcome $k$ of her measurement (which requires $q$ bits), then Bob can compute $f(x,y)$. Thus this constitutes a solution of the communication complexity problem in the entanglement model with half the communication that would be required if they had used the trivial mapping based on teleportation. More importantly, the correlations $P(k,\ell|x,y)$ are non-local, since they could not be obtained in a classical model with shared randomness without at least $c-q>0$ bits of classical communication.
\end{quote}

\subsection{Non-local version of the Distributed Deutsch-Jozsa problem}

The above mapping can be applied to the Distributed Deutsch-Jozsa problem from Section~\ref{secdistributeddj}. We describe here the result of the mapping.
\begin{quote}
{\bf Non-local DJ problem:}
Alice and Bob receive $n$-bit inputs $x$ and $y$ that satisfy the DJ promise:
either $x=y$, or $x$ and $y$ differ in exactly $n/2$ positions. The task is for Alice and Bob to provide outputs $a,b\in \{0,1\}^{\log n}$ such that
when $x=y$ then $a=b$, and when $x$ and $y$ differ in exactly $n/2$ positions then $a\neq b$.
\end{quote}
They achieve this as follows
\begin{enumerate}
\item Alice and Bob share the maximally entangled state
$\displaystyle\frac{1}{\sqrt{n}}\sum_{i=0}^{n-1}\ket{i}\ket{i}$.
\item Alice and Bob both apply locally a conditional phase to obtain:
$\displaystyle\frac{1}{\sqrt{n}}\sum_{i=0}^{n-1} (-1)^{x_i}(-1)^{y_i}\ket{i}\ket{i}$.
\item Alice and Bob both apply a Hadamard transform:
$\displaystyle\frac{1}{n\sqrt{n}}\sum_{a=0}^{n-1}\sum_{b=0}^{n-1}  \left (\sum_{i=1}^n (-1)^{x_i + y_i + i\cdot(a\oplus b)}\right)\ket{a}\ket{b}$.
\item Alice and Bob measure in the computational basis.
\end{enumerate}
For every $a$, the probability that both Alice and Bob obtain the same result $a$ is:
\[
\left|\frac{1}{n\sqrt{n}}\sum_{i=0}^{n-1} (-1)^{x_i + y_i}\right|^2,
\]
which is $1/n$ if $x=y$ and 0 otherwise. Hence this solves the problem.

Note that if Alice then communicated the result of her measurement to Bob
(using $\log n$ bits), he could solve the Distributed Deutsch-Jozsa problem since he could then check whether $k=\ell$ or $k\neq\ell$. But we know that solving the Distributed Deutsch-Jozsa problem requires at least $0.007n$ bits.
Thus we have a non-locality problem that can be solved if Alice and Bob share $\log n$ ebits, but which requires about $0.007n$ bits to be solved in a classical model with shared randomness and classical communication. Note that this very large lower bound on the amount of classical communication would disappear in the bounded-error setting where we allow the correlations $P(a,b|x,y)$ to differ slightly from the ideal correlations.

\subsection{Non-local version of the Hidden Matching problem}

The same mapping can be applied to the Hidden Matching problem to yield a non-locality problem.
\begin{quote}
{\bf Non-local HM problem:}
Assume that $n=2^m$, so we can index the numbers between 1 and $n$ with $m$-bit strings.\\
Alice receives a string $x\in \01^n$. Bob receives a perfect matching $M$ on $\{1,\ldots,n\}$
(i.e.~a partition into $n/2$ disjoint pairs).\\
Alice must give as output some $k\in \01^m$. Bob must give as output a matching $(i,j)\in M$ and $\ell\in \01^m$.\\
Alice and Bob's output must satisfy $i\cdot(k\oplus \ell)) +
j\cdot(k\oplus\ell) = x_i + x_j$ mod 2 (recall that $a\cdot b=\sum_i a_ib_i$ is the
  inner product between bitstrings $a$ and $b$, and $a\oplus b$ is
  the bitwise XOR of $a$ and $b$: the $i$th bit of $a \oplus b$ is $ a_i \oplus b_i$).
\end{quote}

Note that if at the end of the protocol, Alice sends $k$ to Bob at a cost of $m=\log n$ classical bits,
then Bob has enough information to compute the triple $(i,j,x_i\oplus x_j)$, i.e., to solve the Hidden Matching problem as defined in Section~\ref{HMp}. But we know that classical one-way communication from Alice to Bob needs about $\sqrt{n}$ bits to solve the Hidden Matching problem. Therefore the correlations in the non-local HM problem themselves can only be reproduced if Alice sends Bob at least about $\sqrt{n}$ bits of communication (if we are restricted to one-way).

Let us show that Alice and Bob can obtain the correlations of the non-local HM problem using local measurements on $m=\log n$ ebits.
The initial state is:
$$
\frac{1}{\sqrt{n}}\sum_{i\in\01^m}\ket{i}\ket{i}.
$$
Alice adds the phases $(-1)^{x_i}$.
Bob views $M$ as an orthogonal decomposition of the space $\mathbb{C}^n$ into $n/2$
2-dimensional subspaces. For instance, the projector for the subspace corresponding
to $(i,j)\in M$ would be $P_{ij}=\ketbra{i}{i}+\ketbra{j}{j}$.  Bob applies this measurement on
the state he received, and obtains the label of some random $(i,j)\in M$. This projects the joint state to
$$
\frac{1}{\sqrt{2}}\left((-1)^{x_i}\ket{i}\ket{i}+(-1)^{x_j}\ket{j}\ket{j}\right).
$$
Now they both apply Hadamard transforms to each of their qubits.  This gives the state
$$
\frac{1}{\sqrt{2}}\left(\frac{(-1)^{x_i}}{n}\sum_{k,\ell\in\01^m}(-1)^{i\cdot k+i\cdot \ell}\ket{k}\ket{\ell}+
\frac{(-1)^{x_j}}{n}\sum_{k,\ell\in\01^m}(-1)^{j\cdot k+j\cdot \ell}\ket{k}\ket{\ell}\right)
$$
$$
=\frac{1}{n\sqrt{2}}\sum_{k,\ell\in\01^m}\left((-1)^{x_i+i\cdot(k\oplus \ell)}+(-1)^{x_j+j\cdot(k\oplus\ell)}\right)\ket{k}\ket{\ell}.
$$
Both parties measure their half of the state in the computational basis.
They obtain $m$-bit strings $k$ and $\ell$, respectively, satisfying
$x_i +  i\cdot(k\oplus \ell)= x_j +  j\cdot(k\oplus\ell)$ (modulo 2), since the other $k,\ell$-pairs have amplitude 0.
This gives: $i\cdot(k\oplus \ell) + j\cdot(k\oplus\ell) = x_i + x_j$ (modulo 2).

\section{Quantum Fingerprinting and the Simultaneous Message Passing Model}\label{sec:SMP}
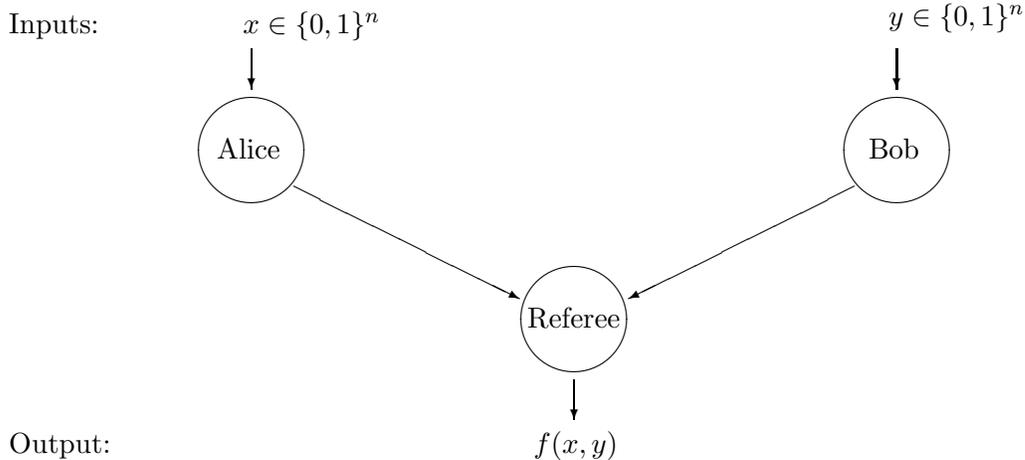
\begin{figure}[h]
\setlength{\unitlength}{100000sp}
\vspace{2mm}
\begin{center}
\begin{picture}(280,115)(0,-45)
\put(60,35){\circle{30}}
\put(51.5,33){Alice}
\put(58,64){$x \in \{0,1\}^n$}
\put(60,60){\vector(0,-1){10}}
\put(70.5,26){\vector(2,-1){56}}
\put(209.5,26){\vector(-2,-1){56}}
\put(140,-7){\circle{30}}
\put(128.5,-9){Referee}
\put(130,-40){$f(x,y)$}
\put(140,-22){\vector(0,-1){10}}
\put(220,35){\circle{30}}
\put(213,33){Bob}
\put(218,66){$y \in \{0,1\}^n$}
\put(220,60){\vector(0,-1){10}}
\put(0,64){Inputs:}
\put(0,-40){Output:}
\end{picture}
\end{center}
\caption{The \text{simultaneous message passing} variant of the
communication complexity scenario:
Alice and Bob receive $n$-bit strings, $x$ and $y$ respectively,
as input and their communication is restricted to each sending
one message to a third party, called the Referee.
From these messages, the Referee computes some function $f(x,y)$
as the output of the protocol.
There are tasks of this form where communication in terms of quantum messages
is exponentially more efficient than communication in terms of classical
messages.}\label{fig:smp}
\end{figure}

We now describe a model, called the \textit{simultaneous message passing (SMP)}
model, that is neither a non-locality test nor the full-fledged
communication complexity scenario, yet that is relevant to both.
The basic structure is illustrated in Fig.~\ref{fig:smp}.
Alice and Bob each receive an $n$-bit input ($x$ and $y$, respectively).
In this scenario, they do not have any shared resources like shared
randomness or an entangled state, but they do have local randomness.
They each are required to send a single message to a third party, called
the Referee.
The Referee, upon receiving message $m_A$ from Alice and $m_B$ from Bob,
should output the value of some (Boolean) function $f(x,y)$.
The goal is to compute $f(x,y)$ with a minimum amount of communication
from Alice and Bob to the Referee.
This scenario was introduced by Yao~\cite{yao:distributive} for the
setting where $m_A$ and $m_B$ are classical messages consisting of bits.
We compare this classical model to the corresponding quantum version,
where $m_A$ and $m_B$ consist of qubits.
We will see that for the very natural problem of equality,
where $f(x,y) =1$ if and only if $x=y$, there is an exponential savings
in communication when qubits are used instead of classical bits.
Classically, the problem of the bounded-error communication complexity of
equality in the SMP model was open for almost twenty years, until Newman and
Szegedy~\cite{newman&szegedy:1round} exhibited a lower bound of about $\sqrt{n}$ bits.
This is tight, since Ambainis~\cite{ambainis:3computer} constructed a bounded
error protocol for this problem where the messages are
$O(\sqrt{n})$ bits long (we describe a slightly less efficient classical
protocol in Section~\ref{sseceqclassical}).
In contrast, Buhrman, Cleve, Watrous, and de Wolf~\cite{bcww:fp} showed that
in the quantum setting this problem can be solved with very little communication:
only $O(\log{n})$ qubits suffice.

\subsection{Quantum fingerprints}
In order to construct the efficient quantum SMP protocol for equality,
we need to borrow ideas from the efficient classical randomized communication
complexity protocol for equality from
Section~\ref{sec:setting}. Recall that in that protocol, Alice interprets her
input $x$ as a polynomial $p_x(t) = \sum_{i=1}^n x_i t^{i-1}$ over
some finite field $\mathbb{F}$ of size $m$ (about $3n$), and then she picks a random point
$a \in \mathbb{F}$ and sends $a$ and $p_x(a)$ to Bob.
The pair $a,p_x(a)$ is called a ``fingerprint'' of $x$, since it describes
characteristics of $x$ that can aid in identifying it.
Carrying out this fingerprinting procedure in superposition results in
a \emph{quantum fingerprint} of $x$:
\[
\ket{F_x} = \frac{1}{\sqrt{m}}\sum_{a \in \mathbb{F}} \ket{a}\ket{p_x(a)}.
\]
Note that $\ket{F_x}$ consists of only $2\log{m} = 2\log{n} + O(1)$ qubits.

\subsection{Classical protocol for equality}\label{sseceqclassical}
A nearly optimal\footnote{Ambainis's protocol from~\cite{ambainis:3computer} gets rid of the $\log{n}$ factor.}
$O(\sqrt{n}\log{n})$ classical protocol for equality in the SMP model goes
as follows. Alice produces a list of $k=O(\sqrt{n})$ random points $a_1,\ldots,a_k$ in $\mathbb{F}$
and sends the list $\{(a_i, p_x(a_i))\}_{i=1}^{k}$ to the Referee.
Bob does the same with respect to $y$, sending $\{(b_i, p_y(b_i))\}_{i=1}^{k}$ to the Referee.
By the birthday paradox (see the footnote in
Section~\ref{foot:birthday}), with constant probability there exist $i$ and $j$ such that both $a_i$ and $b_j$
equal the same field element $d$. In this case the Referee can compare $p_x(d)$ with $p_y(d)$.
If $x=y$ then $p_x=p_y$, and hence $p_x(d) = p_y(d)$.  On the other hand, if $x \neq y$,
then since $p_x$ and  $p_y$ are \emph{different} polynomials of degree at most $n-1$, with probability
 $\geq 2/3$, we have $p_x(d) \neq p_y(d)$. The protocol for
 the Referee is now clear: if the lists of Alice and Bob have a
 point $d$ in common, then the Referee outputs $1$ if and only if
 $p_x(d)=p_y(d)$. If there is no point in common (which happens
 only with small probability) or if $p_x(d)\neq p_y(d)$, then the Referee outputs~0.

\subsection{Quantum protocol for equality}\label{ssecqfpprotocol}
We now have everything in place to describe the quantum protocol for
equality. Alice sends  state $\ket{F_x}$ to the Referee and Bob sends
$\ket{F_y}$. Note that if the Referee now  measures $\ket{F_x}$ in the computational
basis, then he will find a random point $a$ and the value $p_x(a)$, just
like the classical protocol described above. The Referee thus needs to do something smarter.
The key observation is the following about the inner products between fingerprints:
\begin{equation}
\inp{F_x}{F_y} =
\begin{cases}
1 & \mbox{if $x=y$} \\
\leq \frac{1}{3} & \mbox{if $x \not = y$}
\end{cases}
\end{equation}
If $x=y$ then clearly $\inp{F_x}{F_y} =1$.  If $x
\neq y$ then
$$
\inp{F_x}{F_y} = \frac{1}{m}\sum_{i,j \in \mathbb{F}}
\bra{p_x(i)}\inp{i}{j}\ket{p_y(j)} = \frac{1}{m}\sum_{i \in
 \mathbb{F}} \inp{p_x(i)}{p_y(i)}.
$$
Since $p_x$ and $p_y$ are different polynomials of degree at most $n-1$,
they have the same value $p_x(i)=p_y(i)$ for at most $n-1$ values of $i$.
Hence the inner product is at most $\frac{n-1}{m}\leq \frac{1}{3}$.

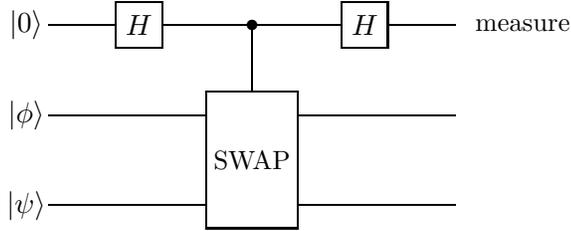
\begin{figure}[hbt]
\centering
\setlength{\unitlength}{0.3mm}
\begin{picture}(200,105)
\put(-10,80){\makebox(20,20){$\ket{0}$}}
\put(-10,40){\makebox(20,20){$\ket{\phi}$}}
\put(-10,0){\makebox(20,20){$\ket{\psi}$}}
\put(210,80){\makebox(20,20){\small measure}}
\put(10,90){\line(1,0){30}}
\put(40,80){\framebox(20,20){$H$}}
\put(140,80){\framebox(20,20){$H$}}
\put(60,90){\line(1,0){80}}
\put(160,90){\line(1,0){30}}
\put(100,90){\line(0,-1){30}}
\put(100,90){\circle*{5}}
\put(80,0){\framebox(40,60){\small SWAP}}
\put(10,50){\line(1,0){70}}
\put(10,10){\line(1,0){70}}
\put(120,50){\line(1,0){70}}
\put(120,10){\line(1,0){70}}
\end{picture}
\caption{Quantum circuit to test if $\ket{\phi}=\ket{\psi}$ or
$|\inp{\phi}{\psi}| \le \frac{1}{3}$.}\label{fig:swaptest}
\end{figure}

When Alice and Bob send their quantum fingerprints to the Referee, he
has to determine  the inner product between the two states he receives.
The following test (Figure~\ref{fig:swaptest}), sometimes called
the SWAP-test, accomplishes this task with a small error probability.

This circuit first applies a Hadamard transform to a qubit that is initially $\ket{0}$, then SWAPs
the other two registers conditioned on the value of the first qubit being $\ket{1}$,
then applies another Hadamard transform to the first qubit and measures it.
Here SWAP is the operation that swaps the states $\ket{\phi}$ and
$\ket{\psi}$: $\ket{\phi}\ket{\psi} \mapsto \ket{\psi}\ket{\phi}$. The
Referee receives $\ket{\phi}$ from Alice and $\ket{\psi}$ from Bob and applies the test to these two states. An easy calculation reveals that the outcome of the
measurement is $1$ with probability $(1-|\inp{\phi}{\psi}|^2)/2$.
Hence if $\ket{\phi}=\ket{\psi}$ then we observe a 1 with probability 0,
but if $|\inp{\phi}{\psi}| \le \frac{1}{3}$ then this probability
is $\ge \frac{4}{9}$. Repeating this procedure with several individual
fingerprints can make the error probability arbitrary close to $0$.

\subsection{Subsequent work in the SMP model}
After the quantum fingerprinting scheme showed the power of quantum communication in the SMP model,
a number of further results appeared. Yao~\cite{yao:qfp} exhibited an efficient protocol for testing if
the inputs $x$ and $y$ are at some constant Hamming distance $d$, while Gavinsky et al.~\cite{gkw:fingerprinting}
related quantum fingerprinting to a technique from machine learning which brings out its weaknesses.
One can also study the variant of the SMP model where Alice and Bob start with a shared entangled state, but can only send
\emph{classical} messages to the Referee.
Gavinsky et al.~\cite{gkrw:identification} exhibited a problem based on the Hidden Matching problem
and a quantum protocol that solves it with $O(\log n)$ ebits and $O(\log n)$
classical bits of communication, while any quantum SMP protocol without prior entanglement needs
to send at least about $(n/\log n)^{1/3}$ qubits.
This shows that entanglement can reduce communication (even quantum communication!) exponentially,
at least for relational problems in the SMP model.\footnote{Recently, Gavinsky~\cite{gavinsky:interactionvsnonlocality}
extended this to a similar separation in the more standard two-way model.}
Finally, Gavinsky, Regev, and de Wolf~\cite{grw:qcsmp} showed that if Alice's message to the referee
is allowed to be quantum, while Bob's message can only be classical,
then the quantum advantages over purely classical protocols mostly disappear.
In particular, the equality problem requires communication at least $\sqrt{n/\log n}$ in this hybrid case.

\section{Other Aspects of Quantum Non-Locality}\label{sec:otherQNL}

\subsection{Non-local boxes}\label{sec:nlbox}
In previous sections we studied a hierarchy of resources. In particular, we discussed and compared the correlations $P(a,b|x,y)$ that can be obtained using only shared randomness, by local measurements on entangled states, and finally those that can be obtained if communication between the parties is allowed. In this section we discuss an interesting set of correlations that lie between the last two classes.

To understand these new correlations, let us note that any correlations $P(a,b|x,y)$ obtained in a local hidden variable model or by local measurements on an entangled state must obey the following properties:
\begin{eqnarray}
&\mbox{\textit{Positivity: }}& P(a,b|x,y)\geq 0;\label{positivity}\\
&\mbox{\textit{Normalization: }}& \sum_{a,b}P(a,b|x,y) = 1;\label{normalization}\\
&\mbox{\textit{No Signalling: }}& \sum_{b}P(a,b|x,y) = P(a|x) \mbox{ is independent of $y$,}\nonumber\\
&& \sum_{a}P(a,b|x,y) = P(b|y)\mbox{is independent of $x$.}\label{nosignalling}
\end{eqnarray}
The last condition expresses the fact that Bob cannot transmit any information about his input $y$ to Alice, and similarly Alice cannot communicate to Bob
any information about her input $x$.
We are interested here in correlations that obey the above three conditions, but that cannot be obtained
from local measurements on entangled states.

To illustrate this idea, suppose that Alice and Bob each have some kind of device (introduced independently in~\cite{khalfitsirelson} and in~\cite{PopescuRohrlich:NLbox}) such that Alice can provide an input $x\in\{0,1\}$ to her device and obtain an output $a\in\{0,1\}$; and Bob can provide an input $y\in\{0,1\}$ to his device and obtain an output $b\in\{0,1\}$, and such that the probabilities of the outputs given the inputs obey
\begin{equation}
P(a,b|x,y)=\left\{
\begin{array}{ll}
\frac{1}{2} & \mbox{ if }a\oplus b = x \wedge y\\
0           & \mbox{ otherwise.}
\end{array}\right.
\label{PRbox}
\end{equation}
Note that, much like the correlations that
can be established by use of quantum entanglement, this
device is atemporal: Alice gets her output as soon as she
feeds in her input, regardless of if and when Bob feeds in
his input, and vice versa. Also inspired by entanglement,
this is a one-shot device: the correlation appears only as a
result of the first pair of inputs fed in by Alice and Bob.
This device obeys the conditions 1 to 3 above, so it cannot be used to signal. We call it a \textit{non-local (NL) box} (other terminology in use is \textit{Popescu-Rohrlich (PR) box},
in reference to~\cite{PopescuRohrlich:NLbox}).

With this device Alice and Bob always obtain $a\oplus b = x \wedge y$, whereas we know that for local measurements on entangled quantum states this relation can only be satisfied with probability at most $\cos^2(\pi/8)$ under the uniform distribution on the inputs $x$ and $y$ (see Section~\ref{Tsirelson} for a proof).
Thus this is an ``imaginary'' device in the sense that it cannot be realized physically without Alice and Bob's devices being connected by some kind of communication channel.
It is, however, an interesting resource to consider, since it is ``stronger'' than correlations that can be obtained from local measurements on entangled states, but ``weaker'' than actual communication.

A systematic study of the properties of correlations obeying the above three conditions was initiated in~\cite{BLMPPR:NLbox}, and it was shown that they obey properties that one thinks of as genuinely quantum, such as monogamy and no-cloning~\cite{MAG:NLbox}. They also allow for secure key distribution~\cite{bhk:crypto}.

Because of the apparent ``reasonableness'' of the non-local box, Popescu and Rohrlich raised the question (in~\cite{PopescuRohrlich:NLbox}, and in fact well before this) why such correlations cannot be realized in nature without communication between the parties.
The most straightforward answer is the technical proof in Section~\ref{Tsirelson}; however, one might seek a more intuitive or philosophical explanation.
One possible approach is provided by communication complexity. It was shown by van Dam~\cite{vanDam:thesis,WvanDam:NLbox}, and also noted by one of the authors of the present review (Cleve), that if Alice and Bob have an unlimited amount of non-local boxes then all communication complexity problems become trivial:
\begin{quote}
Suppose Alice and Bob have an unlimited supply of non-local boxes, as described in Eq.~(\ref{PRbox}).
Suppose Alice receives input $x\in\{0,1\}^n$ and Bob receives input $y\in\{0,1\}^n$.
Then communication complexity becomes trivial, in the sense that the value of any Boolean function
$f(x,y)\in\{0,1\}$ can be computed with certainty with a \emph{single} bit of communication from Alice to Bob.
\end{quote}

To prove this, consider an arbitrary function $f:\{0,1\}^n\times\{0,1\}^n \rightarrow \{0,1\}$. 
It can be expressed as a boolean circuit consisting of \textsc{not} and $\wedge$ (\textsc{and}) gates, with inputs $x_1,\dots,x_n$ and $y_1,\dots,y_n$.
The idea is to represent the value of each gate of this circuit in terms of two \textit{shares},
one possessed by Alice and the other by Bob.
For a bit $a$, its representation as shares is any $(a',a'')$ where
$a = a' \oplus a''$.
Until the end of the protocol, Alice's information about each gate will be just 
the first bit of its share and Bob's information will be the second bit.
They start by constructing shares of the input bits: $(x_i,0)$ for each of Alice's input 
bits $x_i$ (Bob does not need to know $x_i$ to construct his share $0$); and similarly 
$(0,y_i)$ for each of Bob's input bits $y_i$.
For each gate in the circuit, if Alice and Bob collectively know the input bits as shares
then they can produce the shares for the output bit without any communication.
For each \textsc{not} gate, Alice merely negates her share (and Bob does nothing to his share).
For each $\wedge$ gate, assume that the shares of inputs are $(a',a'')$ and $(b',b'')$.
The shares of the output should be $(c',c'')$ such that
\begin{equation}
c' \oplus c''
 = (a' \oplus a'') \wedge (b' \oplus b'') 
 = (a' \wedge b') \oplus (a' \wedge b'')
\oplus (a'' \wedge b') \oplus (a'' \wedge b'')\, .
\end{equation}
Consider the four terms arising above.
Since Alice possesses $a'$ and $b'$, she can easily compute $a' \wedge b'$, and similarly Bob can compute $a'' \wedge b''$.
The difficult terms are $a' \wedge b''$ and $a'' \wedge b'$ because they contain bits that are spread between Alice and Bob---and this is where the non-local boxes are used.
Alice and Bob use one non-local box to obtain bits $d'$ and $d''$ so that 
$d' \oplus d'' = a' \wedge b''$.
They use a second non-local box to obtain $e'$ and $e''$ so that
$e' \oplus e'' = a'' \wedge b'$.
Then Alice sets her share to $c' = (a' \wedge b') \oplus d' \oplus e'$ and Bob sets his share to 
$c'' = (a'' \wedge b'') \oplus d'' \oplus e''$.
Clearly,
\begin{equation}
c' \oplus c''
= (a'\!\wedge b') \oplus (d'\!\oplus e'') \oplus (d'\!\oplus e'') \oplus (a''\!\wedge b'') 
= (a'\!\wedge b') \oplus (a'\!\wedge b'') \oplus (a''\!\wedge b') \oplus (a''\!\wedge b'') 
\, ,
\end{equation}
as required.
At the end, Alice and Bob possess shares for the value of $f$, and Alice sends her one-bit 
share to Bob, enabling him to compute the value of $f$.

%

Is this result specific to the non-local boxes of the form Eq.~(\ref{PRbox}) (in which case it could be viewed as some kind of anomaly in the space of all possible no-signalling correlations), or does it hold for other no-signalling correlations? In particular, does it hold for \emph{noisy} correlations? It was shown in~\cite{BBLMTU:NLbox} that the latter is the case, if one slightly adapts the definition of what it means for communication complexity to be trivial:

\begin{quote}
Suppose Alice and Bob have an unlimited supply of noisy non-local boxes whose outputs
satisfy Eq.~(\ref{PRbox}) with probability $p\geq \frac{3 + \sqrt{6}}{6} \approx 90.8\%$.
Then communication complexity becomes trivial, in the sense that
there exists $q>1/2$ (possibly depending on $p$, but on no other parameter) such that, for any $n\geq 0$, if
Alice receives input $x\in\{0,1\}^n$ and Bob receives input $y\in\{0,1\}^n$, then they can find with probability at least $q$ the value of any Boolean function
$f(x,y)\in\{0,1\}$ with a single bit of communication from Alice to Bob.
\end{quote}

Note that this result does not hold if Alice and Bob share entangled states instead of (noisy) non-local boxes. Indeed this follows from the result of~\cite{cdnt:ip}, discussed in Section~\ref{sect:InnerProduct}, that computing the inner product of two $n$-bit strings
with success probability $q>1/2$ requires $O(n)$ bits of communication, even if
Alice and Bob have an unlimited supply of entangled particles.

Thus the fact that communication complexity is not trivial (i.e., that
some communication complexity problems are hard whereas others are
easy) can be viewed as a partial characterization of the non-local
correlations that can be obtained by local measurements on entangled
particles. Is this a complete characterization? In particular, what is
the exact noise threshold $p$ where non-local boxes with noise $p$
render communication complexity trivial? The current bounds on $p$ are:
$85.4\% \approx \frac{2 + \sqrt{2}}{4} \leq  p \leq \frac{3 + \sqrt{6}}{6} \approx 90.8\%$.
If the lower bound is the correct one, we would have an interesting answer to the question 
raised by Popescu and Rohrlich.
We leave this as an open problem.

Another related open question arising by analogy with the process of entanglement purification~\cite{PhysRevLett.76.722}, is whether it is possible to ``purify'' non-local boxes? That is, given a supply of non-local boxes that work correctly with probability $p$, is it possible to produce, using only local operations, a non-local box with a success probability greater than $p$? For a first step in this direction, see~\cite{forster2008dnl}.


\subsection{Bell inequalities and Tsirelson bounds}

As discussed in the previous section, there are correlations, such as the non-local box, that cannot be reproduced by local measurements on entangled particles, but that nevertheless obey the conditions of positivity, normalisation and no-signalling Eqs.~(\ref{positivity}, \ref{normalization}, \ref{nosignalling}).
More generally, we would like to understand  within the space of all possible correlations ${\bf P}= \{ P(a,b|x,y)\}$ which ones can be obtained by using only shared randomness (i.e., by local hidden variable models), which ones can be realised by carrying out  local measurements on entangled particles, and what are the ultimate limits set by Eqs.~(\ref{positivity}, \ref{normalization}, \ref{nosignalling}).

Answering this question would address the question raised by Popescu and Rohrlich mentioned above, and would give us basic insights into communication complexity. Indeed it would allow us to understand quantitatively the differences between shared randomness, shared entanglement, and non-local correlations, each of which can be viewed as a different resource for communication complexity.
For instance answering this question can have immediate implications for communication complexity in the entanglement model, at least in the case where Alice and Bob use only one round of communication.

Before addressing this question it is useful to understand better the geometry of non-local correlations. To this end we introduce Bell expressions, that is linear combinations of the correlations
\begin{equation}\label{eq:Bell}
C({\bf P})=\sum_{abxy}c_{abxy} P(a,b|x,y)
\end{equation}
where $c_{abxy}$ are real numbers.
It is easy to show that the space of correlations that can be reproduced using local hidden variables (i.e., using only shared randomness) is a polytope. That is, it can be characterised by a finite number of inequalities, called Bell inequalities,  of the form
\begin{equation}
C({\bf P})\leq C_{\mbox{\scriptsize\sc lhv}} \ .
\end{equation}
To compute the maximum value allowed by local hidden variable (LHV) models, we can restrict ourselves to deterministic models, where $a=a(x)$ is a function of input $x$, and $b=b(y)$ is a function of $y$. We then have
$$C_{\mbox{\scriptsize\sc lhv}} = \max_{a(x),b(y)} \sum_{xy}c_{a(x)b(y)xy}\ .$$

If we consider local measurements on entangled quantum states, then we have bounds of the form
\begin{equation}
C({\bf P})\leq C_{\mbox{\scriptsize\sc qm}} \ .
\end{equation}
where
$$
C_{\mbox{\scriptsize\sc qm}} = \max \sum_{abxy}c_{abxy} \langle \psi | \Pi_a(x) \otimes \Pi_b(y) |\psi\rangle$$
where the maximum is taken over all states $|\psi\rangle$, and over all projective measurements
$\{ \Pi_a(x) \}$ (depending only on $x$) and projective measurements $\{ \Pi_b(y) \}$ (depending only on $y$). (By projective measurements, we mean a set of projectors $\Pi_a =\Pi_a^2$ that sum to the identity $\sum_a \Pi_a = I$).
Recently it has been shown how the quantum value $C_{\mbox{\scriptsize\sc lhv}}$ could be bounded by a hierarchy of semidefinite programs~\cite{navascues2007bsq,navascues2008chs,doherty2008qmp}, although the issue of whether this hierarchy converges remains open~\cite{scholz2008tsp}.

If we impose only the no-signalling conditions, then we will have
\begin{equation}
C({\bf P})\leq C_{\mbox{\scriptsize no-signalling}} \ .
\end{equation}
where the right hand side is the maximum of Eq.~(\ref{eq:Bell}) subject to
Eqs.~(\ref{positivity}, \ref{normalization}, \ref{nosignalling}). Note that Eqs.~(\ref{positivity}, \ref{normalization}, \ref{nosignalling}) 
define another polytope, the no-signalling polytope, and the maximum value of $C({\bf P})$ will be attained at a vertex of the polytope.

Let us illustrate the above concepts by a specific kind of Bell expression, called XOR non-local games~\cite{cleve2004cal}. In this particular case, the
outputs $a,b\in\{0,1\}$ are bits and we wish them to come as close as possible to satisfying a condition of the form
\begin{equation}\label{eq:XOR}
a\oplus b = f(x,y)
\end{equation}
for all $x,y$.
The most celebrated example is the CHSH case, where $x,y$ are also bits and the condition is
$a \oplus b = x\wedge y$, see Eq.~(\ref{eq:chsh}).

In the case of XOR games, we take the constants $c_{abxy}$ in Eq.~(\ref{eq:Bell}) to have the form:
\begin{equation}\label{eq:cc}
c_{abxy} = w_{xy}(-1)^{a\oplus b \oplus f(x,y)}=m_{xy}(-1)^{a\oplus b}
\end{equation}
where $w_{xy}\geq 0$ can be thought of as the weight we give to the pair of inputs $x,y$, and
$m_{xy}=w_{xy} (-1)^{f(x,y)}$.
In the particular case of the CHSH expression, we take
$m_{xy} = (-1)^{x\wedge y}$, resulting in the famous CHSH inequality.

When considering LHV theories, it is convenient to define new variables $A_x = (-1)^{a(x)}$ and $B_y=(-1)^{b(y)}$, whereupon the maximum value of the Bell expression reachable by LHV theories is
$$
C_{\mbox{\scriptsize\sc lhv}}=\max_{A_x,B_y\in\{+1,-1\}} \sum_{xy} m_{xy} A_x B_y
$$

In the case of local measurements carried out on entangled quantum states, we can write
$$\sum_{a,b} P(a,b|x,y) (-1)^a (-1)^b
= \langle \psi | A_x \otimes B_y |\psi \rangle$$
where $|\psi\rangle$ is the quantum state shared by Alice and Bob, and $A_x, B_y$ are Hermitian operators with eigenvalues in $\{+1,-1\}$.
We now use the following result of Tsirelson~\cite{tsirelson87}:
\begin{quote}
Suppose Alice and Bob measure observables $A_x$ and $B_y$, both with eigenvalues in $\{+1,-1\}$, on a pure quantum state $|\psi\rangle \in \mathbb{C}^d\otimes \mathbb{C}^d$,
then there are real unit vectors $\alpha(x) , \beta(y) \in\mathbb{R}^{2d^2}$
such that for all $x$ and $y$,
$\langle \psi | A_x \otimes B_y |\psi \rangle = \alpha(x)\cdot \beta(y)$.
\end{quote}

Thus we can re-express the maximal value of $C$ attainable by quantum mechanics as
$$C_{\mbox{\scriptsize\sc qm}} = \max_{\alpha(x) , \beta(y) \in \mathbb{R}^{n}}
\sum_{xy} m_{xy} \alpha(x)\cdot \beta(y)\ .$$

If we impose only the no-signalling conditions, then it is possible to satisfy Eq.~(\ref{eq:XOR}) for all $x,y$ by choosing $P(ab|xy) = 1/2$ if $a\oplus b = f(x,y)$, $P(ab|xy) = 0$ if $a\oplus b \neq f(x,y)$.
Hence the maximum value of the game is
$$C_{\mbox{\scriptsize no-signalling}} = \sum_{xy} |m_{xy}|\ .$$

As illustration, in the case of the CHSH inequality, the results of Section~\ref{sec:CHSH} can be re-expressed as stating that $C_{\mbox{\scriptsize\sc lhv}}=2$ and 
$C_{\mbox{\scriptsize\sc qm}}=2 \sqrt{2}$ and
$C_{\mbox{\scriptsize no-signalling}}=4$.

Interestingly, the ratio between the LHV values and the quantum value can be bounded
independently of the number of inputs $x,y$ and the choice of matrix $m_{xy}$
by Grothendiek's constant $K_G$, as first noted by Tsirelson~\cite{tsirelson87}:
$$
C_{\mbox{\scriptsize\sc qm}} \leq  K_G C_{\mbox{\scriptsize\sc lhv}}\ .
$$
A recent development of this line of work is the realisation that for certain Bell inequalities, a violation larger than a critical value $C({\bf P}) > C_d$ guarantees that if the correlations are obtained by local measurements on an entangled quantum state, then the  state belongs to a Hilbert space of dimension at least $d^2$ (i.e., Alice and Bob's space each have dimension at least $d$)~\cite{brunner2008tdh,wehner2008lbd,briet2009ggi}). These Bell inequalities can thus be thought of as ``dimension witnesses''.

\subsection{Classical simulation of quantum correlations and quantum communication}

Consider a non-locality experiment in which Alice and Bob share an entangled quantum state and carry out local measurements on this state; or consider a quantum communication protocol in which Alice and Bob carry out several rounds of quantum communication and then carry out measurements on the quantum states. How much classical resources are required to reproduce these quantum experiments?
The results from Sections~\ref{sec:cc} and~\ref{sec:NLandCC} show that the classical resources must sometimes be larger, even exponentially larger, than the quantum resources. Is this the worst one can expect? What are good protocols to simulate the quantum experiments with classical resources? In this section we review progress on these questions. Note that we are of course not claiming that Nature works as in these simulations, but rather we are studying how one could mimic Nature with these alternative resources.

\subsubsection{When no communication is needed.}

When states are very noisy, it may be possible to simulate local measurements on them using only shared randomness, even though the states are entangled. Werner's discovery of a family of states, now known as Werner states, for which such a simulation is possible~\cite{PhysRevA.40.4277} is one of the results of quantum information. Werner's model was restricted to local projective measurements. Later improvements include \cite{acin2006gsc}, and \cite{PhysRevA.65.042302} where it was shown that simulations using only shared randomness can also exist when considering the more general case of local Positive Operator Valued Measures\footnote{A Positive Operator Valued Measure (POVM) is a set $\{A_k\}$ of positive-semidefinite matrices that sum to identity: $\sum_k A_k=I$.
When applied to quantum system in state $\rho$, the probability of obtaining measurement outcome $k$ is $\Tr(A_k\rho)$.} (POVMs), which are the most general kind of measurement allowed by quantum mechanics.

\subsubsection{One-way quantum communication.}

Let us first consider the very simple scenario where Alice wants to communicate a single qubit to Bob and Bob wants to carry out a projective measurement on the qubit. We can formalise this simple scenario as follows:
\begin{quote}
{\bf Simulation of one-way communication of a single qubit and subsequent projective measurement.}
Alice receives as input a normalized vector $\vec{x}\in\mathbb{R}^3$, with length $\norm{\vec x}=1$, which describes the quantum state $\rho = \frac{I}{2}+\vec x \cdot \frac{ \vec \sigma}{2}$ where $\vec\sigma=(X,Y,Z)$ is the vector of non-trivial Pauli matrices from Eq.~(\ref{eqpaulis}); Bob receives as input a
normalized vector $\vec y\in\mathbb{R}^3$, which describes his projective measurement $\vec y \cdot \vec \sigma$. Bob must output a bit $b$, with probabilities satisfying
$P(b=0|\vec x \vec y)-P(b=1|\vec x \vec y)=\Tr(\rho  \vec y \cdot \vec \sigma)$.
\end{quote}

We can generalise this to the case where Alice sends $n$ qubits to Bob, and Bob carries out a POVM on the $n$ qubits:
\begin{quote}
{\bf Simulation of one-way communication of $n$ qubits.}
Alice receives as input the classical description of a quantum state $|\psi\rangle$, for instance by giving her the values of the coefficients $c_i$ of the state in a standard basis $\ket{\psi} =\sum_i c_i \ket{i}$. And Bob is given the classical description of a measurement, for instance by giving him the matrix elements of the POVM elements $A_k$ in the standard basis. The task is for Bob to provide an outcome $k$, such that the probability of outcome $k$ occurring is $P(k|\psi)=\bra{\psi}A_k\ket{\psi}$.
\end{quote}

These are communication complexity scenarios where Alice and Bob's
inputs are infinite-dimensional. If one allows for slight
imperfections in the simulation, then one can truncate the description
of the matrix elements of $\ket{\psi}$ and $A_k$, and make the number
of input bits finite. For instance on Alice's side, if $|\psi\rangle$ corresponds to the quantum state of $n$ qubits, then one can truncate the number of inputs to $O(n2^n)$ bits (by describing each coefficient $c_i$ with $O(n)$ bits of precision). If Alice then sends her truncated input to Bob, then we have, up to a small error, a classical simulation (using $O(n2^n)$ bits) of any one-way quantum communication protocol in which $n$ qubits are sent from Alice to Bob. One cannot hope to do much better than this,
since the {\bf HM} problem of Section~\ref{HMp} exhibits an $n$ versus $2^{\Omega(\sqrt{n})}$ gap between the quantum and classical one-way communication complexity (and this was further strengthened to two-way classical communication complexity
in \cite{gavinsky:interactionvsquantum}).

\subsubsection{Entanglement simulation}

We can also consider the case where Alice and Bob want to simulate local measurements on entangled quantum particles.
The simplest non-locality scenario occurs when Alice and Bob carry out projective measurements on a single ebit:
\begin{quote}
{\bf Simulation of projective measurements on a single ebit.}
Alice and Bob each receive as input a normalized vector in $\mathbb{R}^3$, $\vec x$, $\vec y$ with $\norm{\vec x}=\norm{\vec y}=1$, which describe
their projective measurements $\vec x \cdot \vec \sigma$, $\vec y \cdot \vec \sigma$.
Alice and Bob must each output a bit ($a, b$, respectively) such that the correlations obey
$$
P(a=b|\vec x, \vec y) - P(a\neq b|\vec x, \vec y) = -\vec x \cdot \vec y =  \langle \psi_-| \vec x \cdot \vec \sigma \otimes
\vec y \cdot \vec \sigma|\psi_-\rangle,
$$
where $|\psi_-\rangle=(|0\rangle |1\rangle - |1\rangle |0\rangle )/\sqrt{2}$,
and such that the marginals, $P(a|\vec x, \vec y)$ and $P(b|\vec x, \vec y)$, are uniform (i.e.,
$P(a=0|\vec x, \vec y)=P(a=1|\vec x, \vec y)=1/2$, etc.).
\end{quote}

This can be generalized to the case where Alice and Bob carry out POVM's on arbitrary entangled states of $n$ qubits:
\begin{quote}
{\bf Simulation of entangled states of dimension $2^n$.}
 Alice and Bob share a classical description of a pure entangled quantum state $\ket{\psi}_{AB}$, where Alice and Bob's systems are each of dimension $2^n$. Alice and Bob receive as inputs $x,y$ the classical (infinite-dimensional) descriptions of the measurements they should do (for instance the inputs could consist in the matrix elements of the POVM elements in a standard basis). Alice and Bob must provide outputs $a,b$ such that the joint probability $P(a,b|x,y)$ equals the probability of getting measurement outcomes $a$ and $b$ when measurements $x$ and $y$ are carried out on state  $\ket{\psi_{AB}}$.
\end{quote}

If we have a simulation of one-way quantum communication, then we can transform it into a simulation of entanglement.
To see this, note that one can rewrite the joint probabilities as $P(a,b|x,y)=P(a|x)P(b|x,y,a)$.
The simulation is then as follows: Alice chooses $a$ according to the probability distribution $P(a|x)$; she then sends Bob sufficient information so that he can choose an output $b$ distributed according to $P(b|x,y,a)$. It is easy to show that for this second task (producing $b$ distributed according to $P(b|x,y,a)$) it suffices for Alice to send Bob the measurement outcome, and to describe to him the state onto which his system is projected after Alice's measurement.\footnote{We can assume without loss of generality that Alice's POVM elements all have rank~1, which implies that conditional on the measurement outcome, Bob's state is pure.}
Using this correspondence, we thus have a protocol which provides, up to a small error, a classical simulation (using $O(n2^n)$ bits of one-way communication) of any measurement on entangled states of $n$ qubits.

\subsubsection{Exact classical simulations}

Remarkably it is also possible, at least in some cases, to {\textit{perfectly}} simulate the quantum communication or quantum entanglement scenarios with {\textit{finite}} classical communication. In such perfect simulations we do not tolerate any error. Of course such exact simulations are in principle not necessary if one wants to interpret the results of real experiments, as any real experiment will always have small imperfections. But these exact simulations are interesting for at least two reasons. On the one hand they show that perfectly simulating quantum systems is not much more costly than approximately simulating them. On the other hand, these exact simulations have quite interesting structures. One can hope that understanding these structures will help us understand the power (and limitations) of quantum communication.

Exact classical simulations of quantum correlations were first independently reported in~\cite{maudlin:simulations},~\cite{PhysRevLett.83.1874}
and~\cite{steiner:simulations}. Here we review briefly the subsequent works on this topic.

We first consider a weak model, where the \emph{average} amount of classical
communication is bounded (but in the worse case the amount of
classical communication may be infinite). This model was first used
in~\cite{maudlin:simulations,steiner:simulations} in the context of
classical simulation of a single ebit. In~\cite{PhysRevA.63.052305}
this approach was generalized to the simulation  of communicating $n$
qubits, or the simulation of POVM measurements on $n$ ebits, using $O(n2^n)$ bits of two-way classical communication on average.

A stronger and more interesting model is when the amount of classical communication is bounded (even in the worst case). This model was introduced in~\cite{PhysRevLett.83.1874}. The simulations were improved, and in~\cite{PhysRevLett.91.187904} it was shown that the classical simulation of projective measurements on a single ebit could be realized with a single bit of classical communication from Alice to Bob, and the communication of a single qubit could be simulated with 2 bits of communication. Note that these simulations use an infinite amount of shared randomness, a requirement that was shown in~\cite{PhysRevA.63.052305} to be necessary when the amount of communication is bounded (in the worst case).

An even stronger model for the simulation of entanglement is for Alice and Bob to use as resource non-local boxes, rather than classical communication. Indeed, as discussed in Section~\ref{sec:nlbox}, one bit of classical communication can be used to realize a non-local box, but a non-local box cannot be used to communicate. It was shown in~\cite{cgmp:simulations} that simulating projective measurements on a single ebit could be carried out with the use of a single non-local box.
A unified approach to protocols simulating a single ebit with one bit of communication or with one non-local box was presented in~\cite{dlr:simulations}.

\section{Implementations}\label{sec:impl}

\subsection{Inefficient detectors}\label{DetectIneff}

\subsubsection{The detection loophole}

In this section we put a constraint on the quantum model. We will suppose that any measurement on a quantum system gives the results predicted by quantum mechanics with probability $\eta$, and does not give any result with probability $1-\eta$.

The motivation for considering this model is that most quantum communication experiments use photons. Photons are very practical because they can be quite easily produced, manipulated, transmitted over long distances, and measured. Unfortunately photons get absorbed during transmission (in commercial optical fibers, photons have approximately 50\% probability of being absorbed after travelling 15km), and single-photon detectors have limited efficiency: they will sometimes not detect a photon even though it is present. These effects can be described by the above model.

In most experiments to date, the detector efficiency $\eta$ was so low that the correlations could be explained by a classical model using shared randomness and no communication (a local hidden variable model).
This is called the \textit{Detection Loophole}~\cite{pearle:detectionloophole}.
Thus for instance in the CHSH experiment, the correlations can be
explained by a local hidden variable model if $\eta\leq 2/(\sqrt{2} + 1) \simeq 0.8284$.
A detector efficiency better than this bound has not (yet) been  achieved in experiments involving only photons.

One solution to the above problem is technological: one should use a
quantum system on which measurements can be carried out with high
efficiency. In this respect atoms or ions are particularly interesting,
because measurements on these systems can be carried out with
essentially $100\%$ efficiency. Thus experiments involving two entangled ions have been
carried out in which the detection loophole was closed~\cite{rkmsimw:detectionloophole,mmmom:atomatom}. However these
experiments have not yet allowed both the detection loophole and the
locality loophole (i.e., carrying out both measurements at spatially
separated locations) to be closed simultaneously.

Instead of (or in addition to) improved technology,
another solution to this problem is to develop new non-locality tests
that demonstrate non-locality with low detector efficiency. As we shall
see in the following, the communication problems and protocols
developed in the previous sections can be used to build such tests.

\subsubsection{Communication complexity and the detection loophole}

Communication complexity suggests that by increasing the dimension $d$ of the entangled system
under study, one can decrease exponentially (in $d$) the required
efficiency of the detectors. Indeed, it appears that in many cases the
minimum number $c$ of bits of classical communication required to
reproduce the quantum correlations is related to the minimum
efficiency of the detectors required for the correlations to be
non-local by $\eta \geq 2^{-O(c)}$. That there should be a relation between $c$ and $\eta$ was first noted in~\cite{GisinGisin} and further studied in~\cite{Massardetection,bhmr:nonlocality,bhmr:imperfections}.

To understand this relation we will compare two classical schemes:
\begin{itemize}
\item
In the first scheme, which was discussed at length in Sections~\ref{sec:nl} and~\ref{sec:NLandCC}, the detectors have $100\%$ efficiency, the parties have shared randomness and may exchange up to $c$ bits of classical communication.
\item
In the second scheme, the parties have shared randomness, and each party has a detector of efficiency $\eta$. This means that each party will with probability $\eta$ give an output, and with probability $1-\eta$ produce no output. The detectors are assumed to be independent, so that the probability that both detectors give an output is $\eta^2$. In the physics terminology this would be called a local hidden variable model with detector efficiency $\eta$. (We will also consider below the case where one of the detectors has efficiency $\eta$, and the other always gives a result, i.e., is $100\%$ efficient.)
\end{itemize}
These two schemes can be related in a number of ways. The simplest relation is:
\begin{quote}
Any classical protocol with $c$ bits of communication can be mapped into a classical protocol with no communication but with detector efficiency $\eta^2= 2^{-c}$.
\end{quote}
This mapping is very simple: Alice and Bob use shared randomness $r$ which is uniformly distributed over all possible conversations. Each party checks whether $r$ is a conversation that is consistent with their input. If it is then they give the corresponding output, if it is not then they don't give any output. The probability that both Alice and Bob give an output is at least $2^{-c}$.

This protocol is not perfect since the probability that the parties give an output may differ from one party to the other, or from one input to the other. What is interesting is that in a number of cases the converse holds: if the quantum correlations cannot be reproduced with less than $c$ bits of communication, then they can be reproduced without communication only if the detector efficiency $\eta$ is less than $2^{-\Omega(c)}$.

A first example where this converse occurs, is when bounds on $c$ and on the minimum detection efficiency $\eta$ can be obtained from the size of monochromatic rectangles (see Appendix~\ref{appclassicallowerbounds} for a brief presentation of this notion). This approach was  implicit in~\cite{Massardetection} where it was shown that the correlations of the distributed Deutsch-Jozsa problem could not be reproduced by a local hidden variable model if $\eta \geq O(n^{3/4}) 2^{-0.0035 n}$ when the inputs consist of $n$-bit strings, and hence the parties use a maximally entangled system of dimension $n$.  Using the size of monochromatic rectangles was exploited more fully in \cite{bhmr:nonlocality} in the context of a multipartite communication complexity problem, and then extended in~\cite{bhmr:imperfections} to take into account the possibility of errors.
In particular, in~\cite{bhmr:imperfections} it was shown how one could
obtain a lower bound $c \geq B_R$ on the minimum amount of communication
required to reproduce the correlations, where $B_R$ is a function of
the size and discrepancy of rectangles. It then followed that the
correlations could be obtained by a local hidden variable model with
detectors of efficiency $\eta$ only if $\eta \leq 2^{-B_R/n}$ (where $n$ is the number of parties). If the
rectangle lower bound on $c$ is close to tight, then this implies the
relation we mentioned above between $c$ and $\eta$.

\subsubsection{Asymmetric detection loophole}
\label{ADEsect}

Another interesting example arises if we suppose that Alice's detector is inefficient, but that Bob's detector is perfect.
This situation is motivated by the experimental situation reported in~\cite{mmbm:atomphoton}, where an ion is entangled with a photon. As discussed above, the measurements on the ion can be done with $100\%$ efficiency, whereas those on the photon will be inefficient. The problem in which Alice's detector is inefficient and Bob's detector is perfect was previously investigated from the point of view of the detection loophole in~\cite{cl:assymdetectloophole,bgss:assymdetectloophole} for entangled systems of dimension 2.

We prove in Appendix~\ref{ADE} that the Hidden Matching problem is particularly well adapted to this asymmetric scenario. Namely we show that
\begin{quote}
Suppose Alice and Bob try to implement the Hidden Matching problem using $\log n$ ebits, as discussed in Section~\ref{HMp}. Suppose that Alice's detector has efficiency $\eta$ whereas Bob's detector has $100\%$ efficiency. Then the correlations obtained by measuring the ebits cannot be reproduced by a classical model without communication if $\eta \geq 2^{-\Omega(\sqrt{n})}$, even allowing for a small error probability.
\end{quote}
To our knowledge, this is the first time it is shown that an exponentially small detection efficiency can be tolerated when allowing for a small error probability.

\subsection{Present and future experiments}

\subsubsection{Experimental quantum non-locality}

During the past decades there have been many experiments that studied the correlations exhibited by measurements on entangled quantum particles. Their main aim was to test quantum mechanics by comparing its predictions with those of hidden variable models. The short result is that the predictions of quantum mechanics have always
been verified to very high precision. However, up to now some ``loopholes'' have always been left open, which allow the possibility of explaining the data with---admittedly contrived---local hidden variable models.

We very briefly review how experiments on quantum non-locality have been improved during the past decades. We then discuss how the insights from communication complexity suggest new experimental challenges. We also discuss experimental realizations of quantum communication complexity.

After the initial experiment by Freedman and Clauser~\cite{PhysRevLett.28.938} on the correlations exhibited by entangled photons, the first qualitative advance was the experiments of Aspect that used time-varying analyzers in order to close the locality loophole. Indeed in previous experiments the measurements where kept fixed for long periods of time while experimental results were accumulated, then the measurements were changed and a new set of data was acquired for the new measurement setting. In Aspect's experiment~\cite{AspectDR82:Bell} the measurement settings changed periodically in time. In the later experiment of Weihs et al.~\cite{WJSWZ:Bell}, the measurement settings were chosen at random using a quantum random number generator.

Another important advance of the experiments of Aspect et al.~\cite{AspectGR81:Bell,AspectGR82:Bell} was a very precise check that the measured correlations coincide with the correlations $P_{QM}(ab|xy)$ predicted by quantum mechanics for local measurements on a maximally entangled state of two particles (earlier experiments were much more imprecise).

Some other noteworthy advances:
\begin{itemize}
\item Non-locality experiments in which the two particles were separated by a large distance of 10 km~\cite{TBZG:Bell} and 50 km~\cite{MRTZLG:Bell};
\item Non-locality experiments on bipartite entangled systems of dimension 3~\cite{PhysRevLett.89.240401,PhysRevLett.93.010503};
\item Non-locality experiments on entangled states of three~\cite{PBDWZ:GHZ,RNOBBRH:GHZ} and four particles~\cite{SKKLMMRTIWM:4part,PhysRevLett.91.180401}.
\end{itemize}

In all the above experiments the detection loophole was not closed. This means that the raw data acquired during the experiment could be explained by a local hidden variable model. It was only by making the (physically very reasonable) assumption that the events in which the detector gives a click are independent of the measurement settings and measurement results (known in the physics literature as the ``fair sampling assumption") that these experiments could be assumed to be in contradiction with local hidden variable models.

\begin{figure}{\includegraphics[height=60mm]{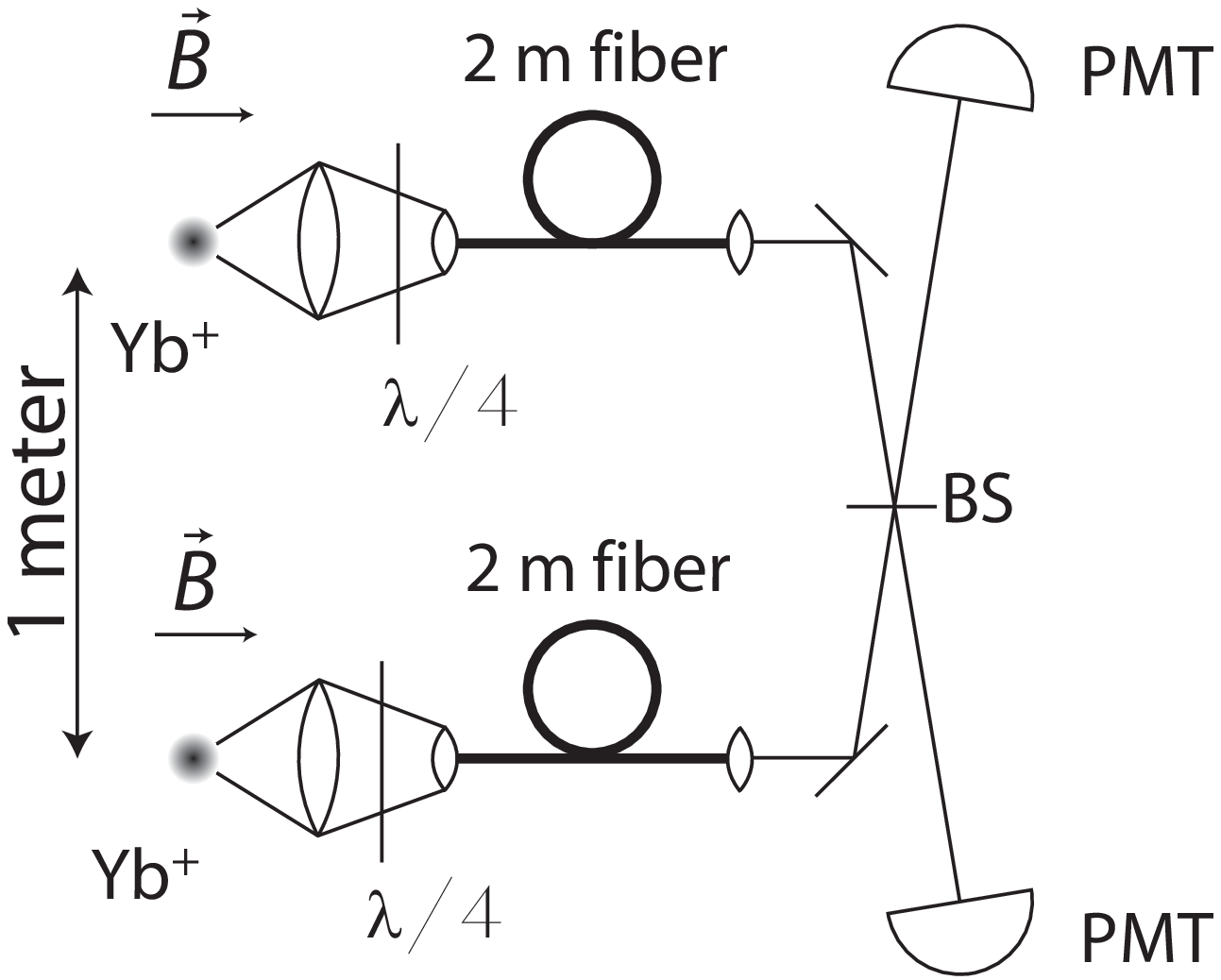}}
\quad\quad\quad
{\includegraphics[height=60mm]{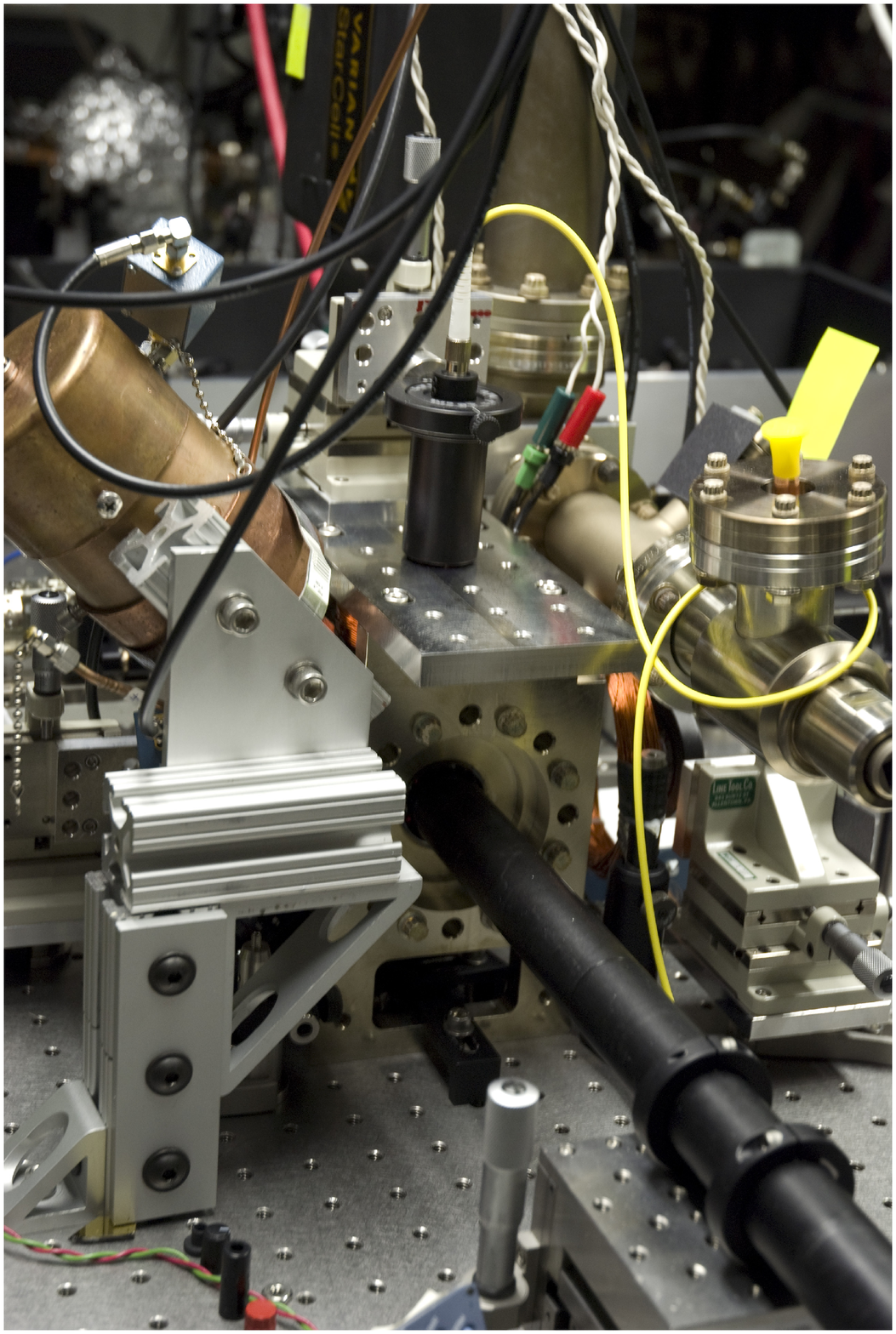}}
\caption{\textit{Bell inequality with two remote atomic qubits.}
The left-hand side is a schematic description of the experiment reported in \cite{mmmom:atomatom} in which the internal states of two ions separated by about one meter were entangled. Measurements on the two ions then allowed the violation of the CHSH inequality with the detection loophole closed. A series of laser pulses simultaneously excite both Yb$^+$ ions in such a way that when they deexcite, they emit a photon whose polarization is entangled with the ion.  A lens is used to couple the photons into optical fibers. The wave plate ($\lambda/4$) is used for convenience to convert circular polarization into linear polarization. The two photons interfere on a Beam Splitter (BS) and are detected by Photo Multiplier Tubes (PMT). Simultaneous detection of a photon by the two PMT's signals that the photons were  in a Bell state, thereby realizing entanglement swapping: the two ions are now entangled. The internal states of the ions are then measured, enabling a violation of the CHSH inequality. Note that there are many inefficiencies in this experiment: only a fraction of emitted photons are coupled into the optical fibers, and only a fraction of the photons reaching the PMT's are detected. But when two photons are detected, one knows with certainty that the two ions are entangled. The right hand side is a photograph of one of the ion traps. The other trap is similar, and located about one meter away on the same optical table. (Both figures courtesy of S.~Monroe and D.~Matsukevich; left-hand side panel copyright American Physical Society).
 }
 \label{figure:twoions}\end{figure}

As mentioned above, there have now been two experiments involving ions in which the detection loophole has been closed. In the first, the two entangled ions were separated by about 3 $\mu m$~\cite{rkmsimw:detectionloophole}, in the second, presented in more detail in Fig.~\ref{figure:twoions}, the two entangled ions were separated by about a meter~\cite{mmmom:atomatom}. In view of these advances, closing both the locality and detection loopholes simultaneously does not seem out of reach.

From the point of view of communication complexity, closing the detection loophole is more important than closing the locality loophole. Indeed, if the detection loophole is not closed, it means that the raw data can be explained by a model without communication. On the other hand, if the detection loophole is closed, then, by sharing the entanglement, the parties have a resource that  could only be reproduced classically by communication between the parties. The same is true in other applications of quantum non-locality: closing the detection loophole (but not necessarily the locality loophole) allows one to increase the security of quantum key distribution~\cite{ABGMPS:qkd}.

\subsection{Future non-locality experiments}\label{sec:FNLE}

The progress in quantum communication complexity points the way towards new tests of quantum non-locality which use not one ebit, as in the CHSH test, but many ebits.
Ideas for these new tests come from the entanglement-based Deutsch-Jozsa problem discussed in Section~\ref{sec:NLandCC},
the entanglement-based Hidden Matching problem discussed in Section~\ref{HMp}, recent work of Gavinsky~\cite{gavinsky:interactionvsnonlocality}, and also the (non-constructive) results on three-party correlations reported in~\cite{PGWPVJ}.
There are at least two motivations for such experiments. First of all they could be more robust against experimental imperfections (such as the detection loophole or errors) than non-locality tests used at present. Second they could illustrate the efficiency of quantum mechanics over classical mechanics, as experiments on a small number $e$ of ebits could only be reproduced classically using an exponentially large (in $e$) amount $c$ of classical communication.

These non-locality experiments on a many ebits can be characterized by several parameters. In particular these would include the number $e$ of ebits involved, or equivalently the dimension $d= 2^e$ of the entangled quantum system; the minimum detector efficiency $\eta$ required for the correlations to be non-local; the amount $\epsilon$ of errors that can be tolerated; and the amount $c$ of classical communication that would be required to reproduce the quantum correlations. In general, for any given non-locality test, we can expect tradeoffs between $\eta$, $\epsilon$ and $c$.

An important point to note is that the proposals inspired by communication complexity typically are asymptotic results that deal with the limit where the number of ebits tends towards infinity: $e\to \infty$. However real experiments will deal with small values of $e$.
For instance, if we think of the detection loophole, one should recall that this is only a problem for experiments dealing with entangled photons. On the other hand, the Hilbert space of a single photon can be larger than~2. One can thus effectively manipulate more than one qubit, while manipulating only a single photon. This is potentially an interesting opportunity.
Indeed it would be very interesting to devise non-locality experiments that tolerate inefficient detectors (say $\eta < 10\%$) in Hilbert spaces of moderate dimension (say $d=10$). If one could devise such a non-locality experiment, there would be a strong incentive to realize it experimentally. Indeed whereas experiments involving entangled atoms or ions may be the short-term solution to solving the detection loophole, such experiments are much slower and much more expensive than experiments involving photons only. Numerical searches for such a non-locality experiment have been undertaken, but unsuccessfully so far~\cite{PhysRevA.66.052112}.

In summary, quantum communication complexity suggests the possibility of new non-locality experiments on a moderate number of ebits that either are very resistant to imperfections, or require very large amounts of classical resources to reproduce classically. Realizing such experiments will require further progress on the theoretical and experimental side.

\subsubsection{Experimental communication complexity}

The experimental situation concerning communication complexity proper is less advanced.
Indeed, in order to carry out any nontrivial experimental demonstration of communication complexity, one needs to take into account the limited efficiency of detectors which has been such a plague for non-locality experiments. In this respect, the first convincing communication complexity experiment to date is that reported in~\cite{TSBBZW:commcompl} in which 6 parties, materialized by waveplates along a beam on an optical table, carried out the communication complexity problem proposed in~\cite{cleve&buhrman:subs,bdht:multiparty,bcd:qecc}, but in the version proposed in~\cite{PhysRevA.65.012318}, which does not use entanglement. In this experiment the limited efficiency of detectors was explicitly taken into account.
Experiments that studied the entanglement-based version of this problem while explicitly taking into account the limited efficiency of the detectors have also been reported~\cite{ZBCYCP:GMN}, based on the proposal of~\cite{CL-T:GMN}.

Another protocol which has been studied experimentally is quantum fingerprinting
which in the SMP model performs exponentially better than classical protocols (see Section~\ref{ssecqfpprotocol}). The possibility of realizing such an experiment at a small scale involving one or a few photons has been discussed in~\cite{PhysRevA.69.022307,Massar:fingerprinting}, and later performed using photons~\cite{HBMLS:fingerprinting} and in NMR~\cite{DZPOKE:fingerprinting}.

In the future we may expect further proof-of-principle experiments of quantum communication complexity involving the exchange of more qubits and larger distance between the parties. Good candidates for such experiments are Raz's communication complexity problem, the Hidden Matching problem and its extensions, and quantum fingerprinting.

\section{Conclusion}\label{sec:concl}

\subsection{Open questions}

Quantum communication complexity and quantum non-locality are by now mature fields. But many questions remain open. Here we collect a few.

\begin{enumerate}

\item{\bf Additional natural problems in quantum communication complexity.} Find additional problems---if possible natural problems that could have potential applications---for which quantum communication is much more efficient than classical commmunication.

\item{\bf How much entanglement is needed to get a reduction of communication: equivalence of quantum communication and entanglement models of communication complexity.} In the entanglement model of communication complexity, the parties have an unlimited supply of entanglement and use it to reduce the amount of classical communication. How much entanglement is really needed? In classical communication complexity with shared randomness Newman's Theorem~\cite{newman:random} states that, if we allow a small increase in the error probability, the parties need only have $O(\log n)$ shared random bits (where $n$ is the size of the inputs). Does something similar hold when we replace shared randomness by entanglement? Answering this question would essentially establish whether the quantum communication and the entanglement models of communication complexity are equivalent

\item{\bf Are most quantum states useful for communication complexity?} It was recently shown in \cite{grossflammiaeisert2009} that most $n$-qubit states (with respect to the uniform measure) are not useful---they are typically too entangled---in the measurement-based version of quantum computation. Are most states useful for communication complexity? For two parties the answer is yes, as they can work in the Schmidt basis. But consider three parties sharing a random state of $3n$ qubits (each party having $n$ qubits). How useful are most states for communication complexity (asymptotically as $n$ tends to infinity)?

\item{\bf Find new non-local games, qualitatively different from existing ones.} In particular consider the following more specific subquestions:
\begin{itemize}
\item For two-party XOR games, the ratio between the classical and the quantum value of the game is bounded by a constant. However in \cite{PGWPVJ} it was shown---using a non-constructive proof---that this is not the case for three-party games. Can one exhibit an explicit example of this type?
\item
Find Bell inequalities involving rather small systems, say where the dimension of each party's Hilbert space is less than $d=10$, which allow for very small detector efficiencies.
\end{itemize}

\item{\bf Non-local boxes and communication complexity.} As discussed in Section~\ref{sec:nlbox}, non-local boxes are an interesting resource to consider from the point of view of communication complexity. In this regard, two interesting questions are:
\begin{itemize}
\item
First, what is the noise threshold below which non-local boxes make communication complexity trivial (see \cite{BBLMTU:NLbox} for a formulation of this problem). Is this threshold the maximum value $p=(2 + \sqrt{2})/4$ attainable by local measurements on entangled quantum systems?
\item
Second, is it possible to amplify non-local correlations, in the sense that given a large number of devices that will produce correlations $P(ab|xy)$ corresponding to PR boxes with noise $p$, is it possible to use the devices in such a way as to produce correlations with a lower value of $p$? A first result in this direction can be found in \cite{forster2008dnl}.
\end{itemize}

\item{\bf Simulation of quantum correlations and quantum communication.} In this context, some questions that come to mind are:
\begin{itemize}
\item
Exact simulation of more than one qubit or ebit using bounded classical communication (in the worse case) or Non-Local Boxes.  Some preliminary results on this topic have been obtained in the particular case where Alice and Bob carry out measurements with binary outcomes~\cite{dlr:simulations07,rt:simulations}.
\item
The simulation of non-maximally entangled states using non-local boxes. This appears to be much harder than the simulation of maximally entangled states, see~\cite{bgs:simulations,bgps:simulations} for some first results.
\item
The simulation of multipartite non-local correlations.
\end{itemize}

\end{enumerate}

\subsection{What have we learnt from quantum communication complexity?}

Communication complexity is  a task for which quantum information can beat classical information. Such tasks are rare, and finding more potential applications of quantum information is very important.

Unfortunately most quantum communication complexity problems are
either extremely sensitive to noise, highly contrived, or do not offer
exponential gains over the best classical protocols (in which case the
advantages of quantum communication will probably be more than offset
by the lower cost and higher speed of classical communication). The
most interesting proposal so far is maybe the SMP model without shared
randomness (a somewhat contrived model) where equality (a very natural
problem) can be solved exponentially more efficiently using quantum communication.
Thus there is the tantalizing possibility that some time in the future, quantum communication complexity could be used in practical applications.

Independently of whether quantum communication complexity ever finds some real-world applications, the results
obtained so far have important conceptual implications.
First of all they offer new insights into the power of quantum information, and in particular of quantum computing. Indeed
the basic aim of computer science, taken in a wide sense, is to accomplish a task by using the minimum amount of resources. In the usual formulation, the resource that we want to minimize is the running time of the computer. This is the most important application of quantum computing as Shor's algorithm suggests that a quantum computer would allow exponential speedups. But in this context it is very difficult---if not impossible---to prove that quantum computers are more powerful than classical computers. The advantage of quantum computation can however be proven in simpler contexts such as the black-box model of quantum computing, where the resource that is quantified is the number of calls to an oracle; or communication complexity where the resource that is quantified is the amount of communication. The existence of these models where it can be rigorously shown that quantum information offers important advantages over classical information reinforces our confidence that quantum computers are much more powerful than classical computers for certain tasks.

Second, the study of quantum communication complexity has led to the
proposal of new tests of quantum mechanics. Indeed from Bell onwards
it was known that if one wants to replace quantum mechanics by a
classical model, this classical model would have to use faster than
light signalling. The discovery of fast quantum algorithms suggested
that such a classical model would use an exponentially large number of
resources. Quantum communication complexity has now advanced to the
point where it may be possible to propose experiments in which one can
prove that a classical simulation would require exponentially
more resources than are used quantum mechanically.

In summary, quantum communication complexity is now a mature field that has led to some fundamental insights into
the nature of computation and the foundations of physics.

\bibliographystyle{plain}

\begin{thebibliography}{100}

\bibitem{aaronson&ambainis:search}
S.~Aaronson and A.~Ambainis.
\newblock Quantum search of spatial regions.
\newblock In {\em Proceedings of 44th IEEE FOCS}, pages 200--209, 2003.
\newblock quant-ph/0303041.

\bibitem{ABGMPS:qkd}
A.~Acin, N.~Brunner, N.~Gisin, S.~Massar, S.~Pironio, and V.~Scarani.
\newblock Device-independent security of quantum cryptography against
  collective attacks.
\newblock {\em Physical Review Letters}, 98:230501, 2007.

\bibitem{acin2006gsc}
A.~Ac{\'\i}n, N.~Gisin, and B.~Toner.
\newblock {Grothendieck's constant and local models for noisy entangled quantum
  states}.
\newblock {\em Physical Review A}, 73:062105, 2006.

\bibitem{ambainis:3computer}
A.~Ambainis.
\newblock Communication complexity in a 3-computer model.
\newblock {\em Algorithmica}, 16(3):298--301, 1996.

\bibitem{Aravind02}
P.~K. Aravind.
\newblock A simple demonstration of {B}ell's theorem involving two observers
  and no probabilities or inequalities.
\newblock quant-ph/0206070, 2002.

\bibitem{AspectDR82:Bell}
A.~Aspect, J.~Dalibard, and G.~Roger.
\newblock Experimental test of {B}ell's inequalities using time- varying
  analyzers.
\newblock {\em Physical Review Letters}, 49:1804, 1982.

\bibitem{AspectGR81:Bell}
A.~Aspect, Ph. Grangier, and G.~Roger.
\newblock Experimental tests of realistic local theories via {B}ell's theorem.
\newblock {\em Physical Review Letters}, 47:460, 1981.

\bibitem{AspectGR82:Bell}
A.~Aspect, Ph. Grangier, and G.~Roger.
\newblock Experimental realization of {E}instein-{P}odolsky-{R}osen-{B}ohm
  {G}edankenexperiment: A new violation of {B}ell's inequalities.
\newblock {\em Physical Review Letters}, 49:91, 1982.

\bibitem{bjk:q1way}
Z.~{Bar-Yossef}, T.~S. Jayram, and I.~Kerenidis.
\newblock Exponential separation of quantum and classical one-way communication
  complexity.
\newblock In {\em Proceedings of 36th ACM STOC}, pages 128--137, 2004.

\bibitem{PhysRevA.65.042302}
J.~Barrett.
\newblock Nonsequential positive-operator-valued measurements on entangled
  mixed states do not always violate a {B}ell inequality.
\newblock {\em Physical Review A}, 65(4):042302, Mar 2002.

\bibitem{bhk:crypto}
J.~Barrett, L.~Hardy, and A.~Kent.
\newblock No signaling and quantum key distribution.
\newblock {\em Physical Review Letters}, 95(1):010503, Jun 2005.

\bibitem{BLMPPR:NLbox}
J.~Barrett, N.~Linden, S.~Massar, S.~Pironio, S.~Popescu, and D.~Roberts.
\newblock Nonlocal correlations as an information-theoretic resource.
\newblock {\em Physical Review A}, 71:022101, 2005.

\bibitem{bell:epr}
J.~S. Bell.
\newblock On the {E}instein-{P}odolsky-{R}osen paradox.
\newblock {\em Physics}, 1:195--200, 1965.

\bibitem{teleporting}
C.~Bennett, G.~Brassard, C.~Cr{\'e}peau, R.~Jozsa, A.~Peres, and W.~Wootters.
\newblock Teleporting an unknown quantum state via dual classical and
  {Einstein-Podolsky-Rosen} channels.
\newblock {\em Physical Review Letters}, 70:1895--1899, 1993.

\bibitem{superdense}
C.~Bennett and S.~Wiesner.
\newblock Communication via one- and two-particle operators on
  {Einstein-Podolsky-Rosen} states.
\newblock {\em Physical Review Letters}, 69:2881--2884, 1992.

\bibitem{bbps:entanglementconcentration}
C.~H. Bennett, H.~J. Bernstein, S.~Popescu, and B.~Schumacher.
\newblock Concentrating partial entanglement by local operations.
\newblock {\em Physical Review A}, 53(4):2046--2052, Apr 1996.

\bibitem{bb84}
C.~H. Bennett and G.~Brassard.
\newblock Quantum cryptography: Public key distribution and coin tossing.
\newblock In {\em Proceedings of the IEEE International Conference on
  Computers, Systems and Signal Processing}, pages 175--179, 1984.

\bibitem{PhysRevLett.76.722}
C.~H. Bennett, G.~Brassard, S.~Popescu, B.~Schumacher, J.~A. Smolin, and W.~K.
  Wootters.
\newblock Purification of noisy entanglement and faithful teleportation via
  noisy channels.
\newblock {\em Physical Review Letters}, 76(5):722--725, Jan 1996.

\bibitem{Bhatia:97a}
R.~Bhatia.
\newblock {\em Matrix Analysis}.
\newblock Number 169 in {Graduate Texts in Mathematics}. Springer-Verlag, {New
  York}, 1997.

\bibitem{brassard:qcc}
G.~Brassard.
\newblock {Quantum communication complexity}.
\newblock {\em Foundations of Physics}, 33(11):1593--1616, 2003.
\newblock quant-ph/0101005.

\bibitem{brassard2005qpt}
G.~Brassard, A.~Broadbent, and A.~Tapp.
\newblock {Quantum pseudo-telepathy}.
\newblock {\em Foundations of Physics}, 35(11):1877--1907, 2005.

\bibitem{BBLMTU:NLbox}
G.~Brassard, H.~Buhrman, N.~Linden, A.~A. M{\'e}thot, A.~Tapp, and F.~Unger.
\newblock Limit on nonlocality in any world in which communication complexity
  is not trivial.
\newblock {\em Physical Review Letters}, 96:250401, 2006.

\bibitem{PhysRevLett.83.1874}
G.~Brassard, R.~Cleve, and A.~Tapp.
\newblock Cost of exactly simulating quantum entanglement with classical
  communication.
\newblock {\em Physical Review Letters}, 83(9):1874--1877, Aug 1999.
\newblock arXiv:quant-ph/9901035.

\bibitem{bhmt:countingj}
G.~Brassard, P.~H{\o}yer, M.~Mosca, and A.~Tapp.
\newblock Quantum amplitude amplification and estimation.
\newblock In {\em Quantum Computation and Quantum Information: A Millennium
  Volume}, volume 305 of {\em AMS Contemporary Mathematics Series}, pages
  53--74. 2002.
\newblock quant-ph/0005055.

\bibitem{briet2009ggi}
J.~Bri{\"e}t, H.~Buhrman, and B.~Toner.
\newblock {A generalized Grothendieck inequality and entanglement in XOR
  games}.
\newblock {\em Arxiv preprint arXiv:0901.2009}, 2009.

\bibitem{PhysRevLett.92.127901}
{\v{C}}.~Brukner, M.~{\.{Z}}ukowski, J.-W. Pan, and A.~Zeilinger.
\newblock Bell\char39{}s inequalities and quantum communication complexity.
\newblock {\em Physical Review Letters}, 92(12):127901, Mar 2004.

\bibitem{bgps:simulations}
N.~Brunner, N.~Gisin, S.~Popescu, and V.~Scarani.
\newblock Simulation of partial entanglement with no-signaling resources.
\newblock arXiv:0803.2359, 2008.

\bibitem{bgs:simulations}
N.~Brunner, N.~Gisin, and V.~Scarani.
\newblock Entanglement and non-locality are different resources.
\newblock {\em New Journal of Physics}, 7:88, 2005.

\bibitem{bgss:assymdetectloophole}
N.~Brunner, N.~Gisin, V.~Scarani, and C.~Simon.
\newblock Detection loophole in asymmetric {B}ell experiments.
\newblock {\em Physical Review Letters}, 98:220403, 2007.

\bibitem{brunner2008tdh}
N.~Brunner, S.~Pironio, A.~Acin, N.~Gisin, A.A. M{\'e}thot, and V.~Scarani.
\newblock {Testing the dimension of Hilbert spaces}.
\newblock {\em Physical Review Letters}, 100(21):210503--210503, 2008.

\bibitem{bcd:qecc}
H.~Buhrman, R.~Cleve, and W.~{van} Dam.
\newblock Quantum entanglement and communication complexity.
\newblock {\em SIAM Journal on Computing}, 30(8):1829--1841, 2001.
\newblock quant-ph/9705033.

\bibitem{bcww:fp}
H.~Buhrman, R.~Cleve, J.~Watrous, and R.~{de} Wolf.
\newblock Quantum fingerprinting.
\newblock {\em Physical Review Letters}, 87(16), September 26, 2001.
\newblock quant-ph/0102001.

\bibitem{BuhrmanCleveWigderson98}
H.~Buhrman, R.~Cleve, and A.~Wigderson.
\newblock Quantum vs.~classical communication and computation.
\newblock In {\em Proceedings of 30th ACM STOC}, pages 63--68, 1998.
\newblock quant-ph/9802040.

\bibitem{bdht:multiparty}
H.~Buhrman, W.~{van} Dam, P.~H{\o}yer, and A.~Tapp.
\newblock Multiparty quantum communication complexity.
\newblock {\em Physical Review A}, 60(4):2737--2741, 1999.
\newblock quant-ph/9710054.

\bibitem{bhmr:nonlocality}
H.~Buhrman, P.~H{\o}yer, S.~Massar, and H.~R{\"o}hrig.
\newblock Combinatorics and quantum nonlocality.
\newblock {\em Physical Review Letters}, 91(047903), 2003.
\newblock quant-ph/0209052.

\bibitem{bhmr:imperfections}
H.~Buhrman, P.~H{\o}yer, S.~Massar, and H.~R{\"o}hrig.
\newblock Multipartite nonlocal quantum correlations resistant to
  imperfections.
\newblock {\em Physical Review A}, 73(012321), 2006.
\newblock quant-ph/0410139.

\bibitem{cabello01B}
A.~Cabello.
\newblock {All versus Nothing} inseparability for two observers.
\newblock {\em Physical Review Letters}, 87(1):010403, Jun 2001.

\bibitem{cabello01A}
A.~Cabello.
\newblock Bell's theorem without inequalities and without probabilities for two
  observers.
\newblock {\em Physical Review Letters}, 86(10):1911--1914, Mar 2001.

\bibitem{cabello2005}
A.~Cabello.
\newblock Stronger two-observer all-versus-nothing violation of local realism.
\newblock {\em Physical Review Letters}, 95(21):210401, 2005.

\bibitem{cl:assymdetectloophole}
A.~Cabello and J.-A. Larsson.
\newblock Minimum detection efficiency for a loophole-free atom-photon {B}ell
  experiment.
\newblock {\em Physical Review Letters}, 98:220402, 2007.

\bibitem{CL-T:GMN}
A.~Cabello and A.~J. L{\' o}pez-Tarrida.
\newblock Proposed experiment for the quantum guess my number protocol.
\newblock {\em Physical Review A}, 71:020301(R), 2005.

\bibitem{cgmp:simulations}
N.~J. Cerf, N.~Gisin, S.~Massar, and S.~Popescu.
\newblock Simulating maximal quantum entanglement without communication.
\newblock {\em Physical Review Letters}, 94:220403, 2005.

\bibitem{Tsirelson80}
B.~S.~Tsirelson (Cirel{'}son).
\newblock Quantum generalizations of {B}ell's inequality.
\newblock {\em Letters in Mathematical Physics}, 4(2):93--100, 1980.

\bibitem{chsh}
J.~F. Clauser, M.~A. Horne, A.~Shimoney, and R.~A. Holt.
\newblock Proposed experiment to test local hidden-variable theories.
\newblock {\em Physical Review Letters}, 23(15):880--884, 1969.

\bibitem{cleve&buhrman:subs}
R.~Cleve and H.~Buhrman.
\newblock Substituting quantum entanglement for communication.
\newblock {\em Physical Review A}, 56(2):1201--1204, 1997.
\newblock quant-ph/9704026.

\bibitem{cdnt:ip}
R.~Cleve, W.~{van} Dam, M.~Nielsen, and A.~Tapp.
\newblock Quantum entanglement and the communication complexity of the inner
  product function.
\newblock In {\em Proceedings of 1st NASA QCQC conference}, volume 1509 of {\em
  Lecture Notes in Computer Science}, pages 61--74. Springer, 1998.
\newblock quant-ph/9708019.

\bibitem{cleve2004cal}
R.~Cleve, P.~H{\o}yer, B.~Toner, and J.~Watrous.
\newblock {Consequences and limits of nonlocal strategies}.
\newblock In {\em 19th IEEE Annual Conference on Computational Complexity,
  2004. Proceedings}, pages 236--249, 2004.

\bibitem{collins2004rtq}
D.~Collins and N.~Gisin.
\newblock {A relevant two qubit Bell inequality inequivalent to the CHSH
  inequality}.
\newblock {\em Journal of Physics A-Mathematical and General},
  37(5):1775--1788, 2004.

\bibitem{cglmp}
D.~Collins, N.~Gisin, N.~Linden, S.~Massar, and S.~Popescu.
\newblock Bell inequalities for arbitrarily high-dimensional systems.
\newblock {\em Physical Review Letters}, 88(4):40404--40404, 2002.

\bibitem{vanDam:thesis}
{W. van} Dam.
\newblock {\em Nonlocality \& Communication Complexity}.
\newblock PhD thesis, University of Oxford, Department of Physics, 2000.

\bibitem{WvanDam:NLbox}
{W. van} Dam.
\newblock Implausible consequences of superstrong nonlocality, 2005.
\newblock arXiv:quant-ph/0501159v1.

\bibitem{vandam2005ssn}
{W. van} Dam, R.~D. Gill, and P.~D. Gr{\"u}nwald.
\newblock The statistical strength of nonlocality proofs.
\newblock {\em {IEEE} Transactions on Information Theory}, 51(8):2812--2835,
  2005.

\bibitem{PhysRevA.69.022307}
J.~Niel de~Beaudrap.
\newblock One-qubit fingerprinting schemes.
\newblock {\em Physical Review A}, 69(2):022307, Feb 2004.

\bibitem{dlr:simulations}
J.~Degorre, S.~Laplante, and J.~Roland.
\newblock Simulating quantum correlations as a distributed sampling problem.
\newblock {\em Physical Review A}, 72:062314, 2005.

\bibitem{dlr:simulations07}
J.~Degorre, S.~Laplante, and J.~Roland.
\newblock Classical simulation of traceless binary observables on any bipartite
  quantum state.
\newblock {\em Physical Review A}, 75:012309, 2007.

\bibitem{deutsch&jozsa}
D.~Deutsch and R.~Jozsa.
\newblock Rapid solution of problems by quantum computation.
\newblock In {\em Proceedings of the Royal Society of London}, volume A439,
  pages 553--558, 1992.

\bibitem{doherty2008qmp}
A.C. Doherty, Y.C. Liang, B.~Toner, and S.~Wehner.
\newblock {The quantum moment problem and bounds on entangled multi-prover
  games}.
\newblock In {\em Proceedings of 23rd IEEE Conference on Computational
  Complexity}, pages 199--210, 2008.

\bibitem{DZPOKE:fingerprinting}
J.~Du, P.~Zou, X.~Peng, D.~K. Oi, L.~C. Kwek, and A.~Ekert.
\newblock Experimental quantum multimeter and one-qubit fingerprinting.
\newblock {\em Physical Review A}, 74:042319, 2006.

\bibitem{ehlich&zeller:schwankung}
H.~Ehlich and K.~Zeller.
\newblock Schwankung von {P}olynomen zwischen {G}itterpunkten.
\newblock {\em Mathematische Zeitschrift}, 86:41--44, 1964.

\bibitem{epr}
A.~Einstein, B.~Podolsky, and N.~Rosen.
\newblock Can quantum-mechanical description of physical reality be considered
  complete?
\newblock {\em Physical Review}, 47:777--780, 1935.

\bibitem{forster2008dnl}
M.~Forster, S.~Winkler, and S.~Wolf.
\newblock {Distilling Non-Locality}.
\newblock {\em Physical review letters}, 102:120401, 2009.

\bibitem{frankl&rodl:forbidden}
P.~Frankl and V.~R{\"o}dl.
\newblock Forbidden intersections.
\newblock {\em Transactions of the American Mathematical Society},
  300(1):259--286, 1987.

\bibitem{PhysRevLett.28.938}
S.~J. Freedman and J.~F. Clauser.
\newblock Experimental test of local hidden-variable theories.
\newblock {\em Physical Review Letters}, 28(14):938--941, Apr 1972.

\bibitem{PhysRevA.65.012318}
E.~F. Galv\~ao.
\newblock Feasible quantum communication complexity protocol.
\newblock {\em Physical Review A}, 65(1):012318, Dec 2001.

\bibitem{gavinsky:interactionvsquantum}
D.~Gavinsky.
\newblock Classical interaction cannot replace a quantum message.
\newblock In {\em Proceedings of 40th ACM STOC}, pages 95--102, 2008.
\newblock quant-ph/0703215.

\bibitem{gavinsky:interactionvsnonlocality}
D.~Gavinsky.
\newblock Classical interaction cannot replace quantum nonlocality.
\newblock arXiv:0901.0956, 2008.

\bibitem{gkkrw:1way}
D.~Gavinsky, J.~Kempe, I.~Kerenidis, R.~Raz, and R.~{de} Wolf.
\newblock Exponential separations for one-way quantum communication complexity,
  with applications to cryptography.
\newblock In {\em Proceedings of 39th ACM STOC}, pages 516--525, 2007.
\newblock quant-ph/0611209.

\bibitem{gkrw:identification}
D.~Gavinsky, J.~Kempe, O.~Regev, and R.~{de} Wolf.
\newblock Bounded-error quantum state identification and exponential
  separations in communication complexity.
\newblock In {\em Proceedings of 38th ACM STOC}, pages 594--603, 2006.
\newblock quant-ph/0511013.

\bibitem{gkw:fingerprinting}
D.~Gavinsky, J.~Kempe, and R.~{de} Wolf.
\newblock Strengths and weaknesses of quantum fingerprinting.
\newblock In {\em Proceedings of 21st IEEE Conference on Computational
  Complexity}, pages 288--295, 2006.
\newblock quant-ph/0603173.

\bibitem{grw:qcsmp}
D.~Gavinsky, O.~Regev, and R.~{de} Wolf.
\newblock Simultaneous communication protocols with quantum and classical
  messages.
\newblock {\em Chicago Journal of Theoretical Computer Science}, 7, 2008.
\newblock quant-ph/0807.2758.

\bibitem{GisinGisin}
N.~Gisin and B.~Gisin.
\newblock A local hidden variable model of quantum correlation exploiting the
  detection loophole.
\newblock {\em Physics Letters A}, 260:323--327, 1999.

\bibitem{greenberger:ghz}
D.~M. Greenberger, M.~Horne, and A.~Zeilinger.
\newblock Going beyond {B}ell's theorem.
\newblock In M.~Kafatos, editor, {\em Bell's Theorem, Quantum Theory, and
  Conceptions of the Universe}, pages 69--72. Kluwer Academic, 1989.

\bibitem{grossflammiaeisert2009}
D.~Gross, S.~T. Flammia, and J.~Eisert.
\newblock Most quantum states are too entangled to be useful as computational
  resources.
\newblock {\em Physical Review Letters}, 102:190501, 2009.

\bibitem{grover:search}
L.~K. Grover.
\newblock A fast quantum mechanical algorithm for database search.
\newblock In {\em Proceedings of 28th ACM STOC}, pages 212--219, 1996.
\newblock quant-ph/9605043.

\bibitem{holevo}
A.~S. Holevo.
\newblock Bounds for the quantity of information transmitted by a quantum
  communication channel.
\newblock {\em Problemy Peredachi Informatsii}, 9(3):3--11, 1973.
\newblock English translation in {\it Problems of Information Transmission},
  9:177--183, 1973.

\bibitem{HBMLS:fingerprinting}
R.~T. Horn, S.~A. Babichev, K.-P. Marzlin, A.~I. Lvovsky, and B.~C. Sanders.
\newblock Single-qubit optical quantum fingerprinting.
\newblock {\em Physical Review Letters}, 95:150502, 2005.

\bibitem{Hromkovic97}
Juraj Hromkovi\v{c}.
\newblock {\em Communication complexity and parallel computing}.
\newblock Springer-Verlag, New York, 1997.

\bibitem{ks:disj}
B.~Kalyanasundaram and G.~Schnitger.
\newblock The probabilistic communication complexity of set intersection.
\newblock {\em SIAM Journal on Discrete Mathematics}, 5(4):545--557, 1992.
\newblock Earlier version in Structures'87.

\bibitem{khalfitsirelson}
L.~A. Khalfi and B.~S. Tsirelson.
\newblock Quantum and quasi-classical analogs of {B}ell inequalities.
\newblock In P.~Lahti and P.~Mittelstaedt, editors, {\em Symposium on the
  Foundations of Modern Physics}, pages 441--460. World Scientific, Singapore,
  1985.

\bibitem{ksw:dpt-siam}
H.~Klauck, R.~{\v{S}}palek, and R.~{de} Wolf.
\newblock Quantum and classical strong direct product theorems and optimal
  time-space tradeoffs.
\newblock {\em SIAM Journal on Computing}, 36(5):1472--1493, 2007.
\newblock Earlier version in FOCS'03. quant-ph/0402123.

\bibitem{knuth:cm}
D.~E. Knuth.
\newblock Combinatorial matrices.
\newblock In {\em Selected Papers on Discrete Mathematics}, volume 106 of {\em
  CSLI Lecture Notes}. Stanford University, 2003.

\bibitem{kremer:thesis}
I.~Kremer.
\newblock Quantum communication.
\newblock Master's thesis, Hebrew University, Computer Science Department,
  1995.

\bibitem{kushilevitz&nisan:cc}
E.~Kushilevitz and N.~Nisan.
\newblock {\em Communication Complexity}.
\newblock Cambridge University Press, 1997.

\bibitem{linial&shraibman:cc}
N.~Linial and A.~Shraibman.
\newblock Lower bounds in communication complexity based on factorization
  norms.
\newblock In {\em Proceedings of 39th ACM STOC}, pages 699--708, 2007.

\bibitem{MRTZLG:Bell}
I.~Marcikic, H.~de~Riedmatten, W.~Tittel, H.~Zbinden, M.~Legr{\'e}, and
  N.~Gisin.
\newblock Distribution of time-bin entangled qubits over 50 km of optical
  fiber.
\newblock {\em Physical Review Letters}, 93:180502, 2004.

\bibitem{MAG:NLbox}
Ll. Masanes, A.~Acin, and N.~Gisin.
\newblock General properties of nonsignaling theories.
\newblock {\em Physical Review A}, 73:012112, 2006.

\bibitem{Massardetection}
S.~Massar.
\newblock Nonlocality, closing the detection loophole, and communication
  complexity.
\newblock {\em Physical Review A}, 65:032121, 2002.

\bibitem{Massar:fingerprinting}
S.~Massar.
\newblock Quantum fingerprinting with a single particle.
\newblock {\em Physical Review A}, 71:012310, 2005.

\bibitem{PhysRevA.63.052305}
S.~Massar, D.~Bacon, N.~J. Cerf, and R.~Cleve.
\newblock Classical simulation of quantum entanglement without local hidden
  variables.
\newblock {\em Physical Review A}, 63(5):052305, Apr 2001.

\bibitem{PhysRevA.66.052112}
S.~Massar, S.~Pironio, J.~Roland, and B.~Gisin.
\newblock Bell inequalities resistant to detector inefficiency.
\newblock {\em Phys. Rev. A}, 66(5):052112, Nov 2002.

\bibitem{mmmom:atomatom}
D.~N. Matsukevich, P.~Maunz, D.~L. Moehring, S.~Olmschenk, and C.~Monroe.
\newblock {B}ell inequality violation with two remote atomic qubits.
\newblock {\em Physical Review Letters}, 100:150404, 2008.

\bibitem{maudlin:simulations}
T.~Maudlin.
\newblock Bell's inequality, information transmission, and prism models.
\newblock In D.~Hull, M.~Forbes, and K.~Okruhlik, editors, {\em PSA:
  Proceedings of the Biennial Meeting of the Philosophy of Science
  Association}, volume~1, pages 404--417, 1992.

\bibitem{Mermin90}
N.~D. Mermin.
\newblock Simple unified form for the major no-hidden-variables theorems.
\newblock {\em Physical Review Letters}, 65:3373--3376, 1990.

\bibitem{mermin:reality}
N.~D. Mermin.
\newblock What's wrong with these elements of reality?
\newblock {\em Physics Today}, 43:9--11, 1990.

\bibitem{Mermin93}
N.~D. Mermin.
\newblock Hidden variables and the two theorems of {J}ohn {B}ell.
\newblock {\em Reviews of Modern Physics}, 65(3):803--815, 1993.

\bibitem{mmbm:atomphoton}
D.~L. Moehring, M.~J. Madsen, B.~B. Blinov, and C.Monroe.
\newblock Experimental {B}ell inequality violation with an atom and a photon.
\newblock {\em Physical Review Letters}, 93:090410, 2004.

\bibitem{navascues2007bsq}
M.~Navascues, S.~Pironio, and A.~Acin.
\newblock Bounding the set of quantum correlations.
\newblock {\em Physical Review Letters}, 98(1):10401, 2007.

\bibitem{navascues2008chs}
M.~Navascues, S.~Pironio, and A.~Ac{\'\i}n.
\newblock {A convergent hierarchy of semidefinite programs characterizing the
  set of quantum correlations}.
\newblock {\em New Journal of Physics}, 10(073013):073013, 2008.

\bibitem{nayak:qfa}
A.~Nayak.
\newblock Optimal lower bounds for quantum automata and random access codes.
\newblock In {\em Proceedings of 40th IEEE FOCS}, pages 369--376, 1999.
\newblock quant-ph/9904093.

\bibitem{nayak&salzman:entanglement}
A.~Nayak and J.~Salzman.
\newblock On communication over an entanglement-assisted quantum channel.
\newblock In {\em Proceedings of 34th ACM STOC}, pages 698--704, 2002.

\bibitem{newman:random}
I.~Newman.
\newblock Private vs.~common random bits in communication complexity.
\newblock {\em Information Processing Letters}, 39(2):67--71, 1991.

\bibitem{newman&szegedy:1round}
I.~Newman and M.~Szegedy.
\newblock Public vs.~private coin flips in one round communication games.
\newblock In {\em Proceedings of 28th ACM STOC}, pages 561--570, 1996.

\bibitem{nielsen&chuang:qc}
M.~A. Nielsen and I.~L. Chuang.
\newblock {\em Quantum Computation and Quantum Information}.
\newblock Cambridge University Press, 2000.

\bibitem{PBDWZ:GHZ}
J.-W. Pan, D.~Bouwmeester, M.~Daniell, H.~Weinfurter, and A.~Zeilinger.
\newblock Experimental test of quantum nonlocality in three-photon
  {G}reenberger-{H}orne-{Z}eilinger entanglement.
\newblock {\em Nature}, 403:515, 2000.

\bibitem{paturi:degree}
R.~Paturi.
\newblock On the degree of polynomials that approximate symmetric {B}oolean
  functions.
\newblock In {\em Proceedings of 24th ACM STOC}, pages 468--474, 1992.

\bibitem{pearle:detectionloophole}
P.~M. Pearle.
\newblock Hidden-variable example based upon data rejection.
\newblock {\em Physical Review D}, 2:1418, 1970.

\bibitem{PGWPVJ}
D.~Perez-Garcia, M.M. Wolf, C.~Palazuelos, I.~Villanueva, and M.~Junge.
\newblock Unbounded violation of tripartite {B}ell inequalities.
\newblock {\em Communications in Mathematical Physics}, 279:455, 2008.
\newblock arXiv:quant-ph/0702189.

\bibitem{PopescuRohrlich:NLbox}
S.~Popescu and D.~Rohrlich.
\newblock Quantum nonlocality as an axiom.
\newblock {\em Foundations of Physics}, 24:379, 1994.

\bibitem{RNOBBRH:GHZ}
A.~Rauschenbeutel, G.~Nogues, S.~Osnaghi, P.~Bertet, M.~Brune, J.-M. Raimond,
  and S.~Haroche.
\newblock Step-by-step engineered multiparticle entanglement.
\newblock {\em Science}, 288:5473, 2000.

\bibitem{raz:qcc}
R.~Raz.
\newblock Exponential separation of quantum and classical communication
  complexity.
\newblock In {\em Proceedings of 31st ACM STOC}, pages 358--367, 1999.

\bibitem{razborov:disj}
A.~Razborov.
\newblock On the distributional complexity of disjointness.
\newblock {\em Theoretical Computer Science}, 106(2):385--390, 1992.

\bibitem{razborov:qdisj}
A.~Razborov.
\newblock Quantum communication complexity of symmetric predicates.
\newblock {\em Izvestiya of the Russian Academy of Sciences, mathematics},
  67(1):159--176, 2003.
\newblock quant-ph/0204025.

\bibitem{rt:simulations}
O.~Regev and B.~Toner.
\newblock Simulating quantum correlations with finite communication.
\newblock In {\em Proceedings of 48th Annual IEEE Symposium on Foundations of
  Computer Science (FOCS'07)}, pages 384--394, 2007.

\bibitem{rivlin&cheney:approx}
T.~J. Rivlin and E.~W. Cheney.
\newblock A comparison of uniform approximations on an interval and a finite
  subset thereof.
\newblock {\em SIAM Journal on Numerical Analysis}, 3(2):311--320, 1966.

\bibitem{rkmsimw:detectionloophole}
M.~A. Rowe, D.~Kielpinski, V.~Meyer, C.A. Sackett, W.~M. Itano, C.~Monroe, and
  D.~J. Wineland.
\newblock Experimental violation of a {B}ell's inequality with efficient
  detection.
\newblock {\em Nature}, 409:791, 2001.

\bibitem{SKKLMMRTIWM:4part}
C.~A. Sackett, D.~Kielpinski, B.~E. King, C.~Langer, V.~Meyer, C.~J. Myatt,
  M.~Rowe, Q.~A. Turchette, W.~M. Itano, D.~J. Wineland, and C.~Monroe.
\newblock Experimental entanglement of four particles.
\newblock {\em Nature}, 404:256, 2000.

\bibitem{scholz2008tsp}
V.~B. Scholz and R.~F. Werner.
\newblock {Tsirelson's Problem}.
\newblock {\em Arxiv preprint arXiv:0812.4305}, 2008.

\bibitem{schrodinger:35}
E.~Schr{\"o}dinger.
\newblock Discussion of probability relations between separated systems.
\newblock {\em Proceedings of the Cambridge Philosophical Society},
  31:555--563, 1935.

\bibitem{schrodinger:36}
E.~Schr{\"o}dinger.
\newblock Probability relations between separated systems.
\newblock {\em Proceedings of the Cambridge Philosophical Society},
  32:446--451, 1936.

\bibitem{schumacher:95}
B.~Schumacher.
\newblock Quantum coding.
\newblock {\em Physical Review A}, 51(4):2738--2747, Apr 1995.

\bibitem{shannon:communication}
C.~E. Shannon.
\newblock A mathematical theory of communication.
\newblock {\em Bell System Technical Journal}, 27:379--423, 623--656, 1948.

\bibitem{sherstov:qcclower}
A.~Sherstov.
\newblock The pattern matrix method for lower bounds on quantum communication.
\newblock In {\em Proceedings of 40th ACM STOC}, pages 85--94, 2008.

\bibitem{shor:factoring}
P.~W. Shor.
\newblock Polynomial-time algorithms for prime factorization and discrete
  logarithms on a quantum computer.
\newblock {\em SIAM Journal on Computing}, 26(5):1484--1509, 1997.
\newblock Earlier version in FOCS'94. quant-ph/9508027.

\bibitem{steiner:simulations}
M.~Steiner.
\newblock Towards quantifying non-local information transfer: finite-bit
  non-locality.
\newblock {\em Physics Letters A}, 270:239--244, 2000.
\newblock arXiv:quant-ph/9902014.

\bibitem{PhysRevLett.93.010503}
R.~T. Thew, A.~Ac\'in, H.~Zbinden, and N.~Gisin.
\newblock Bell-type test of energy-time entangled qutrits.
\newblock {\em Physical Review Letters}, 93(1):010503, Jul 2004.

\bibitem{TBZG:Bell}
W.~Tittel, J.~Brendel, H.~Zbinden, and N.~Gisin.
\newblock Violation of {B}ell inequalities by photons more than 10 km apart.
\newblock {\em Physical Review Letters}, 81:3563, 1998.

\bibitem{PhysRevLett.91.187904}
B.~F. Toner and D.~Bacon.
\newblock Communication cost of simulating {B}ell correlations.
\newblock {\em Physical Review Letters}, 91(18):187904, Oct 2003.

\bibitem{TSBBZW:commcompl}
P.~Trojek, C.~Schmid, M.~Bourennane, C.~Brukner, M.~{\.{Z}}ukowski, and
  H.~Weinfurter.
\newblock Experimental quantum communication complexity.
\newblock {\em Physical Review A}, 75:050305(R), 2005.

\bibitem{tsirelson87}
B.~S. Tsirelson.
\newblock Quantum analogues of the {B}ell inequalities. the case of two
  spatially separated domains.
\newblock {\em Journal of Soviet Mathematics}, 36:557--570, 1987.

\bibitem{PhysRevLett.89.240401}
A.~Vaziri, G.~Weihs, and A.~Zeilinger.
\newblock Experimental two-photon, three-dimensional entanglement for quantum
  communication.
\newblock {\em Physical Review Letters}, 89(24):240401, Nov 2002.

\bibitem{wehner2008lbd}
S.~Wehner, M.~Christandl, and A.C. Doherty.
\newblock {Lower bound on the dimension of a quantum system given measured
  data}.
\newblock {\em Physical Review A}, 78(6), 2008.

\bibitem{WJSWZ:Bell}
G.~Weihs, T.~Jennewein, C.~Simon, H.~Weinfurter, and A.~Zeilinger.
\newblock Violation of {B}ell's inequality under strict {E}instein locality
  conditions.
\newblock {\em Physical Review Letters}, 81:5039, 1998.

\bibitem{PhysRevA.40.4277}
R.~F. Werner.
\newblock Quantum states with {Einstein-Podolsky-Rosen} correlations admitting
  a hidden-variable model.
\newblock {\em Physical Review A}, 40(8):4277--4281, Oct 1989.

\bibitem{werner2001amb}
R.~F. Werner and M.~M. Wolf.
\newblock {All-multipartite {B}ell-correlation inequalities for two dichotomic
  observables per site}.
\newblock {\em Physical Review A}, 64:032112, 2001.

\bibitem{werner2001bia}
R.~F. Werner and M.~M. Wolf.
\newblock {Bell inequalities and entanglement}.
\newblock {\em Quantum Information and Computation}, 1(3):1--25, 2001.
\newblock Arxiv preprint quant-ph/0107093.

\bibitem{yao:distributive}
A.~C-C. Yao.
\newblock Some complexity questions related to distributive computing.
\newblock In {\em Proceedings of 11th ACM STOC}, pages 209--213, 1979.

\bibitem{yao:qcircuit}
A.~C-C. Yao.
\newblock Quantum circuit complexity.
\newblock In {\em Proceedings of 34th IEEE FOCS}, pages 352--360, 1993.

\bibitem{yao:qfp}
A.~C-C. Yao.
\newblock On the power of quantum fingerprinting.
\newblock In {\em Proceedings of 35th ACM STOC}, pages 77--81, 2003.

\bibitem{ZBCYCP:GMN}
J.~Zhang, X.-H. Bao, T.-Y. Chen, T.~Yang, A.~Cabello, and J.-W. Pan.
\newblock Experimental quantum guess my number protocol using multiphoton
  entanglement.
\newblock {\em Physical Review A}, 75:022302, 2007.

\bibitem{PhysRevLett.91.180401}
Z.~Zhao, T.~Yang, Y.-A. Chen, A.-N. Zhang, M.~{\.{Z}}ukowski, and J.-W. Pan.
\newblock Experimental violation of local realism by four-photon
  {G}reenberger-{H}orne-{Z}eilinger entanglement.
\newblock {\em Physical Review Letters}, 91(18):180401, Oct 2003.

\bibitem{Zukowski:all-Bell}
M.~{\.{Z}}ukowski and {\v{C}}.~Brukner.
\newblock Bell's theorem for general n-qubit states.
\newblock {\em Physical Review Letters}, 88(21):210401, May 2002.

\end{thebibliography}

\appendix

\section{Nayak's Proof of a Consequence of Holevo's Bound}\label{nayakbound}

Here we prove that if we are encoding $n$ bits in $d$-dimensional quantum states,
then the average recovery probability is at most $d/2^n$.
Therefore, an exact procedure requires $d \ge 2^n$, and thus at least $n$ qubits.

Let $\rho_0,\dots,\rho_{2^n-1}$ be the $d$-dimensional states that encode the
elements of $\{0,1\}^n$ (which we identify with $\{0,1,\ldots,2^n-1\}$ in
the obvious way).
Let $E_0,\dots,E_{2^n-1}$ be the measurement operators
applied for decoding (they sum to the $d$-dimensional identity).
The probability of successfully recovering $x \in \{0,1\}^n$ from its encoding
is $\Tr(E_x\rho_x)$.
Therefore, we can bound the success probability for a uniformly
random $x \in \{0,1\}^n$ by
\begin{eqnarray}
\frac{1}{2^n}\sum_{x=0}^{2^n-1} \Tr(E_x \rho_x) & \leq &
\frac{1}{2^n}\sum_{x=0}^{2^n-1} \Tr(E_x) \nonumber \\
& = & \frac{1}{2^n}\Tr\left(\sum_{x=0}^{2^n-1} E_x\right) \nonumber \\
& = & \frac{1}{2^n}\Tr(I) \nonumber \\
& = & \frac{d}{2^n}.
\end{eqnarray}
The first inequality follows because the density operator $\rho_x$ is
positive semi-definite and has trace~1, therefore it can be unitarily
diagonalized: $U^*\rho_x U =  D $, where $D$ is  diagonal with
 diagonal entries that  are non-negative and sum to $1$.
Because the trace is invariant under cyclic permutations of the matrices,  we now have $\Tr(E_x \rho_x) = \Tr(U^* E_x U U^*
\rho_x U) = \Tr(U^* E_x U D) \leq \Tr( U^* E_x U I) = \Tr(E_x)$.

\section{Rectangles and the Lower Bound for Distributed Deutsch-Jozsa}\label{appclassicallowerbounds}

Separations between quantum and classical communication complexity always require two things: an efficient quantum protocol for some problem,
and a \emph{lower bound} on the communication of all classical protocols solving that same problem.
In this appendix we will give some tools for lower bounding classical communication complexity,
leading eventually to the lower bound on classical protocols for the Distributed Deutsch-Jozsa problem
that we mentioned in Section~\ref{secdistributeddj}.

\subsection{Rectangles}\label{ssecrectangles}

Consider some communication complexity problem $f:X\times Y\rightarrow\01$,
where Alice starts with an input $x\in X$ and Bob starts with an input $y\in Y$.
We start by introducing the crucial combinatorial notion for classical lower bounds.
A \emph{rectangle} is a set $R\subseteq X\times Y$ that is of the form $R=A\times B$
with $A\subseteq X$ and $B\subseteq Y$. For example, if $n=2$ and $A=\{00,01\}$, $B=\{01,10\}$
then $R=A\times B=\{(00,01),(00,10),(01,01),(01,10))\}$ is a rectangle.
The following result is a fundamental property of classical deterministic protocols.

\begin{lemma}
If a deterministic protocol has communication $c$, then there exist $2^c$ rectangles $R_1,\ldots,R_{2^c}$ that partition $X\times Y$, such that the protocol gives the same output $a_i$ for each $(x,y)\in R_i$.
\end{lemma}

We omit the easy proof of this lemma, which is by induction on $c$.
For example, suppose there is only one $k$-bit message $m$ going from Alice to Bob and
then Bob returns the 1-bit output. Then the $2^{k+1}$ rectangles would be of the form
$R_{m,a}=A_m\times Y_{m,a}$, with $m\in\01^k$ and $a\in\01$,
where $A_m$ is the set of $x$'s for which Alice sends $k$-bit message $m$,
and $Y_{m,a}$ is the set of $y$'s for which Bob returns output $a$ when receiving message $m$.
Note that if our protocol computes $f$ correctly, then the rectangles are ``monochromatic'':
the protocol returns the same answer $f(x,y)$ for all $(x,y)\in R_i$.

As a simple application of this we prove the so-called ``rank lower bound''.
Consider some communication complexity problem $f:X\times Y\rightarrow\01$.
Let $M_f$ be the $|X|\times|Y|$ matrix whose entries are defined by $M_f(x,y)=f(x,y)$.
This is called the \emph{communication matrix} of $f$.  It can be viewed as a 2-dimensional truth table.
We use $\rank(f)$ to denote the rank of this matrix over the field of real numbers.
For example, the communication matrix for the equality function is the $2^n\times 2^n$
identity matrix, which has 1s on its diagonal and 0s elsewhere. Hence $\rank(\EQ)=2^n$.

Suppose we have some $c$-bit deterministic protocol that computes $f$.
We know that this partitions the input space $X\times Y$ into rectangles $R_1,\ldots,R_{2^c}$.
Since each 1-input $(x,y)$ occurs in exactly one 1-rectangle, we have
$$
M_f=\sum_{i:a_i=1} R_i,
$$
where we view $R_i$ as a $2^n\times 2^n$ matrix with 1s on its elements and 0s elsewhere.
Note that $R_i$ is a matrix of rank 1.
Hence, using $\rank(A+B)\leq\rank(A)+\rank(B)$, we get
$$
\rank(M_f)=\rank\left(\sum_{i:a_i=1} R_i\right)\leq \sum_{i:a_i=1} \rank(R_i)=\sum_{i:a_i=1}1\leq 2^c.
$$
But that means that a lower bound on the rank of $M_f$ implies a lower bound on the communication!
In particular, it follows that for the equality problem, the communication $c$ needs to be at least $n$.

\subsection{Randomized protocols}\label{app:randprot}

In a randomized protocol, Alice and Bob may flip coins and  the protocol has to
output the right value $f(x,y)$ with probability $\geq 2/3$ for all $(x,y)\in\mathcal{D}$.
We can fix these coins to obtain a deterministic protocol.
Suppose randomized protocol $A$ uses $c$ bits of communication and has success probability $2/3$ on all inputs.
Let $A(x,y,r_A,r_B)=1$ if the protocol gives the correct output $f(x,y)$ on input $x,y$ using coin flips
$r_A$ for Alice and $r_B$ for Bob, and $A(x,y,r_A,r_B)=0$ otherwise.
For each $x,y$, a good randomized protocol satisfies
$$
\Exp_{r_A,r_B}[A(x,y,r_A,r_B)]\geq 2/3,
$$
where the expectation is taken over uniformly chosen strings $r_A$ and $r_B$.
Now let $\mu:\01^n\times\01^n\rightarrow[0,1]$ be an input distribution.  Then also
$$
\Exp_{\mu,r_A,r_B}[A(x,y,r_A,r_B)]\geq 2/3,
$$
where the expectation is taken over $r_A,r_B$, and $x,y$ picked according to $\mu$.
By the averaging principle, there exists a way to fix $r_A$ and $r_B$ such that the success
probability (under $\mu$) of the resulting \emph{deterministic} protocol is at least $2/3$.
Accordingly, if we want to lower bound the randomized communication complexity of a function,
it suffices to find some ``hard'' input distribution $\mu$, and to show that all \emph{deterministic}
protocols that have error at most $1/3$ under that distribution, need a lot of communication.

The reason why the step to deterministic protocols is helpful, is that deterministic protocols
partition the input space into rectangles as we've seen before.  Suppose we can show that
all ``large'' rectangles in the communication matrix
have roughly as many 0s as 1s in them (weighed according to $\mu$).
Then the protocol will make a large error on all large rectangles.  Conversely, if we know
the protocol does not make a large error, most of its rectangles must have been ``small''.
But that can only be if there are many rectangles. Since the number of rectangles is $2^c$,
the communication $c$ must have been large.
This idea leads to the following lower bound method.
The \emph{discrepancy} of rectangle $R=A\times B$ under $\mu$ is
the difference between the weight of the 0s and the 1s in that rectangle:
$$
\delta_\mu(R)=\left|\mu(R\cap f^{-1}(1))-\mu(R\cap f^{-1}(0))\right|
$$
The discrepancy of $f$ under $\mu$ is the maximum over all rectangles:
$$
\delta_\mu(f)=\max_R\delta_\mu(R).
$$
If $f$ has small discrepancy, that means that all ``large'' rectangles are roughly balanced.
Suppose a deterministic protocol partitions the input space into rectangles $R_1,\ldots,R_{2^c}$.
Suppose it has success probability $1/2+\eps$. The best bias (difference between success and failure probabilities)
that the protocol can achieve on rectangle $R_i$, is $\delta_\mu(R_i)$, by giving the output with highest weight in that rectangle.
The success probability is $\sum_i\mu(R_i\cap f^{-1}(a_i))$ and the
error probability is $\sum_i\mu(R_i\cap f^{-1}(1-a_i))$, where $a_i$
is the majority value of $f$ on the pairs $(x,y) \in R_i$, weighted
according to $\mu$.
Hence we have
$$
2\eps\leq\sum_{i=1}^{2^c}\mu(R_i\cap f^{-1}(a_i)) - \sum_{i=1}^{2^c}\mu(R_i\cap f^{-1}(1-a_i))\leq \sum_{i=1}^{2^c}\delta_\mu(R_i)\leq 2^c\delta_\mu(f).
$$
This is a lower bound on the communication: $c\geq \log(2\eps/\delta_\mu(f))$.
Accordingly, a distribution $\mu$ where $\delta_\mu(f)$ is small gives a lower bound on the communication of deterministic protocols for $f$ under $\mu$, and then the same lower bound applies to randomized protocols.

\subsection{Discrepancy of the inner product function}\label{ssecdiscip}

To illustrate the discrepancy lower bound technique,
we now consider the inner product function, defined by $\IP(x,y)=x\cdot y\pmod 2$.
We will show that its discrepancy under the uniform distribution is very small.
We analyze the $2^n\times 2^n$ matrix $M$ whose $(x,y)$ entry is $(-1)^{x\cdot y}$.
This is just the communication matrix for $\IP$, with 0s replaced by 1s, and 1s replaced by $-1$s.
Lindsey's lemma shows that large rectangles in $M$ are quite balanced:

\begin{lemma}[Lindsey]
For every rectangle $R=A\times B$, the absolute value of the
sum of the $M$-entries in that rectangle is at most $\sqrt{|A|\cdot|B|\cdot 2^n}$.
\end{lemma}

\begin{proof}
It is easy to see that $M$ is symmetric and $M^2=2^n I$.  This implies, for any vector $v$,
$$
\norm{M v}^2=v^TM^TM v=2^n v^T v=2^n\norm{v}^2,
$$
where the norm is the usual Euclidean vector length.
Let $v_A\in\01^{2^n}$ and $v_B\in\01^{2^n}$ be the
characteristic (column) vectors of the sets $A$ and $B$.
The sum of the $M$-entries in $R$ is $\sum_{a\in A, b\in B}M_{ab}=v_A^T M v_B$.
We can bound this using Cauchy-Schwarz:
$$
|v_A^T M v_B|\leq \norm{v_A}\cdot \norm{M v_B}=\norm{v_A}\cdot \sqrt{2^n}\norm{v_B}=\sqrt{|A|\cdot|B|\cdot 2^n}.
$$
\end{proof}

Let $\mu(x,y)=1/2^{2n}$ be the uniform input distribution.  Note that the discrepancy
of the rectangle $R$ under $\mu$ is exactly the difference of $+1$'s and $-1$'s in $R$,
divided by $2^{2n}$. By Lindsey's lemma, this is $\delta_\mu(R)\leq \sqrt{|A|\cdot|B|}/2^{3n/2}$.
Because $|A|,|B|\leq 2^n$, it follows that the discrepancy of the inner product function
under the uniform distribution is $\delta_\mu(\IP)\leq 2^{-n/2}$.
Hence we get a $n/2$ lower bound on the randomized communication complexity of \IP.

\subsection{The lower bound for the Distributed Deutsch-Jozsa problem}\label{app:DJ}

Recall the Distributed Deutsch-Jozsa problem from Section~\ref{secdistributeddj}.
Buhrman, Cleve, and Wigderson~\cite{BuhrmanCleveWigderson98} used a combinatorial result
of Frankl and R\"odl~\cite{frankl&rodl:forbidden} to prove the following classical lower bound:

\begin{theorem}
Every deterministic classical protocol that solves the Distributed Deutsch-Jozsa problem, needs
to communicate at least $0.007 n$ bits.
\end{theorem}

\begin{proof}
Suppose there is a $c$-bit deterministic classical protocol for the problem.
Each $c$-bit conversation corresponds to a rectangle $R=A\times B$, with $A,B\subseteq\01^n$,
such that the protocol has the same conversation and output
if, and only if, $(x,y)\in R$.
Since there are at most $2^c$ possible conversations,
the protocol partitions $\01^n\times\01^n$ in at most $2^c$
different such rectangles.
Now consider all $n$-bit strings $x$ with Hamming weight $n/2$ (i.e., $n/2$ ones and $n/2$ zeroes).
There are ${n\choose n/2}\approx 2^n/\sqrt{n}$ of those.
Since every $(x,x)$-pair must occur in some rectangle and
there are only $2^c$ rectangles, there is a rectangle $R=A\times B$
that contains at least $2^n/(\sqrt{n}2^c)$ different such $(x,x)$-pairs.
Let $S=\{x : |x|=n/2, \ (x,x)\in R\}$ be the set of such $x$.
Since $R$ contains some $(x,x)$-pairs (on which the protocol outputs 1)
and the protocol has the same output for all inputs in $R$,
$R$ cannot contain any 0-inputs.
This implies that the Hamming distance of every pair $x,y\in S$
is different from $n/2$, for otherwise $(x,y)$ would be a 0-input in $R$.
Viewing the strings $x$ in $S$ as characteristic vectors of sets, it is easy to
see that the size of the intersection of $x,y\in S$ is never $n/4$.
Thus we have a set system $S$ of at least $2^n/\sqrt{n}2^c$ sets over an $n$-element universe,
such that the size of the intersection of any two sets in $S$ is not $n/4$.
However, by Corollary~1.2 of~\cite{frankl&rodl:forbidden},
such a set system can have at most $1.99^n$ elements, so we have
$$
\frac{2^n}{\sqrt{n}2^c}\leq |S|\leq 1.99^n.
$$
This implies $c\geq\log(2^n/\sqrt{n}1.99^n)\geq 0.007\ n$.
\end{proof}

\section{Razborov's Lower Bound for the Quantum Communication Complexity of Intersection}\label{appquantumlowerbounds}

While the previous section discussed some basic methods for lower bounding \emph{classical} communication complexity,
here we focus on methods to lower bound \emph{quantum} communication complexity (sometimes with prior entanglement).

\subsection{The Kremer-Razborov-Yao lemma and its consequences}

The following lemma is due to Razborov~\cite[Proposition~3.3]{razborov:qdisj}
and is similar to earlier statements by Yao~\cite{yao:qcircuit} and Kremer~\cite{kremer:thesis}.
It can intuitively by viewed as a quantum analogue of the rectangle-decomposition
of classical protocols that we explained in Section~\ref{ssecrectangles}.
We skip the easy proof, which is by induction on $q$.

\begin{lemma}[Kremer-Razborov-Yao]\label{lemkremeryao}
Let $\ket{\Psi}$ denote the (possibly entangled) starting state of a quantum protocol that communicates $q$ qubits of communication and has binary output.
For all inputs $x$ of Alice and $y$ of Bob, there exist linear operators $A_h(x),B_h(y)$, for all $h\in\01^{q-1}$,
each with operator norm (i.e., largest singular value) at most~1, such that the acceptance probability (i.e., probability of output~`1') of the protocol is
$$
P(x,y)=\norm{\displaystyle\sum_{h\in\01^{q-1}} (A_h(x)\otimes B_h(y))\ket{\Psi}}^2,
$$
where the norm is the usual Euclidean vector length.
\end{lemma}

Consider the special case where the protocol starts without entanglement, so we can write $\ket{\Psi}=\ket{\Psi_A}\ket{\Psi_B}$.
In this case we can rewrite the acceptance probabilities as
\begin{eqnarray*}
P(x,y) & = & \norm{\displaystyle\sum_{h\in\01^{q-1}} (A_h(x)\otimes B_h(y))\ket{\Psi_A}\ket{\Psi_B}}^2\\
       & = & \bra{\Psi_A}\bra{\Psi_B}\left(\sum_{h\in\01^{q-1}} (A_h(x)\otimes B_h(y))\right)^*\cdot\left(\sum_{h'\in\01^{q-1}} (A_{h'}(x)\otimes B_{h'}(y))\right)\ket{\Psi_A}\ket{\Psi_B}\\
       & = & \sum_{h,h'\in\01^{q-1}}\bra{\Psi_A}A_h(x)^*A_{h'}(x)\ket{\Psi_A}\cdot\bra{\Psi_B} B_h(y)^*B_{h'}(y)\ket{\Psi_B}.
\end{eqnarray*}
Let $a(x)$ be the $2^{2q-2}$-dimensional row vector with $(h,h')$-entry equal to $\bra{\Psi_A}A_h(x)^*A_{h'}(x)\ket{\Psi_A}$,
and similarly define column vector $b(y)$ with entries $\bra{\Psi_B} B_h(y)^*B_{h'}(y)\ket{\Psi_B}$, then the last expression is just the scalar product $a(x)b(y)$.
If we now define $A$ to be the $|X|\times 2^{2q-2}$ matrix with rows $a(x)$,
and $B$ the $2^{2q-2}\times|Y|$ matrix with columns $b(y)$, then we have proved the following lemma.

\begin{lemma}\label{lemqaccprobs}
Consider a quantum communication protocol (without prior entanglement) on input-set $X\times Y$, that communicates $q$ qubits,
with acceptance probabilities denoted by $P(x,y)$, and $P$ the corresponding $|X|\times|Y|$ matrix.
There exist $|X|\times 2^{2q-2}$ matrix $A$ and $2^{2q-2}\times|Y|$ matrix $B$, both with entries of absolute value at most~1,
such that $P=AB$.
\end{lemma}

Note that the rank of matrix $P$ is at most $2^{2q-2}$, since $\rank(AB)\leq\min(\rank(A),\rank(B))$.
This allows us to generalize the classical rank lower bound from Section~\ref{ssecrectangles} to the quantum domain.
If we have a $q$-qubit protocol that computes some function $f:X\times Y\rightarrow\01$ with success probability~1,
then $P(x,y)$ equals $f(x,y)$, and the $|X|\times|Y|$ matrix $P$ is actually the communication matrix $M_f$, whose $(x,y)$ entry is $f(x,y)$.
Hence we obtain a lower bound $q\geq\frac{\rank(P)}{2}+1=\frac{\rank(M_f)}{2}+1$ on the quantum communication of protocols with success probability~1.
Similarly, one can obtain lower bounds on the bounded-error quantum communication complexity by lower bounding the
rank needed for a matrix $P$ that is close to the matrix of function values at each entry
(since an $\eps$-error protocol satisfies $|P(x,y)-f(x,y)|\leq\eps$ for all inputs).

Finally, let us note without proof that one can also use the discrepancy method (Section~\ref{app:randprot})
to lower bound quantum communication complexity~\cite{kremer:thesis},
even for protocols with prior entanglement~\cite{linial&shraibman:cc}.
Since the Inner Product function has very small discrepancy (Section~\ref{ssecdiscip}),
we thus have another way of showing a linear lower bound for it,
different from the one explained in Section~\ref{sect:InnerProduct}.

\subsection{Translation from protocols to polynomials}

The following key lemma is implicit in Razborov's paper~\cite{razborov:qdisj}; the presentation we give here is taken from~\cite{ksw:dpt-siam}.
It allows us to translate the average acceptance probability of a $q$-qubit protocol
(as a function of the intersection size $i$ of the inputs $x$ and $y$, viewed as subsets of $\{1,\ldots,n\}$)
to a polynomial in $i$ of degree roughly $q$. Accordingly, efficient protocols give low-degree polynomials.

Razborov's proof relies on the following linear algebraic notions.
The \emph{operator norm} $\norm{A}$ of a matrix $A$ is its largest singular value $\sigma_1$
(not to be confused with the Euclidean vector norm of Lemma~\ref{lemkremeryao}).
The \emph{trace inner product}---also known as Hilbert--Schmidt inner product---between $A$ and $B$ is $\inpc{A}{B}=\Tr(A^*B)$.
The \emph{trace norm} is $\norm{A}_{tr}=\max\{|\inpc{A}{B}|:\norm{B}=1\}=\sum_i\sigma_i$,
the sum of all singular values of $A$.
The \emph{Frobenius norm} is
$\norm{A}_F=\sqrt{\smash{\sum_{i j}|A_{i j}|^2}\vphantom{\sum_i}}=\sqrt{\sum_i\sigma_i^2}$.

\begin{lemma}\label{lemrazborov}
Consider a quantum communication protocol (without prior entanglement) on $n$-bit inputs $x$ and $y$, that communicates $q$ qubits,
with acceptance probabilities denoted by $P(x,y)$.
Define
\[
P(i)=\Exp_{|x|=|y|=n/4, |x\wedge y|=i}[P(x,y)],
\]
where the expectation is taken uniformly over all $x,y$
that each have weight $n/4$ and that have intersection $i$.
For every $d\leq n/4$ there exists a degree-$d$ polynomial $q$ such
that $|P(i)-q(i)|\leq 2^{2q-(d/4)}$ for all $i\in\{0,\ldots,n/8\}$.
\end{lemma}

\begin{proof}
We only consider the ${\cal N}={n\choose n/4}$ strings of weight $n/4$.
Let $P$ denote the ${\cal N}\times {\cal N}$ matrix of the
acceptance probabilities on these inputs.
By Lemma~\ref{lemqaccprobs}, we can write $P=A B$,
where $A$ is an ${\cal N}\times 2^{2q-2}$ matrix with each
entry at most 1 in absolute value, and similarly for $B$.
Note that $\norm{A}_F,\norm{B}_F\leq\sqrt{{\cal N}2^{2q-2}}$.
By the Cauchy-Schwarz inequality for unitarily invariant norms~\cite[p.~95]{Bhatia:97a}, we have
$$
\norm{P}_{tr}\leq\norm{A}_F\cdot\norm{B}_F\leq {\cal N}2^{2q-2}.
$$
Let $\mu_i$ denote the ${\cal N}\times {\cal N}$ matrix corresponding
to the uniform probability distribution on $\{(x,y) : |x\wedge y|=i\}$.
These ``combinatorial matrices'' have been well studied~\cite{knuth:cm}.
Note that $\inpc{P}{\mu_i}$ is the expected acceptance
probability $P(i)$ of the protocol under that distribution.
One can show that the different $\mu_i$ commute; thus they have
the same eigenspaces $E_0,\ldots,E_{n/4}$ and can be simultaneously
diagonalized by some orthogonal matrix $U$.
For $t\in\{0,\ldots,n/4\}$, let $(U P U^T)_t$ denote the block of
$U P U^T$ corresponding to $E_t$, and let $a_t=\Tr((U P U^T)_t)$ be its trace.
Then we have
$$
\sum_{t=0}^{n/4} |a_t|\leq\sum_{j=1}^{{\cal N}} \left|(U P U^T)_{jj}\right|
\leq\norm{U P U^T}_{tr}=\norm{P}_{tr}\leq {\cal N}2^{2q-2},
$$
where the second inequality is a property of the trace norm.

Let $\lambda_{it}$ be the eigenvalue of $\mu_i$ in eigenspace $E_t$.
Knuth~\cite{knuth:cm} gives an exact combinatorial expression for $\lambda_{it}$.
We will not state this explicitly here, but just note that $\lambda_{it}$ is a degree-$t$ polynomial in $i$,
and that $|\lambda_{it}|\leq 2^{-t/4}/{\cal N}$ for $i\leq n/8$.
Now consider the high-degree polynomial $p$ defined by
$$
p(i)=\sum_{t=0}^{n/4} a_t\lambda_{it}.
$$
This satisfies
$$
p(i)=\sum_{t=0}^{n/4} \Tr((U P U^T)_t)\lambda_{it}=\inpc{U P U^T}{U\mu_i U^T}=\inpc{P}{\mu_i}=P(i).
$$
Let $q$ be the degree-$d$ polynomial obtained by removing the high-degree parts of $p$:
$$
q(i)=\sum_{t=0}^d a_t\lambda_{it}.
$$
Then $P$ and $q$ are close on all integers $i$ between 0 and $n/8$:
$$
|P(i)-q(i)|=|p(i)-q(i)|=\left|\sum_{t=d+1}^{n/4} a_t\lambda_{it}\right|\leq
\frac{2^{-d/4}}{{\cal N}}\sum_{t=0}^{n/4}|a_t|\leq 2^{-d/4+2q}.
$$
\end{proof}

\subsection{The quantum lower bound for Intersection}

Now suppose we have a $q$-qubit protocol for the Intersection problem,
say with error probability at most $1/3$ on every input $x,y$.
Our goal is to show that $q$ is at least about $\sqrt{n}$.
Since the protocol outputs 1 with high probability if, and only if,
$x$ and $y$ intersect in at least one point, we know the following
about the quantity $P(i)=\Exp_{|x|=|y|=n/4, |x\wedge y|=i}[P(x,y)]$:
$P(0)\in[0,1/3]$ and $P(i)\in[2/3,1]$ if $i\in\{1,\ldots,n\}$.

This $P(i)$ is only defined on integers,
but by Lemma~\ref{lemrazborov} we can approximate it up to some small additive
error $\eps$ using a polynomial $q$ of degree $d=8q+\ceil{4\log(1/\eps)}$.
Then we know $q(0)\in[-\eps,1/3+\eps]$ and $q(i)\in[2/3-\eps,1+\eps]$.
However, the following result of Ehlich and Zeller~\cite{ehlich&zeller:schwankung}
and Rivlin and Cheney~\cite{rivlin&cheney:approx} says that such a polynomial $q$ must have degree about $\sqrt{n}$:

\begin{theorem}[Ehlich \&~Zeller; Rivlin \&~Cheney]\label{thehlichzeller}
Let $p: \mathbb{R}\rightarrow\mathbb{R}$ be a polynomial such that
$b_1\leq p(i)\leq b_2$ for every integer $0\leq i\leq N$, and
the derivative $p'$ satisfies $|p'(x)|\geq c$ for some real $0\leq x\leq N$.
Then the degree of $p$ is at least $\sqrt{cN/(c+b_2-b_1)}$.
\end{theorem}

It thus follows that the original protocol must have communicated at least about $\sqrt{n}$ qubits.
In his paper, Razborov gives essentially tight lower bounds not just for the Intersection problem,
but for any communication problem that depends only on the size of the intersection of the inputs $x$ and~$y$.
This combines Lemma~\ref{lemrazborov} with a polynomial degree lower bound due to Paturi~\cite{paturi:degree}.
The lower bound proof we gave here only applies to quantum protocols that do not start with an entangled state,
but Razborov showed the same lower bound for protocols with prior entanglement, at the expense of some more technical complication.
Recently, an alternative proof was obtained by Sherstov~\cite{sherstov:qcclower}.

\section{Asymmetric Detection Efficiency}\label{ADE}

Here we prove the results stated in Section~\ref{ADEsect} concerning
the connection between asymmetric experiments where a single detector is inefficient, and classical protocols with perfect detectors that use one-way communication, i.e., 
where all the communication takes place from Alice to Bob.

Let us suppose that in order to reproduce the quantum correlations using one-way communication from Alice to Bob and shared randomness, $c_{\epsilon'}$
 bits of communication are required to reproduce the correlations with error $\epsilon'$.  More precisely, the error is measured as the total variational distance between the predictions of quantum theory $P_{QM}(ab|xy)$ and the output $P_{class}(ab|xy)$ of the classical protocol:
$$
\mbox{error} =\max_{xy} \sum_{ab} |P_{class}(ab|xy) - P_{QM}(ab|xy)|
$$

 Let us also suppose that there exists a protocol that uses only shared randomness (a local hidden variable model) in which Alice's detector has efficiency $\eta_{\epsilon}$ and Bob's detector is perfect, and that reproduces the quantum correlations with error $\epsilon$. More precisely the fact that Alice's detector has efficiency $\eta$ means that
 $P(\perp b|xy)=\eta$ independently of $b,x,y$, where $\perp$ corresponds to Alice's detector not giving a result.
 The error is measured as the total variational distance between the predictions of quantum theory $P_{QM}(ab|xy)$ (when the detectors are $100\%$ efficient) and the predictions $P_{LHV}(ab|xy)$ of the LHV model. We divide the latter by $\eta$ to take into account that Alice's detector gives a result with probability $\eta$:
$$
\mbox{error} =\max_{xy} \sum_{ab} \left|\frac{P_{LHV}(ab|xy)}{\eta} - P_{QM}(ab|xy)\right|
$$
Then we have:

\begin{theorem}
With the above hypothesis, we have $\eta_\epsilon \leq O((-\ln \epsilon)2^{-c_{2\epsilon}})$.
\end{theorem}

To prove this, we use the local hidden variable model (LHV) model with detection efficiency $\eta_\epsilon$ to construct a classical protocol with communication.
The LHV uses shared randomness $r$.
 Alice and Bob share $k$ independently chosen instances of the shared randomness $r_1,r_2,\ldots,r_k$.
 Alice  checks whether she should give an output for at least one value of the shared randomness. This occurs with probability $1- (1-\eta)^k$. If so, she sends Bob the index $j$ of the shared randomness $r_j$ for which she gives an output (using $\log k$ bits of communication), and they give the corresponding output. If there is no instance of the shared randomness for which Alice should give an output in the LHV model, Alice gives a random output and sends Bob a random index $j$. This occurs with probability $(1- \eta)^k$, and in this case Alice and Bob's results will most likely be completely different from those predicted by quantum mechanics.
The error probability in the model with communication is thus
$P(error)\leq ( 1- (1-\eta)^k) \epsilon   + (1-\eta)^k \leq \epsilon + (1-\eta)^k $.
Let us take $k = \frac{\ln \epsilon}{\ln(1 - \eta)}$, then the error
is bounded by $P(error) \leq \epsilon + (1-\eta)^{\frac{\ln
    \epsilon}{\ln(1 - \eta)}} =  2\epsilon$.
But we know that to produce the correlations with error $ 2 \epsilon$ we need at least $c_{2 \epsilon}$ bits of one-way communication, hence $k \geq 2^{c_{2\epsilon}}$.
Therefore $-\ln (1 -\eta)\leq
(-\ln \epsilon)2^{-c_{2\epsilon}}$, which implies the result.

(Note that the above mapping does not hold when both Alice and Bob's detectors are inefficient, since if they try the above procedure, they will need to find a value of the shared randomness $r_j$ for which both their detectors produce an output, i.e., solve an instance of the Intersection problem.)

Let us apply this result to the Hidden Matching problem.  As
mentioned in Section~\ref{HMp}, this problem can be solved using $\log n$ ebits and $\log n$ bits of classical communication from Alice to Bob; but if  classical communication from Alice to Bob is considered, then at least $\Omega(\sqrt{n})$ bits of communication are required, even allowing for a small error probability. This implies that the correlations obtained by measuring the ebits can only be reproduced using at least $\Omega(\sqrt{n})$ bits of classical communication from Alice to Bob, even allowing for a small error probability. The above result then shows that these correlations remain non-local (i.e., cannot be reproduced by a classical model without communication) if Bob's detector has $100\%$ efficiency and Alice's detector has efficiency $\eta \geq 2^{-\Omega(\sqrt{n})}$, even allowing for a small error probability.

\end{document}